\documentclass[a4paper,12pt]{elsarticle}
\usepackage{amsmath,bm,graphicx}
\usepackage{subfigure}
\usepackage{amsfonts}
\usepackage{bm,epsfig}
\usepackage{amssymb}
\usepackage{color}
\usepackage{mathabx}

\newtheorem{lem}{Lemma}[section]
\newtheorem{thm}[lem]{Theorem}
\newtheorem{prop}[lem]{Proposition}
\newtheorem{cor}[lem]{Corollary}
\newtheorem{defi}[lem]{Definition}

\newtheorem{exa}[lem]{Example}
\newtheorem{rem}[lem]{Remark}

\begin{document}

\begin{frontmatter}

\title{Exact discrete resonances in the Fermi-Pasta-Ulam-Tsingou system}
\author{M. D. Bustamante}
\ead{miguel.bustamante@ucd.ie}
\author{K. Hutchinson}
\address{School of Mathematics and Statistics,
University College Dublin, Belfield, Dublin 4, Ireland}
\author{Y. V. Lvov}
\address{Department of Mathematical Sciences, Rensselaer Polytechnic Institute, Troy, New York 12180, USA}
\author{M. Onorato}
\address{Dipartimento di Fisica, Universit\`{a} di Torino, Via P. Giuria 1,  Torino, 10125, Italy}
\address{INFN, Sezione di Torino, Via P. Giuria 1, Torino, 10125, Italy}

\begin{abstract} 
In systems of 
$N$ coupled anharmonic oscillators, exact resonant interactions play an important 
role in the exchange of energy between normal modes. In the weakly nonlinear regime, 
those 
interactions may be responsible for 
the equipartition 
of energy in Fourier space. 
 Here we consider analytically resonant wave-wave
 interactions for the celebrated Fermi-Pasta-Ulam-Tsingou (FPUT) system. 
 Using a number-theoretical approach based on  \emph{cyclotomic polynomials}, we show that the problem of finding 
 exact resonances for a system of $N$ particles is equivalent to a Diophantine equation
 whose solutions depend sensitively on the number of particles $N$, in particular on the set of divisors of $N$.
 We provide
 an algorithm to construct {\it all} possible resonances for $N$
 particles, based on two basic methods: pairing-off and cyclotomic, which we introduce and use to build up explicit solutions to the $4$-wave, $5$-wave and $6$-wave resonant conditions. Our results shed some light in the understanding of the long-standing FPUT paradox, regarding the sensitivity of the resonant manifolds with respect to the number of particles $N$ and the corresponding time scale of the interactions leading to an eventual thermalisation. In this light we demonstrate that $6$-wave resonances always exist for any $N$, while $5$-wave resonances exist if $N$ is divisible by $3$ and $N \geq 9$. It is known that in the discrete case $4$-wave resonances do not produce energy mixing across the spectrum, so we investigate whether $5$-wave resonances can produce energy mixing  across a significant region of the Fourier spectrum by looking at the structure of the interconnected network of Fourier modes that can interact nonlinearly via resonances. We obtain that the answer depends on the set of odd divisors of $N$ which are not divisible by $3$: the size of this set determines the number of dynamically independent components, corresponding to independent constants of motion (energies). We show that $6$-wave resonances connect all these independent components, providing in principle a restoring mechanism for  full-scale thermalisation. 
   
\end{abstract}

\end{frontmatter}

\section{Introduction}

During the early fifties, Fermi, Pasta, Ulam and Tsingou were interested in understanding the role 
of a weak nonlinearity in the phenomenon of thermalisation and transport mechanisms exhibited
by real solids \cite{fermi1955studies}. They considered a simplified model 
 consisting  of $N$ identical masses connected by a nonlinear
spring; the elastic force between the masses is  expressed as a power series in the
spring deformation $\Delta x$:
\begin{equation}
F=-\kappa \Delta x+\alpha \Delta x^2 +\beta \Delta x^3 +..., \label{force}
\end{equation}
where $\kappa,\alpha$ and $\beta$ are elastic, spring dependent,
constants. One can recognise that for small displacements the classical Hooke's law is recovered.
They performed a number of 
numerical simulations with initial conditions characterised by energy concentrated 
on a single mode. They were expecting that, after some time, the nonlinearity would 
have played the key role of re-distributing the energy among different modes up to its 
 equipartition.
The results they obtained were somehow disappointing because, after some initial 
sharing of energy between a few modes, energy returned to the initial condition. The recurrence time observed 
has been named \emph{FPU recurrence}, after the work of E. Fermi, J. Pasta and S. Ulam \cite{fermi1955studies}
(recently it has been recognised that also Mary Tsingou took part to the research and sometimes the recurrence is named \emph{FPUT} \cite{dauxois2008fermi}).
This highly remarkable result underlines  the fact that nonlinearity is not sufficient for guaranteeing the phenomenon of equipartition. The results  \cite{fermi1955studies} were followed by a number of developments: in the sixties Zabusky and Kruskal studied the long wave limit of the FPUT chain and, using numerical simulations, they made the exceptional  discovery of solitons \cite{zabusky1965}, which was followed by the discovery of integrable partial differential equations. Those systems by definition do not reach a thermalisation state because the motion in phase space is confined on invariant tori. The knowledge of integrable systems did help only marginally to the understanding of the observed recurrent phenomena in the original FPUT system, the latter being a non integrable system.

Another attempt to explain the numerical results was based on the KAM theorem, presented by Kolmogorov more or less in the same years as the FPUT studies. Loosely speaking, under some specific assumptions, the theorem states that a slightly perturbed integrable Hamiltonian remain 
quasi-periodic. The KAM theorem appeared a natural tool to estimate the time scale of equipartition; however, to our knowledge, no quantitative prediction on the time scale has been made.  Its applicability to the FPUT system has been established only recently by \cite{rink2006proof}.
More precisely, it has been shown that the $\beta$-Fermi-Pasta-Ulam-Tsingou lattice with fixed endpoints and an arbitrary finite number of moving particles has a completely integrable Birkhoff normal form, which is an integrable approximation to the original Hamiltonian function. The result was extended to 
the $(\alpha+\beta)$-FPUT system in \cite{henrici2008results}.

In 1966 a remarkable result was obtained by Izrailev and Chirikov \cite{izrailev1966} who developed the concept of overlap of nonlinear resonances.  The idea behind it is 
that for small nonlinearity, the perturbation to the unperturbed Hamiltonian is small and
motion should be of quasi-periodic nature, as observed in the numerics by Fermi, Pasta, Ulam and Tsingou. However, by increasing the amplitude of the perturbation, one expects that stochasticity appears. Izrailev and Chirikov were able to estimate the limit of stochasticity. This theoretical prediction was successfully tested numerically in different 
papers (see for example \cite{livi1985equipartition} and more recent developments in 
\cite{lvov2018double}).  In \cite{cretegny1998localization}, the authors studied numerically the  $\beta$-FPUT system considering as initial condition the mode corresponding to the highest frequency and found that above an analytically derived energy threshold, such mode is shown to be modulationally unstable; the instability gives rise to the formation of chaotic breathers whose life time is related to the time for the system to reach equipartition. 

All the above scenarios  rely on the existence of an energy threshold above which 
a  thermalisation is observed. 
Recent numerical simulations (see for example \cite{benettin2013fermi,benettin2011time}) have shown that, even for very small nonlinearity, equipartition is reached. The evolution of the spectrum is characterised by two stages: a first one in which the spectrum assumes an exponential decay (this is called metastable state, after \cite{fucito1982approach}) and the second one characterised by the observation of thermalisation. Recently, in \cite{onorato2015route} and \cite{lvov2018double} the wave turbulence approach to the  $\alpha$-FPUT and $\beta$-FPUT  system  was considered and the problem with $N$=16, 32 and 64 masses was addressed. The main idea under such approach is that, in the weakly nonlinear limit, exact resonant interactions are responsible for an irreversible transfer of energy which leads to equipartition (similar results have been obtained for the discrete nonlinear Klein-Gordon equation \cite{pistone2018thermalization}).  By exact resonant interactions between $M$ waves we mean that a wave scattering process characterised by the following conservation laws (momentum, $k$, and energy, $\omega(k)$) 
\begin{equation}
\begin{split}
&k_1\pm k_2\pm...\pm k_M=0\\
&\omega(k_1)\pm \omega(k_2)\pm... \pm \omega(k_M)=0 \label{res_cond}
\end{split}
\end{equation}
is possible. For the specific case of a discrete lattice, wave numbers $k_i$ are integer values and the equality in the first equation should be 
considered in modulo $N$. In  \cite{lvov2018double} and \cite{onorato2015route}  it was shown that in the $\beta$- and $\alpha$-FPUT systems with $N= 32$ the dominant interaction is a six-wave one, i.e. $M=6$.  Based on the kinetic description of waves \cite{falkovich1992kolmogorov},  a nonlinear time scale is associated to each resonant process; consequently, it is possible to estimate the time scale associated with the equipartition. 
In \cite{onorato2015route} it has been also conjectured  that the dominant interaction 
problem is strongly dependent on the number of masses, $N$, of the chain.
Therefore, finding the value of $M$ for a given number of particles $N$ is of fundamental importance for understanding the relaxation dynamics. However, establishing the lowest value of $M$ at which the system of equations  (\ref{res_cond}) has solutions is not an easy task. In 
\cite{onorato2015route} the existence of the solutions and the solution themselves were found numerically; the numerical approach is feasible when the number of particles is small and soon becomes prohibitive as the number of particles increases.

In this paper we attack the problem from an analytical point of view and show that the problem of finding $M$-wave resonances reduces to the solution of  a set of Diophantine equations for which we develop an algorithm for finding the solutions.
The paper is organised as follows: Section \ref{sec:model} is devoted to an introduction to the $(\alpha+\beta)$- FPUT physical model in the so called {\it wave action} variables;
we also give an overview of the wave-wave interaction theory in the weakly-nonlinear limit. In Section
\ref{sec:resint}, the general problem of finding solutions to the resonance conditions (\ref{res_cond}) is presented and discussed: basic definitions of irreducible and composite resonances, elementary results such as forbidden processes ($3$-wave resonances as well as processes converting $M-1$ to $1$ waves, for any $M \geq 3$), and a preliminary discussion advocating that in the weakly-nonlinear limit it is necessary and sufficient to search for $M$-wave resonances for $4\leq M \leq 6$. In Section \ref{sec:pairing-off} we introduce the \emph{pairing-off method}, which provides the simplest type of solutions to the resonance conditions (\ref{res_cond}), valid for $2S$-wave resonances. This method produces the well-known $4$-wave resonances, and  generalises the recently found $6$-wave resonances for $N=2^b$ to the case of $N$ arbitrary. The pairing-off method turns out to be an essential part of the more complex methods discussed in the rest of the paper. In Section \ref{sec:need_for_general} we show, with a simple example of $6$-wave interactions in the case when the number of particles is $N=6$, that not all solutions that exist are in pairing-off form, which serves to motivate our introduction of the  \emph{cyclotomic method} in Sections \ref{sec:RealPolynomials} and 
 \ref{sec:cyclotomic_method}. The main result of Section \ref{sec:RealPolynomials} is to simplify enormously the resonance conditions (\ref{res_cond}) in terms of polynomials in one complex variable $P(x)$ with integer coefficients, which evaluate to real values when $x$ is equal to the primitive $2N$-th root of unity $\zeta = \exp(i \pi/N)$, where $N$ is the number of particles. The resonance conditions reduce to a simple linear Diophantine equation for a set of unknown integers, of easy solution. Sections \ref{sec:cyclotomic_method} and \ref{sec:5-wave-resonances} provide for the first time in the FPUT literature a method to construct $5$-wave resonances. They describe in detail the systematic construction of $5$-wave resonances (also known as resonant quintets), a construction requiring $N > 6$ and $3 \divides N$ ($3$ divides $N$, i.e., $N$ is divisible by $3$). A thorough analysis of the clustering of resonant quintets via common-mode connections is presented, crucially identifying a large number of highly symmetrical clusters with octahedron topology (termed octahedron clusters). Moreover, we identify the set of all wavenumbers involved in some $5$-wave resonant interaction.
 In Section \ref{sec:super-clusters} we proceed to deepen the analysis of these octahedron clusters by analysing the inter-cluster connectivity via common modes across different octahedron clusters, a crucial aspect that provides the basis for the ideas of \emph{ thermalisation} due to nonlinear interactions in the FPUT Hamiltonian system. This defines the \emph{super-clusters}, namely the graphs whose vertices are the octahedron clusters and whose edges represent all possible  inter-cluster connections via common modes. The striking result is that the super-clusters are a disjoint union of connected subgraphs, which means that the universe of available modes separates into independent components that do not interact with each other. Via a couple of theorems we completely classify these components in terms of the set of divisors $d$ of $N$ such that $2\ndivides d$ and $3 \ndivides d$. We analyse the super-clusters for two examples: $N=3\cdot 5^2 = 75$ and $N=2^2 \cdot 3\cdot5\cdot7=420$. In Section \ref{sec:analysis} we discuss the implications of the super-cluster decomposition into independent pieces in regimes dominated by $5$-wave resonant interactions, regarding the hypotheses of  thermalisation. We show that, for general $N$, $6$-wave interactions must be included in order to achieve a full-scale thermalisation. We conclude that the idea of component-wise thermalisation via $5$-wave interactions (faster than $6$-wave interactions) is interesting and deserves attention.
 Finally in Section \ref{sec:underpinnings} we provide the solid mathematical underpinnings for the explicit construction of the general solution to the resonance conditions (\ref{res_cond}), in terms of rings and groups of polynomials with integer coefficients. We conclude with some perspectives and discussions in Section \ref{sec:conclusions}.

 \section{ Background on the $(\alpha$+$\beta)$-FPUT model\label{sec:model}}

The Hamiltonian for a chain of $N$ identical particles of mass
$m$, subject to a force of the type in (\ref{force}) can be expressed as an unperturbed 
Hamiltonian, $H_0$, plus two perturbative terms, $H_3$, $H_4$:
 \begin{equation}
H=H_0+H_3+H_4
\end{equation}
with 
\begin{equation}
\begin{split}
&H_0=\sum_{j=1}^N\left(\frac{1}{2 m}p_j^2+\kappa\frac{1}{2}(q_j-q_{j+1})^2\right),\\
&H_3=\frac{\alpha}{3}\sum_{j=1}^N(q_j-q_{j+1})^3,\\
&H_4=\frac{\beta}{4}\sum_{j=1}^N(q_j-q_{j+1})^4.
\label{H_FPU}
\end{split}
\end{equation}
$q_j(t)$ is the displacement of the particle $j$ from its
equilibrium position and $p_j(t)$ is the associated momentum.
The canonical equations of motion are:
\begin{equation}
\begin{split}
&\dot{q}_j=\frac{\partial H}{\partial p_j}=\frac{p_j}{m},\label{dyn_pq_1}\\
&\dot{p}_j=-\frac{\partial H}{\partial q_j}=\kappa(q_{j-1}-2q_j+q_{j+1})+
\alpha\big[(q_{j+1}-q_j)^2-(q_j-q_{j-1})^2\big]+\\
&+\beta\big[(q_{j+1}-q_j)^3-(q_j-q_{j-1})^3\big]
\end{split}
\end{equation}
with $j=0,1,..,N-1$. 
The  Newton's equation reduces to:
\begin{equation}
m \ddot{q}_j=\kappa(q_{j+1}+q_{j-1}-2q_j)
+\alpha\big[(q_{j+1}-q_j)^2-(q_j-q_{j-1})^2\big]+
\beta\big[(q_{j+1}-q_j)^3-(q_j-q_{j-1})^3\big]
\label{eq:FPUalfabeta}
\end{equation}
This is known as the $\alpha+\beta$-FPUT model.
\subsection{Equation in Fourier space}
Our approach is developed in Fourier space (periodic boundary conditions are assumed, i.e. $q_{N}=q_{0}$) and the
following definitions of the direct and inverse Discrete Fourier
Transform are adopted:
\begin{equation}
Q_k=\frac{1}{N}\sum_{j=0}^{N-1} q_j e^{-i 2\pi k j/N},\;q_j=\sum_{k=0}^{N-1} Q_k e^{ i 2\pi j k/N}, \label{DFT}
\end{equation}
where $Q_k$ are the Fourier amplitudes of the displacement; similarly $P_k$ can be introduced as the Fourier transform of the momentum $p_j$. In terms of degrees of freedom, it will turn out that $Q_0$ is a real and ignorable coordinate, thus $P_0$ (also real) is a constant of motion. Therefore, there are $N-1$ interacting degrees of freedom (or $2N-2$ if we count $P$ and $Q$ independently). From here on we will use the convention that the wavenumbers $k$ parameterizing these degrees of freedom are positive: $1\leq k \leq N-1$. As we will see later, this convention is very useful when dealing with the frequency resonance conditions. Notice that, in this convention, the reality of $q_j$ and $p_j$ is equivalent to the conditions 
\begin{equation}
\label{eq:reality_conditions}
Q_{N-k} = Q_k^*\,, \qquad P_{N-k} = P_k^*\,,\qquad k=1, \ldots, N-1\,.
\end{equation}
The Poisson bracket changes from the canonical one $[q_j ,p_{j'}] = \delta_{j j'}$ to:
$$[Q_k,P_{k'}^*] = \frac{1}{N^2}\sum_{j=0}^{N-1} e^{-i 2\pi k j/N}\sum_{j'=0}^{N-1} e^{i 2\pi k' j'/N} [q_j ,p_{j'}] = \frac{1}{N^2}\sum_{j=0}^{N-1}  e^{i 2\pi (k'-k) j/N}  = \frac{\delta_{k k'}}{N}.$$

Inserting the definition of the Fourier transform in the  Hamiltonian we get:
\begin{equation}
\begin{split}
\nonumber
&\frac{H}{N}=\frac{P_0^2}{2m} + \frac{1}{2m}\sum_{k=1}^{N-1}\left({|P_k|^2}+m^2\,\omega_k^2\,|Q_k|^2\right)+
\frac{1}{3}\sum_{k_1,k_2,k_3=1}^{N-1} \tilde V_{1,2,3}Q_1Q_2Q_3 \delta_{1+2+3}+\\
&+\frac{1}{4}\sum_{k_1,k_2,k_3,k_4=1}^{N-1}  \tilde T_{1,2,3,4}Q_1Q_2Q_3Q_4 \delta_{1+2+3+4},
\label{hamQ}
\end{split}
\end{equation}
where we define the dispersion relation:
\begin{equation}
\label{Dispersion}
\omega_k=\omega(k)={2\sqrt{\frac{\kappa}{m}}\sin(\pi k/{N})}\,,\quad 1 \leq k \leq N-1
\end{equation}
(notice that $\omega_k$ is positive over our conventional range of $k$ values), and we have introduced, in the triple and quadruple sums, the notation:
$Q_i=Q(k_i,t)$, $\delta_{1+2+3}=\delta(k_1+k_2+k_3 \mod N)$, 
which is the Kronecker $\delta$, taking value 1 when the modular-arithmetic condition $k_1+k_2+k_3=0  \mod N$ is satisfied. This condition arises from the spatial discreteness of the original problem. Similarly,  
$\delta_{1+2+3+4} = \delta(k_1+k_2+k_3 +k_4\mod N)$. Finally, $\tilde T_{1,2,3,4}=\tilde T(k_1,k_2,k_3,k_4)$
and $\tilde V_{1,2,3}=\tilde V(k_1,k_2,k_3)$ are two matrices that weigh the nonlinear interactions:
\begin{eqnarray}
\tilde V_{1,2,3}&=&8 i \alpha \mathrm{e}^{i \pi (k_1+k_2+k_3)/N} \sin\frac{\pi k_1}{N} \,\sin\frac{\pi k_2}{N} \,\sin\frac{\pi k_3}{N}\,,\\
\tilde T_{1,2,3,4}&=&{16\beta} \mathrm{e}^{i \pi (k_1+k_2+k_3+k_4)/N} \sin\frac{\pi k_1}{N} \,\sin\frac{\pi k_2}{N} \,\sin\frac{\pi k_3}{N} \,\sin\frac{\pi k_4}{N}.
\end{eqnarray}
The equation of motion takes the following form:
\begin{equation}
\ddot Q_1+\omega_1^2 Q_1=
\frac{1}{m}\sum_{k_2,k_3 } \tilde V_{1,2,3}Q_{2}Q_{3}\delta_{1+2+3} 
+ \frac{1}{m}\sum_{k_2,k_3,k_4}  \tilde T_{1,2,3,4}Q_{2}Q_{3}Q_{4}\delta_{1+2+3+4}, \label{eq:Qequ}
\end{equation}
where all the sums on $k_j$ goes from 1 to $N-1$.
\subsection{Normal modes}
We introduce the normal variable (or wave action variable) $a_k=a(k,t)$ via:
\begin{equation}
\label{NormalMode}
a_k=\frac{1}{\sqrt{2 m \omega_k}}(m\omega_k Q_k+ i P_k),\quad
a_{N-k}^*=\frac{1}{\sqrt{2 m \omega_k}}(m\omega_k Q_k- i P_k),\quad k=1,\ldots,N-1
\end{equation}
%
%
where we remark again that  $\omega_k$ is positive for the selected values of $k$.  
Poisson brackets are now:
$$[a_k,a_{k'}^*] = \frac{1}{{2 m \omega_k}}[(m\omega_k Q_k+ i P_k),(m\omega_{k'} Q_{k'}^*- i P_{k'}^*)] = - \frac{i}{N} \delta_{k k'}\,.$$
The above relations are easy to invert to obtain:
\begin{equation}
\label{inNormalMode}
Q_k=\frac{1}{\sqrt{2 m\omega_k}}(a_k + a_{N-k}^*)
\end{equation}
which can be used directly in the Hamiltonian (\ref{hamQ}) to get (we discard the $P_0$ term from here on):
 \begin{equation}
 \begin{split}
&\frac{H}{N}=\sum_{k=1}^{N-1}\omega_k|a_k|^2+
 \sum_{k_1,k_2,k_3} V_{1,2,3}\left[\frac{1}{3}(a_1a_2 a_3+ c.c.) \delta_{1+2+3}+
(a_1^*a_2 a_3+ c.c.) \delta_{1-2-3}\right]+ \\
&+ \sum_{k_1,k_2,k_3,k_4}  T_{1,2,3,4}
 \big[
(a_1^*a_2 a_3 a_4+c.c.) \delta_{1-2-3-4}+
\frac{3}{2}a_1^*a_2^* a_3 a_4 \delta_{1+2-3-4}+
\\
&+\frac{1}{4}( a_1a_2 a_3 a_4+c.c. )\delta_{1+2+3+4}
\big],
\end{split}
\end{equation}
with
\begin{equation}
\begin{split}
&V_{1,2,3}=\frac{\tilde V_{1,2,3}}{2m^{3/2} 
 \sqrt{2m\omega_{k_1}\omega_{k_2}\omega_{k_3}}} = \frac{i \alpha \mathrm{e}^{i \pi (k_1+k_2+k_3)/N}}
 {{(m \kappa)}^{3/4}}\sqrt{ \sin\frac{\pi k_1}{N} \,\sin\frac{\pi k_2}{N} \,\sin\frac{\pi k_3}{N}} \,,\\
 & T_{1,2,3,4}=\frac{\tilde T_{1,2,3,4}}{4m^2\sqrt{\omega_{k_1}\omega_{k_2}\omega_{k_3}\omega_{k_4}}} = \frac{\beta \mathrm{e}^{i \pi (k_1+k_2+k_3+k_4)/N}}{m \kappa} \sqrt{ \sin\frac{\pi k_1}{N} \,\sin\frac{\pi k_2}{N} \,\sin\frac{\pi k_3}{N}\,\sin\frac{\pi k_4}{N}}.
\end{split}
\end{equation}
The  equation of motion becomes
\begin{equation}
\begin{split}
&i\frac{\partial a_1}{\partial t}=\frac{1}{N}\frac{\delta H}{\delta a_1^*}=\omega_{k_1} a_{1}+
\sum_{k_2,k_3} V_{1,2,3}(a_2a_3 \delta_{1-2-3}+2 a_2^*a_3  \delta_{1+2-3}+
a_2^*a_3^*\delta_{1+2+3})+\\
&+\sum_{k_2,k_3,k_4} T_{1,2,3,4}(
a_2a_3a_4\delta_{1-2-3-4} +
3a_2^*a_3a_4\delta_{1+2-3-4}+
+3a_2^*a_3^*a_4\delta_{1+2+3-4}+
\\&
+a_2^*a_3^*a_4^*\delta_{1+2+3+4})
 \label{eq:fourwaveint}
\end{split}
\end{equation}
We emphasise key aspects of this equation:

\begin{enumerate}
\item It is equivalent to equation (\ref{eq:FPUalfabeta}). The degrees of freedom are the complex modes $a_k, a_k^*, \quad k=1,\ldots, N-1$, so a total of $2N -2$ degrees of freedom, just like originally --after separating the conserved quantity $P_0$ and its conjugate variable $Q_0$.  
\item While it might not be suitable for numerical computations, it is a very good starting point for developing theoretical approaches. In particular, all the possible mechanisms 
 of interaction are highlighted by the presence of the Kronecker $\delta$. 
\end{enumerate}
 
\subsection{Canonical Transformation and reduced Hamiltonian}
Equation (\ref{eq:fourwaveint}) contains three and four wave interactions which are related 
to the quadratic and cubic nonlinearity contained in the Newton's equation.
 If one is interested in the long time behavior of
 the system, then it can be argued that some interactions play a more important role; indeed, nonlinearity in Fourier space 
results in a intrinsic forcing which is efficient if its frequency matches perfectly the 
 frequency of oscillation of the forced wave. In terms of our equation of motion,
  the condition on Kronecker $\delta$ over wave numbers should be accompanied 
 by an analogue condition in frequencies. For example, considering the last 
 nonlinear term in  the first line of equation (\ref{eq:fourwaveint}),  the resonant condition would imply that:
\begin{equation}
\begin{split}
 &k_1+k_2+k_3=0 \pmod{N} \\
 &\omega_1+\omega_2+\omega_3=0,
 \end{split}
 \end{equation}
 Three wavenumbers with $1\leq k_j\leq N-1$ can add up to $N$ or $2N$, therefore it is always possible to satisfy the first equation; however, 
 it is obvious that, because frequencies are always positive, the second
 condition cannot be satisfied. Moreover, it is known \cite{rink2001symmetry} (see also Theorem \ref{thm:forbidden} below) that all $3$-wave processes are
 {\it forbidden}. 
Also, one can show that the leading non-trivial resonant process in equation 
(\ref{eq:fourwaveint}) is the following:
 \begin{equation}
\begin{split}
& k_1+k_2-k_3-k_4=0 \pmod{N}\\
& \omega_1+\omega_2-\omega_3-\omega_4=0.
 \end{split}
 \end{equation}
 This known result (see \cite{rink2001symmetry}), which we will prove again in Section \ref{sec:forbidden}, implies that 
 all other terms can be removed from the Hamiltonian. Following the work
 \cite{krasitskii1994reduced}, a suitable 
 canonical transformation can be used to the purpose. The idea is to impose the 
 following near identity transformation:
 \begin{equation}
 \begin{split}
 &a_1=b_1+
 \sum_{k_2,k_3} \left(A_{1,2,3}^{(1)} b_2 b_3\delta_{1-2-3}+A_{1,2,3}^{(2)} b_2^* b_3\delta_{1+2-3}+
 A_{1,2,3}^{(3)} b_2^* b_3^*\delta_{1+2+3}\right)+
 \\&
 +\sum_{k_2,k_3,k_4} (B_{1,2,3,4}^{(1)} b_2 b_3b_4\delta_{1-2-3-4}+
 B_{1,2,3,4}^{(2)} b_2^* b_3b_4\delta_{1+2-3-4}+
 \\&
 B_{1,2,3,4}^{(3)} b_2^* b_3^* b_4\delta_{1+2+3-4}+
 B_{1,2,3,4}^{(4)} b_2^* b_3^* b_4^*\delta_{1+2+3+4})
 \end{split}
 \end{equation} 
 and select the matrices $A_{1,2,3}^{(i)}$, $B_{1,2,3,4}^{(i)}$ in order to remove the unwanted interactions. Note that the 
 transformation is asymptotic in the sense that it is formally valid only for weak nonlinearity.
The algebra is quite lengthy and we refer to   \cite{krasitskii1994reduced}
and \cite{janssen2009some} for the details. As an example, the coefficients $A_{1,2,3}^{(i)}$ 
turn out to be 
given by 
\begin{equation}
A_{1,2,3}^{(1)}=\frac{V_{1,2,3}}{\omega_3+\omega_2-\omega_1},\;
A_{1,2,3}^{(2)}=\frac{2V_{1,2,3}}{ \omega_3-\omega_2-\omega_1},\;
A_{1,2,3}^{(3)}=\frac{V_{1,2,3}}{-\omega_3-\omega_2-\omega_1}.
\end{equation}
It is interesting to underline the presence of the frequency conditions at the denominator.
 If  these were satisfied, then the canonical transformation would diverge; consequently,
only nonresonant terms can be removed (this is the well known problem of 
small divisors).
After removing the nonresonant terms, the Hamiltonian reduces to:
 \begin{equation}
 \begin{split}
& \frac{H}{N}=\sum_{k=1}^{N-1}\omega_k|b_k|^2+\frac{1}{2} \sum_{k_1,k_ 2, k_3,k_4} \bar T_{1,2,3,4}b_1^*b_2^* b_3 b_4
\delta_{1+2-3-4} + \textrm{higher-order terms} 
\label{hamzak}
\end{split}
\end{equation}
and the equation of motion becomes
\begin{equation}
\begin{split}
i\frac{\partial b_1}{\partial t}=\omega_1 b_1 + \sum_{k_1,k_ 2, k_3,k_4}  \bar T_{1,2,3,4}
b_2^*b_3b_4\delta_{1+2-3-4} + \textrm{higher-order terms.}
\label{eq:zakh}
\end{split}
\end{equation}
This is known as the {\it Zakharov equation}. While removing the nonresonant terms, the canonical transformation has also  the property of naturally introducing  higher order wave-wave interaction terms which have been omitted in  equation (\ref{eq:zakh}). In Section \ref{sec:resint} it will be shown that the $4$-wave interactions are not sufficient for 
spreading energy across the spectrum (see also \cite{henrici2008results,onorato2015route}) and,
depending on the number of interacting particles, either the $5$-wave or
the $6$-wave resonant interactions become the dominant interaction mechanisms.
We remark that the removed $3$-wave interactions do not disappear from the system without leaving trace: a memory of them is contained in the new matrix $\bar T_{1,2,3,4}$, whose analytical form is given in 
 \ref{app:int}. Moreover, most of the $5$-wave resonances, which exist when $N$ is divisible by $3$ and greater than $6$, can be constructed starting from the $3$-wave frequency resonances. We mention that the approach used to reach equation (\ref{eq:zakh}) is analogous to  
the one in the problem of surface gravity waves in infinite water depth, where, despite the fact that the 
system is continuous and the dispersion relation is different, the lowest-order resonant
 interaction is the four-wave one.
 
 It is interesting to note that it is possible to show that the Hamiltonian in  (\ref{hamzak})
 is completely integrable if higher order terms are neglected. This important result was achieved  for the $\alpha+\beta$-FPUT in \cite{henrici2008results}. More specifically,
  it was shown that  any periodic FPUT chain with an odd number $N$ of particles admits a {\it Birkhoff normal} form up to order 4, whereas any periodic FPUT chain with $N$ even admits a {\it resonant normal} form up to order 4. This resonant normal form of order 4 turns out to be completely integrable.

\section{Exact M-wave resonant interactions: Statement of the problem} \label{sec:resint}

\subsection{Basic definitions}

As mentioned an important energy transfer occurs between modes whose wavenumbers and frequencies satisfy  exact resonance conditions. Here we rephrase the problem in a more
precise way, suitable for analytical treatment.
Consider $N$ particles and $M$ wavenumbers, $k_1,k_2, \ldots, k_M$, which are integers satisfying $1 \leq k_j \leq N-1$ for all $j = 1, \ldots, M.$ 
In principle there is no limit on the number of interacting waves $M,$ because this is determined by
 the order of the nonlinear interaction appearing in the extra terms of equation (\ref{eq:zakh}). If $M \geq N$ then the corresponding nonlinear interaction is self-interacting for some mode(s), simply because there are only $N-1$ different wavenumbers and we are picking $M$ of them to construct the resonance. In practical applications, however, $M=5, 6$ will be the most relevant cases regarding thermalisation, with $N$ arbitrary.  
For an $M$-wave nonlinear interaction, wavenumbers satisfy the momentum conservation condition
\begin{equation}
\label{eq:M-wave_k}
k_1 + \ldots + k_S = k_{S+1} + \ldots + k_{S+T} \pmod{N}, 
\end{equation}
where $S + T = M$ and the symbol ``${\rm mod}\;N$'' means equality modulo $N$. This process converts  $S$ waves into $T$ waves.

As previously mentioned, we are interested in the long time behaviour of a weakly nonlinear FPUT chain whose 
dynamics is ruled by exact resonant interactions. Therefore, the condition on the momentum,
equation  (\ref{eq:M-wave_k}),  is accompanied by the following frequency condition:
\begin{equation}
\label{eq:M-wave_omega}
\omega(k_1) + \ldots + \omega(k_S) = \omega(k_{S+1}) + \ldots + \omega(k_{S+T})\,,
\end{equation}
We recall that in the interval  $1 \leq k_j \leq N-1$ the dispersion relation is positive and it reduces to 
\begin{equation}
\label{eq:dispersion_dimensionless}
\omega(k)=2 \sin(\pi k/N)\,,
\end{equation}
where we have set for simplicity $m=\kappa=1$.

Equations (\ref{eq:M-wave_k}), (\ref{eq:M-wave_omega}) and (\ref{eq:dispersion_dimensionless}) with $N, S, T$ natural numbers and $M = S + T,$ determine the general problem of finding $M$-wave exact resonances. The unknowns are the integers $k_1, \ldots, k_M,$ satisfying $1 \leq k_j \leq N-1$ for all $j = 1, \ldots, M.$ The equations are thus a system of Diophantine equations. The rest of this paper is dedicated to finding both general and algorithmic solutions to this problem.  
Note that the results are valid for $\alpha$, $\beta$, $(\alpha+\beta)$ or any other system with linear dispersion relation given by $\omega(k)=2 \sin(\pi k/N)$.

\begin{defi}[$M$-wave resonance]
\label{defi:M-wave resonance}
Let $N$ be the number of particles of the FPUT system. An $M$-wave resonance is a list (i.e., a multi-set) $\{k_1, \ldots, k_S; k_{S+1}, \ldots, k_{S+T}\}$ with $S,T>0$, $S+T=M$ and $1\leq k_j \leq N-1$ for all $j =1, \ldots, M$, that is a solution of the momentum conservation and frequency resonance conditions
\begin{equation}
\label{eq:M-wave_k_omega}
\begin{split}
k_1 + \ldots + k_S &=k_{S+1} + \ldots + k_{S+T} \pmod N,\\
\omega(k_1) + \ldots + \omega(k_S) &= \omega(k_{S+1}) + \ldots + \omega(k_{S+T}),
\end{split}
\end{equation}
where $\omega(k) = 2 \sin (\pi k/N)$.
\end{defi}

\begin{rem}
Notice that this definition contains trivial resonances such as the $2$-wave resonances $k=k$, $\omega(k)=\omega(k)$ valid for any wavenumber $k$, which do not play a dynamical role on their own, although they do play a dynamical role in the composite or reducible resonances defined below. 
\end{rem}

\begin{rem}
Any permutation of the first $S$ wavenumbers and any permutation of the last $T$ wavenumbers lead to the same physical resonance. Thus in many applications we will select one representative, determined by: $k_1 \leq \ldots \leq k_S$ and, when possible, either $k_{S+1} \geq \ldots \geq k_{S+T}$ or $k_{S+1} \leq \ldots \leq k_{S+T}$.
\end{rem}

\begin{rem}
\label{rem:StoT}
Also, interchanging the first list (of $S$ wavenumbers) with the second list (of $T$ wavenumbers) leads to the same physical resonance. To deal with this symmetry we will impose, for example, $S\geq T$ when possible.
\end{rem}

\begin{defi}[Irreducible $M$-wave resonance]
An irreducible $M$-wave resonance is an $M$-wave resonance which does not contain sub-lists that are $M'$-wave resonances, for any $M'<M$. In other words, there are no integers $S',T'$ with $0<S'<S$ and $0<T'<T$ such that
\begin{equation}
\begin{split}
k_1 + \ldots + k_{S'} &= k_{S'+1} + \ldots + k_{S'+T'} \pmod N,\\
\omega(k_1) + \ldots + \omega(k_{S'}) &= \omega(k_{S'+1}) + \ldots + \omega(k_{S'+T'}),
\end{split}
\end{equation}
where $\omega(k) = 2 \sin (\pi k/N)$.
\end{defi}

\begin{defi}[Composite or reducible $M$-wave resonance]
An $M$-wave resonance is called composite or reducible if and only if it is not irreducible.
\end{defi}

\begin{rem}
Given an $M'$-wave resonance and an $M''$-wave resonance, it is easy to construct a composite $(M'+M'')$-wave resonance: let $\{k_1, \ldots, k_{S'}; k_{S'+1}, \ldots, k_{S'+T'}\}$ be an $M'$-wave resonance, and let $\{\ell_1, \ldots, \ell_{S''}; \ell_{S''+1}, \ldots, \ell_{S''+T''}\}$ be an $M''$-wave resonance. Then the following is a composite $(M'+M'')$-wave resonance:
$$\{k_1, \ldots, k_{S'},\ell_1, \ldots, \ell_{S''}; k_{S'+1}, \ldots, k_{S'+T'},\ell_{S''+1}, \ldots, \ell_{S''+T''}\}\,.$$
As a particular case consider $M''=2$, i.e. the trivial irreducible $2$-wave resonance $\{k;k\}$, for any choice of $k$. This gives the composite $(M'+2)$-wave resonance $\{k_1, \ldots, k_{S'},k; k_{S'+1}, \ldots, k_{S'+T'},k\}$. In general, any number of irreducible resonances can be composed in order to generate a composite resonance, since both the momentum conservation equation and the frequency resonance condition are solved by sub-sums.
\end{rem}

From here on we will focus mainly on irreducible $M$-wave resonant interactions.

\subsection{Forbidden $M$-wave processes}
\label{sec:forbidden}

It is well known that $3$-wave resonances do not exist. In \cite{rink2001symmetry} a proof is provided using number theory. Also, an explicit result from the same reference can be used to infer that $4$-wave resonances must transform $2$ waves into $2$ waves (i.e. in our notation $S=T=2$ is necessary).

For any number of particles $N$ we provide a simpler proof of these statements, and a generalisation of these to the case when the number of interacting waves $M$ is arbitrary, by using the subadditivity of the dispersion relation.

\begin{thm}
\label{thm:forbidden}
Resonant processes converting $1$ wave to $M-1$ waves or $M-1$ waves to $1$ wave do not exist, for any $M \neq 2$. Also, resonant processes converting $0$ wave to $M$ waves or $M$ waves to $0$ wave do not exist, for any $M>0$.
\end{thm}
\emph{Proof.}
The function $\omega(k) = 2 |\sin(\frac{\pi k}{N})|$ is strictly subadditive for $k \in \mathbb{R},$ $k \notin  N \mathbb{Z}$: 
$$\omega(k_1+k_2) < \omega(k_1)+\omega(k_2), \quad k_1,k_2  \in \mathbb{R} \setminus  N \mathbb{Z}$$
(see \cite{laatsch1964extensions,bruckner1960minimal,matkowski2011subadditive}). Therefore, for example, resonant processes converting $2$ waves into $1$ wave (or vice versa) are not allowed because this would require $\omega(k_1+k_2) = \omega(k_1)+ \omega(k_2),$ which is not possible. 
Similarly, resonant processes converting $M-1$ waves into $1$ wave (or vice versa) are not allowed because subadditivity implies
$$\omega(k_1+\ldots+k_p) < \omega(k_1)+ \ldots + \omega(k_p), \quad k_1,\ldots, k_p  \in \mathbb{R} \setminus  N \mathbb{Z},$$
for any $p \geq 2.$

Also, for  any number of particles $N$, resonant processes converting $0$ wave to $M$ waves or $M$ waves to $0$ wave do not exist. This is because the function $\omega(k) = 2 |\sin(\frac{\pi k}{N})|$ is positive for $k \in \mathbb{R} \setminus N \mathbb{Z}.$ \qed\\

The case $M=2$ converting $1$ wave to $1$ wave is not covered by this theorem because there is a resonance, of only one type: the so-called `trivial' resonance, given by $k=k$ and $\omega(k) = \omega(k)$, for any wavenumber $k$. This trivial resonance does not play any dynamical role unless it is a part of a composite resonance.

\begin{rem} As $M$ increases, the number of allowed (i.e., not forbidden by the above analysis) $M$-wave resonances increases. For example, $5$-wave resonances can convert $2 \leftrightarrow 3$ waves; $6$-wave resonances can convert $2 \leftrightarrow 4$ and $3 \leftrightarrow 3$ waves; $7$-wave resonances can convert $2 \leftrightarrow 5$ and $3 \leftrightarrow 4$ waves, and so on.
\end{rem}

\subsection{What is new about $4$-wave, $5$-wave and $6$-wave resonances}

In the next Sections we will introduce two new methods to construct $M$-wave resonances: pairing-off method and cyclotomic method.  The appropriate method to be used depends on properties of the number of particles $N$ and the number of waves $M$. The case of $4$-wave resonances has been studied extensively and all solutions are known \cite{rink2001symmetry,rink2006proof,henrici2008results,onorato2015route,rink2001symmetry}. A crucial aspect of $4$-wave resonances is that they are integrable and thus they do not produce energy mixing across the Fourier spectrum: one needs to go to higher orders. The case of $5$-wave resonances is completely new and relies on the existence of cyclotomic polynomials (to be defined below). The case of $6$-wave resonances is also new and both methods (pairing-off and cyclotomic) are used to construct them. It is worth mentioning that $6$-wave resonances were found recently in the particular case when $N$ is a power of two \cite{onorato2015route}. One important corollary of the constructions below will be that, for any $N \geq 4$, there exist $6$-wave resonances. Therefore, we do not need to search for $M$-wave resonances with $M > 6$ because these will provide less relevant corrections to the system's behaviour, in the weakly-nonlinear limit that we are considering. 

\section{Pairing-off method to obtain $2S$-wave resonances converting $S$ waves to $S$ waves}
\label{sec:pairing-off}

The pairing-off method to be introduced in this Section produces the simplest type of resonances in the FPUT system: simple in its statement, its explicit construction, and its applicability to an arbitrary number of particles $N$. In addition to that, it plays a fundamental role in the more complex types of resonances discussed throughout the paper. For example, one of the rigorous results proven in Section \ref{sec:underpinnings}, Corollary \ref{cor:triv}, states that when the number of particles $N$ is an odd prime or a power of $2$, any $M$-wave resonance must be of pairing-off form, and in particular $M$ must be even. For arbitrary $N$, pairing-off solutions to the resonance conditions will exist along with more complex solutions based on the cyclotomic polynomials. 

\subsection{General pairing-off resonances}

Due to the identities $\omega(k) = \omega({N-k}), \quad k=1,\ldots,N-1,$ one can build immediately a set of particular solutions of equations (\ref{eq:M-wave_k}) and (\ref{eq:M-wave_omega}) provided we choose $S=T$, so the number of interacting waves $M (=2S)$ is even. The idea is to ``pair-off'' incoming and outgoing waves, as follows:
$$k_{S+j} = N-k_j\,,\qquad j=1,\ldots,S.$$
In this way, the frequency resonance condition (\ref{eq:M-wave_omega}) is automatically solved ``by pairs'' since $\omega({k_{S+j}}) = \omega({k_j})$. The momentum conservation condition (\ref{eq:M-wave_k}), on the other hand, leads to a single equation:
\begin{equation}
\label{eq:N-wave_k}
k_1 + \ldots + k_S = \frac{N \nu}{2}\,,
\end{equation}
where the integer variable $\nu$ satisfies $2S/N \leq \nu < 3S/2$ and is introduced as a parameterisation of the modular arithmetic condition in (\ref{eq:M-wave_k}). The lower bound in this latter inequality follows from the fact $k_j \geq 1\,,\quad j=1,\ldots,S.$ The upper bound follows from the fact that the symmetry referred to in Remark \ref{rem:StoT} allows us to avoid repeated solutions by choosing a maximum of $\lfloor S/2\rfloor$ wavenumbers with values above $\lfloor N/2 \rfloor$. Therefore we get  $k_1 + \ldots + k_S \leq (N-1) \lfloor S/2\rfloor  + \lfloor N/2 \rfloor (S - \lfloor S/2\rfloor) = (\lceil N/2 \rceil -1) \lfloor S/2\rfloor  + S \lfloor N/2 \rfloor$, so we obtain
\begin{equation}
\label{eq:nu_bound}
\nu \leq \frac{2}{N} \bigg((\lceil N/2 \rceil -1) \lfloor S/2\rfloor  + S \lfloor N/2 \rfloor\bigg).
\end{equation}
 For $N\geq 4$ this upper bound is easily shown to be strictly bounded from above by $3S/2$. 
However, the sharper version (\ref{eq:nu_bound}) of the upper bound for $\nu$ will be useful in practical search applications. For practical purposes we consider explicitly two cases:

\begin{enumerate}
\item[(i)] For $N$ even and $S$ general, the upper bound (\ref{eq:nu_bound}) reads
$$\nu \leq \frac{2}{N} \bigg((N/2 -1) \lfloor S/2\rfloor   + S  N/2 \bigg)= S+\lfloor S/2\rfloor - \frac{2}{N} \lfloor S/2\rfloor.$$

\item[(ii)] For $N$ odd and $S$ general, the upper bound (\ref{eq:nu_bound}) reads
$$\nu \leq \frac{2}{N} \bigg(((N+1)/2 -1) \lfloor S/2\rfloor   + S  (N-1)/2\bigg) = S+\lfloor S/2\rfloor - \frac{1}{N}(S+\lfloor S/2\rfloor).$$
\end{enumerate}

\subsection{A well-known case: Pairing-off $4$-wave resonances}

The case $S= T =2$, corresponding to $4$-wave resonances, has been studied extensively \cite{rink2001symmetry,rink2006proof,henrici2008results,onorato2015route}, due to its relation with integrability and resonant Birkhoff normal forms. An important result from \cite{rink2001symmetry} in our notation is that, for any $N$, any $4$-wave resonance that is not composite must be of the pairing-off form.

Solutions of equation (\ref{eq:N-wave_k}) with $\nu=2$ exist for all $N$: they are just composite resonances, made out of two `trivial' $2$-wave resonances, parameterised as follows, in the notation of Definition \ref{defi:M-wave resonance}: 
$$\{k_1,k_2; k_2, k_1\}\,, \qquad 1\leq k_1 \leq  k_2 \leq N-1\,.$$ 
These resonances produce nonlinear frequency shifts and do not contribute to the energy transfer between modes \cite{onorato2015route}.

Solutions of equation (\ref{eq:N-wave_k}) with $\nu=1$ exist for $N$ even only and are parameterised as follows:
\begin{equation}
\label{eq:4-wave_sol_k}
\{k_1,k_2; k_3,k_4\} = \left\{k_1,  \frac{N}{2} - k_1;  N-k_1, \frac{N}{2}+k_1\right\} \,, \qquad k_1 = 1, \ldots , \lfloor N/4 \rfloor,
\end{equation}
the momentum conservation condition reading $k_1 + k_2 = k_3 + k_4 - N$. In general, each of these $4$-wave resonances or ``quartets'' produce energy mixing across the four modes involved, \emph{but there is no energy mixing across different quartets}. A simple counting shows that the mode with wavenumber $N/2$ is the only one that does not interact in any quartet. As for the other modes, two cases arise: (i) If $N/2$ is an odd integer, then there are $(N/2-1)/2$ such quartets with $4$ distinct modes each, and therefore a total of $N-2$ modes belong to quartets. (ii) If $N/2$ is an even integer, then the case $k_1 = N/4$ in equation (\ref{eq:4-wave_sol_k}) becomes a degenerate quartet made out of a repeated pairing-off  resonance:
$\{N/4, N/4; 3N/4, 3N/4\}$, consisting of $2$ modes only. The remaining $N/4-1$ quartets have $4$ distinct modes each, giving a sub-total of $N-4$ modes. Adding up we get again $N-2$ modes.

\subsection{A new case: Pairing-off  $6$-wave resonances} 
\label{sec:6_wave_pairing_off}
The case $S=T=3$, corresponding to $6$-wave interactions, was discussed for the first time in \cite{onorato2015route} in the context of wave turbulence theory. Although the cited reference looked at the case $N=2^r$ (i.e. a power of two) only, {it is easy to show that solutions exist for any $N$. We will make this explicit now}. The key step is to evaluate the bound (\ref{eq:nu_bound}). For $S=3, N=6$ we get
$2S/N \leq \nu \leq S+\lfloor S/2\rfloor - \frac{2}{N} \lfloor S/2\rfloor \Rightarrow 1\leq \nu \leq 3  + 2/3.$ Since $\nu$ is integer we get $1 \leq \nu \leq 3$. 

\subsubsection{Pairing-off  $6$-wave resonances for $N$ odd}

For $N$ odd one must take $\nu$ even, so we need to consider the case $\nu = 2$ only. To avoid over-counting, we assume that only one wavenumber (say, $k_3$) can take values above $(N-1)/2$. Thus, equation (\ref{eq:N-wave_k}) becomes $k_1 + k_2 + k_3  = N,$ with two-parameter solution
$$k_3 = N - k_1 - k_2, \quad 1\leq k_2 \leq (N-1)/2 , \quad 1 \leq k_1 \leq \min(k_2,N - 2 k_2)\,.$$
The latter inequality arises from the ordering convention $1\leq k_1\leq k_2 \leq k_3$.

The general solution for the $6$ interacting wavenumbers $\{k_1,k_2,k_3;k_4,k_5, k_6\}$ in the case $N$ odd is thus
\begin{eqnarray}
\label{eq:N_odd_6-wave_sol_k}
\{k_1,k_2,k_3;k_4,k_5, k_6\} = \{k_1,k_2,N - k_1 - k_2;N-k_1,N - k_2, k_1 + k_2\}\,, \qquad ~\\
\nonumber 1\leq k_2 \leq (N-1)/2 , \quad 1 \leq k_1 \leq \min(k_2,N - 2 k_2)\,.
\end{eqnarray}
Wavenumber momentum condition reads $k_1 + k_2 + k_3 = k_4 + k_5 + k_6 - N$.

\subsubsection{Pairing-off $6$-wave resonances for $N$ even}

 We take $N\geq 4$ since otherwise solutions are trivial. One can choose $\nu=1,2$ or $3$ in equation (\ref{eq:N-wave_k}). Again, to avoid over-counting we assume only one wavenumber (say, $k_3$) can take values above $N/2$.  

The branch $\nu=1$ in equation (\ref{eq:N-wave_k}), valid for $N \geq 6$, gives $k_1 + k_2 + k_3 = N/2$, with two-parameter solution
$$k_3 = \frac{N}{2} - k_1 - k_2, \quad 1\leq k_2 \leq \frac{N}{4} -\frac{1}{2} , \quad  1 \leq k_1 \leq \min(k_2,\frac{N}{2} - 2 k_2) \,.$$
The upper bounds on the free parameters $k_2$ and $k_1$ above arise from the ordering convention $1\leq k_1\leq k_2 \leq k_3 \leq N-1$. The same comment applies to the following branches.

The next two branches in equation (\ref{eq:N-wave_k}) are valid for $N\geq 4$. The branch $\nu = 2$ gives $k_1 + k_2 + k_3 = N$, with two-parameter solution
$$k_3 = N - k_1 - k_2, \quad 1\leq k_2 \leq \lfloor(N-1)/2\rfloor = N/2-1, \quad 1 \leq k_1 \leq \min (k_2,N - 2 k_2)\,.$$

The branch $\nu = 3$ gives $k_1 + k_2 + k_3 = 3N/2,$ with two-parameter solution
$$k_3 = 3N/2 - k_1 - k_2, \quad \left\lceil \frac{N}{4}+\frac{1}{2}\right\rceil \leq k_2 \leq N/2, \quad 1 \leq k_1\leq k_2\,.$$

In summary, the general solution for the $6$ interacting wavenumbers $k_1, \ldots , k_6$ is:\\

For branch $\nu = 1$, valid for even $N \geq 6$:  
\begin{eqnarray}
\label{eq:N_even_6-wave_sol_k_branch_1}
\{k_1,k_2,k_3;k_4,k_5, k_6\} = \{k_1,k_2,\frac{N}{2} - k_1 - k_2;N-k_1,N - k_2, \frac{N}{2} + k_1 + k_2\}\,, \qquad ~\\
\nonumber 1\leq k_2 \leq \frac{N}{4} -\frac{1}{2} , \quad  1 \leq k_1 \leq \min(k_2,\frac{N}{2} - 2 k_2)\,.
\end{eqnarray}
(wavenumber resonance condition reads $k_1 + k_2 + k_3 = k_4 + k_5 + k_6 - 2 N$).\\

For branch $\nu = 2$, valid for even $N \geq 4$:
\begin{eqnarray}
\label{eq:N_even_6-wave_sol_k_branch_2}
\{k_1,k_2,k_3;k_4,k_5, k_6\} = \{k_1,k_2,N - k_1 - k_2;N-k_1,N - k_2, k_1 + k_2\}\,, \qquad ~\\
\nonumber 1\leq k_2 \leq N/2 - 1 , \quad 1 \leq k_1 \leq \min(k_2,N - 2 k_2)\,.
\end{eqnarray}
(wavenumber resonance condition reads $k_1 + k_2 + k_3 = k_4 + k_5 + k_6 - N$).\\

For branch $\nu = 3$, valid for even $N \geq 4$:
\begin{eqnarray}
\label{eq:N_even_6-wave_sol_k_branch_3}
\{k_1,k_2,k_3;k_4,k_5, k_6\} = \{k_1,k_2,3N/2 - k_1 - k_2;N-k_1,N - k_2, k_1 + k_2 - N/2\}\,, \qquad ~\\
\nonumber \left\lceil \frac{N}{4}+\frac{1}{2}\right\rceil \leq k_2 \leq N/2, \quad 1 \leq k_1\leq k_2\,.
\end{eqnarray}
(wavenumber resonance condition reads $k_1 + k_2 + k_3 = k_4 + k_5 + k_6$).\\

More generally, the case $S \geq 4$ has solutions for any $N \geq 4$. In principle, these can be constructed explicitly as a multi-parameter solution of equation (\ref{eq:N-wave_k}).

\section{The need for a more general method illustrated with the case $N=6$}
\label{sec:need_for_general}
In the case when $N$ (number of particles) is arbitrary, the pairing-off solutions to the $2L$-wave resonant conditions (with $2L \geq 6$) do not exhaust all possible solutions.
For example, in the case $N=6$ (six particles) there is a $6$-wave resonance that is not of pairing-off form:
$$1+1+5+5 = 3+3 \pmod 6\,,\qquad \omega(1) + \omega(1) +\omega(5)+\omega(5)=\omega(3)+\omega(3)\,.$$ 
The frequency resonance condition is satisfied because of the identity
$$\omega(1) + \omega(5) = \omega(3), \quad \mathrm{or} \quad \sin \pi/6 + \sin 5\pi/6 = \sin 3\pi/6\,,$$
which is reminiscent of a triad resonance. What is the origin of this identity? The answer is given in terms of the $2N$-th root of unity and the so-called cyclotomic polynomials. 

\section{Writing the resonance conditions in terms of real polynomials on the $(2N)^{\mathrm{th}}$ root of unity}
\label{sec:RealPolynomials}
Going back to the dimensionless definition of dispersion relation, equation (\ref{eq:dispersion_dimensionless}), it will be useful to write it as complex exponential:
$$\omega(k) \equiv  -i \left(\zeta^{k} - \zeta^{-k}\right),$$
where
\begin{equation}
\label{eq:zeta}
\zeta = \exp\left(\frac{i\,\pi}{N}\right)
\end{equation}
is a primitive $2N$-th root of unity: $\zeta^{2N} = 1.$ 

In terms of $\zeta$, the frequency  resonance condition (\ref{eq:M-wave_omega}) is
\begin{equation}
\label{eq:M-wave_zeta}
\left(\zeta^{k_1} - \zeta^{-k_1}\right) + \ldots + \left(\zeta^{k_S} - \zeta^{-k_S}\right) = \left(\zeta^{k_{S+1}} - \zeta^{-k_{S+1}}\right) + \ldots + \left(\zeta^{k_{S+T}} - \zeta^{-k_{S+T}}\right)\,.
\end{equation}
Recalling that $k_j$ are all positive and that $\zeta$ is a unit complex number, it follows that $\zeta^{-k_j}$ is the complex conjugate of $\zeta^{k_j}$, so relation (\ref{eq:M-wave_zeta}) is equivalent to the statement that a polynomial is real: 
\begin{equation}
\label{eq:polynomial_rho}
\rho(\zeta) \equiv \zeta^{k_1} + \ldots + \zeta^{k_S} - \left(\zeta^{k_{S+1}} + \ldots + \zeta^{k_{S+T}}\right) \in \mathbb{R}\,.
\end{equation}
In other words, solving the frequency resonance conditions is an easy task: it amounts to finding all real polynomials on the variable $\zeta$. Before tackling the general problem we present two examples: 
\begin{exa} The pairing-off resonances shown in a previous subsection correspond to a ``pairing-off'' real polynomial, made out of real binomials: by setting $S=T$ and $k_{S+j} = N - k_j$ we obtain, from equation (\ref{eq:polynomial_rho}),
$$\rho(\zeta) = \zeta^{k_1} + \ldots + \zeta^{k_S} - \left(\zeta^{N - k_1} + \ldots + \zeta^{N - k_S}\right) = \left(\zeta^{k_1} + \zeta^{-k_1}\right) + \ldots + \left(\zeta^{k_S} + \zeta^{-k_S}\right)\,,$$
which is real, pair by pair.
\end{exa}

\begin{exa} The resonance shown in the previous subsection does not correspond to the pairing-off form. Rather, it corresponds to an element of the kernel of map (\ref{eq:polynomial_rho}), since, for $N=6$ we have
$$\zeta = \exp\left(\frac{i\,\pi}{6}\right) \Longrightarrow \zeta + \zeta^5 - \zeta^3 = 0\,.$$
Thus, the resonance corresponds to $S=4$, $T=2$ with wavenumbers $k_1 = k_2 = 1, \quad k_3 = k_4 = 5,$ and  $k_5 = k_6 = 3$ so we obtain, from equation (\ref{eq:polynomial_rho}),
$$\rho(\zeta) = \zeta^{k_1} + \zeta^{k_2} + \zeta^{k_3} + \zeta^{k_4} - \zeta^{k_5}- \zeta^{k_6}= 2\left(\zeta^{k_1} + \zeta^{k_3} - \zeta^{k_5}\right) = 0,$$
again real.
\end{exa}

\begin{defi}[Real FPUT polynomial]
For any number of particles $N$, a real FPUT polynomial is a polynomial 
$\rho(x)$ of degree $N-1$ in the complex variable $x$, with integer coefficients:
\begin{equation}
\label{eq:FPU_polynomial}
\rho(x) = \sum_{k=1}^{N-1} \rho_k x^k\,, \qquad \rho_k \in \mathbb{Z}\,,
\end{equation}
such that $\rho(\zeta) \in \mathbb{R}$, where $\zeta = \exp(i \pi/N)$. 
\end{defi}

\begin{rem}\label{rem:const}
A real polynomial may have a nontrivial constant term: $\rho(0) \neq 0$, but the constant term does not contribute to the wave resonance condition. So, for our present purposes, we 
can always discard the constant term from the real FPUT polynomials.  
\end{rem}

In general, while this interpretation in terms of real polynomials effectively solves  the frequency resonance condition (\ref{eq:M-wave_zeta}), one still has to solve the momentum conservation condition (\ref{eq:M-wave_k}), but this is a linear problem so it is much simpler. It turns out we can re-write the problem of finding $M$-wave resonances entirely in terms of real FPUT polynomials.

\begin{defi}[Resonant FPUT polynomial]
A resonant FPUT polynomial is a real FPUT polynomial $\rho(x)$ that satisfies $\rho'(1) = 0 \pmod N$, where prime denotes differentiation.
\end{defi}

\begin{thm}[$M$-wave resonances in terms of resonant FPUT polynomials]
For any number of particles $N$, the problem of finding $M$-wave resonances, equations (\ref{eq:M-wave_k_omega}), is equivalent to the problem of finding a resonant FPUT polynomial $\rho(x)$, i.e. a polynomial $\rho(x)$ of degree $N-1$ in the complex variable $x$, with integer coefficients, such that 
$$\rho(\zeta) \in \mathbb{R} \qquad \mathrm{and} \qquad \rho'(1) = 0 \pmod N\,,$$
where $\zeta = \exp(i \pi/N)$ and prime denotes differentiation. Moreover, writing the resonant FPUT polynomial as $\rho(x) = \sum_{k=1}^{N-1}\rho_k x^k$ we obtain the number $M$ of resonant waves via the formula
\begin{equation}
\label{eq:M-wave_M=S+T}
M = \sum_{k=1}^{N-1}|\rho_k|\,.
\end{equation}
\end{thm}
\emph{Proof.} First, the frequency resonance condition in (\ref{eq:M-wave_k_omega}) is equivalent to condition (\ref{eq:polynomial_rho}), which in turn can be re-written in terms of the real FPUT polynomial (\ref{eq:FPU_polynomial})  as
\begin{equation}
\label{eq:M-wave_poly}
\rho(\zeta) \equiv \sum_{k=1}^{N-1}\rho_k\zeta^k \in \mathbb{R} \,,
\end{equation}
where the integers $\{\rho_k\}_{k=1}^{N-1}$ are defined by  
\begin{equation}
\label{eq:rho}\rho_k = \sum_{j=1}^{S} \delta_{k}^{k_j} - \sum_{j=S+1}^{S+T} \delta_{k}^{k_j}\,, \quad 1 \leq k \leq N-1\,.
\end{equation}
Notice that in composite resonances where one of the components is a `trivial' $2$-wave resonance, there will be cancelling contributions from the two sums in equation (\ref{eq:rho}). Thus, we will concentrate on the case where no trivial resonances are included in the composite resonances.

Second, the momentum conservation condition in (\ref{eq:M-wave_k_omega}) can be written in the form
\begin{equation}
\label{eq:M-wave_k_poly}
\left.\left(\frac{d\rho(x)}{dx}\right|_{x=1} \right)= \sum_{k=1}^{N-1}k\,\rho_k \, = \, 0   \qquad ({\rm mod}\;N)\,,
\end{equation}
where again $\{\rho_k\}_{k=1}^{N-1}$ are given in (\ref{eq:rho}).

Conversely, suppose given a real FPUT polynomial: $\rho(x)=\sum_{k=1}^{N-1}\rho_k x^k\,,\quad \rho_k\in \mathbb{Z}.$
Then $\sum_{k=1}^{N-1}\rho_k\zeta^k= \sum_{k=1}^{N-1}\rho_k\zeta^{-k}$
and hence 
$$\sum_{k=1}^{N-1}\rho_k(\zeta^k-\zeta^{-k})=0$$ 
gives a solution to the wave resonance condition in (\ref{eq:M-wave_k_omega}), with $M=\sum_{k=1}^{N-1}|\rho_k|$ (to be shown below). The momentum conservation condition (\ref{eq:M-wave_k_poly}) remains to be discussed.

In terms of the new notation, the list (i.e. the set that may include repeated elements) of $S$ `incoming' momenta is defined by
$$\{k_1, \ldots, k_S\} = \{k | 1\leq k \leq N-1, \,\, \rho_k > 0\}.$$
In order to avoid over-counting, we order these as $k_1 \leq  \ldots \leq k_S.$

The list of $T$ `outgoing' momenta is defined similarly by
$$\{k_{S+1}, \ldots, k_{S+T}\} = \{k | 1\leq k \leq N-1, \,\, \rho_k < 0\}.$$
Again in order to avoid over-counting, it is useful to order these as $k_{S+1} \geq  \ldots \geq k_{S+T}.$
 
The momentum conservation condition is
$$k_1 + \ldots + k_S = k_{S+1} + \ldots + k_{S+T}  \qquad ({\rm mod}\;N)$$ 
(all momenta are integers between $1$ and $N-1$). Alternatively, this can be constructed directly from the polynomial $\rho(x)$ as follows: write
$$\rho'(1) = 0  \qquad ({\rm mod}\;N),$$
which gives 
$$\sum_{k=1}^{N-1}k \rho_k = 0  \qquad ({\rm mod}\;N).$$
The interpretation of this is that the positive coefficients provide us the incoming momenta (with multiplicities $\rho_k$) and the negative terms provide us the outgoing momenta. Thus the number $S$ of `incoming' waves and the number $T$ of `outgoing' waves can be computed from the integer coefficients $\rho_k$ of the real FPUT polynomial, as follows:
$$S = \frac{1}{2}\sum_{k=1}^{N-1}|\rho_k| + \rho_k \,, \quad  T = \frac{1}{2}\sum_{k=1}^{N-1}|\rho_k| - \rho_k\,.
$$
Finally, the total number of waves is given by
$M = S+T = \sum_{k=1}^{N-1}|\rho_k|$. \qed

So we have reduced the general problem of finding $M$-wave resonances to the problem of finding resonant FPUT polynomials. Solving the frequency resonance condition (\ref{eq:M-wave_poly}) amounts to finding a basis for the real FPUT polynomials, which is a relatively easy task that requires however a number of mathematical constructions, to be discussed in Section \ref{sec:underpinnings}. The momentum conservation condition (\ref{eq:M-wave_k_poly}) becomes a linear Diophantine problem on the integer coefficients $\{\rho_k\}_{k=1}^{N-1}$.

\begin{rem} To illustrate how easy it is to find resonances, we stress that if $\rho(x)$ is any real FPUT polynomial then the frequency condition is automatically satisfied. A solution of the momentum condition can be found by defining $d = \gcd\left(N, \,\sum_k k \rho_k\right)$ so the new 
polynomial $(N/d) \rho(x)$ is a resonant FPUT polynomial since it satisfies, in addition, the $M$-wave momentum condition, with $M = \frac{N}{d} \sum_{k=1}^{N-1}|\rho_k|.$ 
\end{rem}

The above remark shows that we can find resonances.  Now, when the number of particles $N$ is arbitrary, there are two types of questions that we address in the following Sections:
\begin{enumerate}
\item Algorithmic constructions (Sections \ref{sec:cyclotomic_method}, \ref{sec:5-wave-resonances}, \ref{sec:super-clusters}, \ref{sec:analysis} and \ref{app:6-wave-cyclotomic}): 
\begin{enumerate}
\item $5$-wave resonances, necessarily not of pairing-off form because the number of interacting waves ($M=5$) is odd.
\item $6$-wave resonances that are not of pairing-off form.
\item Connectivity of resonant clusters when combining $4$-, $5$- and $6$-wave resonances, and their implications regarding the dynamics of the FPUT system and its eventual thermalisation.
\end{enumerate}

\item Mathematical underpinnings (Section \ref{sec:underpinnings} and \ref{app:ker}):
\begin{enumerate}
\item Construction of a basis for the real FPUT polynomials.
\item Full understanding of a real FPUT polynomial as a sum of pairing-off terms plus terms that are not of pairing-off form (i.e., based exclusively on cyclotomic polynomials). 
\item General results on the total number of interacting waves $M$. 
\end{enumerate}

\end{enumerate}

\section{The cyclotomic method: constructing resonant FPUT polynomials of short length based on cyclotomic polynomials}
\label{sec:cyclotomic_method}
The results of this Section are motivated by the rigorous construction of the basis for the real FPUT polynomials (Section  \ref{sec:basis_of_kernel}). However this Section stands alone. 
\begin{defi}[Cyclotomic polynomial]
\label{defi:Cyclotomic}
For any $n \in \mathbb{N},$ the cyclotomic polynomial $\Phi_{n}(x)$ is the unique irreducible polynomial with integer coefficients which divides the polynomial $x^{n}-1$ but does not divide the polynomial $x^j - 1$ for any $j<n.$ One has $\Phi_{n}(z_n) = 0$, where $z_n \equiv \exp\left(\frac{2\,i\,\pi}{n}\right)$. More generally, $\Phi_{n}\left(\exp\left(\frac{2\,i\,\pi\,k}{n}\right)\right) = 0$ for any integer $k$ such that $1 \leq k < n$ and $\gcd(k,n)=1$.
\end{defi}

The key ingredient to construct $M$-wave resonances stems from the fact that if an odd prime $p$ is such that $p \divides N$  (in words, if $p$ divides $N$) then the polynomial
$$ g_p(x) \equiv \Phi_{p}(x^{2N/p}) = 1 + x^{2N/p} + x^{2 \cdot (2N/p)} + \ldots + x^{(p-1)\cdot (2N/p)}$$
is a ``real'' polynomial in the sense that, evaluating at $x=\zeta$, it vanishes because $\Phi_{p}(\mathrm{e}^{2 i \pi/p}) = 0$.
However, this polynomial is not yet a real FPUT polynomial, because its degree is greater than $N-1$. But one can reduce its degree to $N-1$ by repeated use of the identity $\zeta^N=-1$: if the coefficient of $x^k$ is nonzero and if $k = t N + r$, with $r \leq N-1$, then replace $x^k$ with $(-1)^t x^r$. This leads to the following real FPUT polynomial:
$$f_p(x) = 1 - x^{N/p} + x^{2N/p} - \ldots + x^{(p-1)N/p}\,.$$
 In fact, this polynomial is in the kernel of the map (\ref{eq:polynomial_rho}), i.e. $f_p(\zeta) = 0$.
This allows us to generate a number of real FPUT polynomials. First, multiply $f_p(x)$ by $x^n$ with $1 \leq n \leq N/p-1$, producing the set of real FPUT polynomials
$$P_n^{(p)}(x) \equiv x^n f_p(x) =  x^n - x^{N/p+n} + x^{2N/p+n} - \ldots + x^{(p-1)N/p+n}\,, \qquad 1 \leq n \leq N/p-1.$$
It is remarkable that these polynomials produce automatically the frequency resonance conditions corresponding to $p$-wave resonances:
$$\omega(n) - \omega(N/p+n) +\omega(2N/p+n) - \ldots +\omega((p-1)N/p+n) = 0 \,, \qquad 1 \leq n \leq N/p-1,$$
which follows directly from the identity $P_n^{(p)}(\zeta) = 0$. The question to be investigated is therefore whether these polynomials satisfy the momentum conservation condition $\left. \frac{d P_n^{(p)}}{dx}\right|_{x=1} = 0 \pmod N$ for some $n$, and if the answer is no, then we need to ask whether adding another real FPUT polynomial to $P_n^{(p)}(x)$ can produce a polynomial that solves also the momentum conservation condition, i.e. a resonant FPUT polynomial.

To illustrate the idea consider the case $p=3$ where $3 \divides N$.  Then all the polynomials
$$P_n^{(3)}(x) = x^n f_3(x) =  x^n - x^{N/3+n} + x^{2N/3+n}\,, \qquad 1 \leq n \leq N/3-1$$
are real FPUT polynomials, i.e. they satisfy automatically the frequency resonance conditions. However, none of these polynomials can be resonant because the number $M$ of interacting waves, which is equal to the sum of the absolute values of the coefficients of the FPUT polynomial as follows from equations (\ref{eq:FPU_polynomial}) and (\ref{eq:M-wave_M=S+T}), is equal to $3$ by inspection. Thus, resonance would imply the existence of $3$-wave resonances, which we know is false from Theorem \ref{thm:forbidden}. It turns out that in order to produce a resonant FPUT polynomial out of $P_n^{(3)}(x)$ a simple recipe is needed: to add to it a pairing-off term of the form $x^q - x^{N-q}$, where $1\leq q \leq N-1$ and $q$ is to be found explicitly.  So we get the $5$-wave resonances (also known as ``resonant quintets'') out of the FPUT resonant polynomials
$$R_{n,q}^{(3)}(x) = x^n - x^{N/3+n} + x^{2N/3+n} + x^q - x^{N-q}\,, \qquad 1 \leq n \leq N/3-1\,, \qquad1\leq q \leq N-1$$
where the frequency resonance is automatically satisfied since $R_{n,q}^{(3)}(\zeta) \in \mathbb{R}$, and the momentum conservation condition $\left. \frac{d R_{n,q}^{(3)}}{dx}\right|_{x=1} = 0 \pmod N$ is a Diophantine equation that is linear in $N, n$ and $q$:
\begin{equation}
\label{eq:5-wave-pair-off}
n + N/3 + 2 q = 0 \pmod N\,,  \qquad 1 \leq n \leq N/3-1\,, \qquad1\leq q \leq N-1\,.\end{equation}
In the alternative notation in terms of wavenumbers these resonances are denoted by $\{n,2N/3+n, q; N/3+n,N-q\}$ as is evident from inspection of the monomials in $R_{n,q}^{(3)}(x)$ above. 

One could repeat this idea by noting that each term in $P_n^{(3)}(x)$ could be replaced by its pairing-off partner, which would produce a new real FPUT polynomial of length $3$. For example, consider instead of $P_n^{(3)}(x)$ the polynomial
$P_n^{(3)}(x) + (x^{N-(N/3+n)} - x^{N/3+n}) = x^{n} - x^{2N/3-n} + x^{2N/3+n}$, and thus the $5$-wave resonances come from FPUT resonant polynomials of the form $x^{n} - x^{2N/3-n} + x^{2N/3+n} + x^{q'} - x^{N-q'}$. The frequency resonance condition is trivially satisfied, since these are real FPUT polynomials. The momentum conservation condition reads 
\begin{equation}
\label{eq:5-wave-pair-off-2}
3 n + 2 q' = 0 \pmod N \,,  \qquad 1 \leq n \leq N/3-1\,, \qquad1\leq q' \leq N-1\,.
\end{equation}
How many different cases do we have? There are $2$ choices of wavenumbers for each of the $3$ terms in $P_n^{(3)}(x)$, leading to a total of $2^3=8$ possible sets of $5$-wave resonances constructed in this way, each with a relation analogous to (\ref{eq:5-wave-pair-off}) and (\ref{eq:5-wave-pair-off-2}). Amongst these $8$ sets of $5$-wave resonances, some cases will be related by symmetries so we need to be aware of these to avoid over-counting. In the next Section we present the explicit solutions of these equations, leading to a family of interconnected $5$-wave resonances.

Finally, these ideas could be extended to the case $p=5$ where one could ask again whether the polynomial $P_n^{(5)}(x)$ is resonant or not, in order to obtain more $5$-wave resonances. The explicit solution is shown in the next Section as well.

We leave for \ref{app:6-wave-cyclotomic} the explicit solutions for $6$-wave resonances that are not of pairing-off form. These new solutions are built by either adding two real FPUT polynomials of length $3$ in order to construct a resonant FPUT polynomial of length $6$, or by adding a real FPUT polynomial of length $3$ with one of length $5$, in such a way that one of the monomials from each polynomial cancel out, leading to a resonant FPUT polynomial of length $6$ again.

\section{Constructing $5$-wave resonances for arbitrary $N (> 6)$}
\label{sec:5-wave-resonances}
We provide results for $5$-wave resonances, also called resonant quintets, in the notation 
$\{k_1,k_2,k_3; k_4,k_5\}$ that was introduced earlier.

First, we provide four results stemming from the existence of cyclotomic polynomials of length $3$. By construction, these resonant quintets satisfy the following relations: \\
(i) The $3$-wave frequency resonance $\omega(k_1) + \omega(k_2) = \omega(k_4)$. \\
(ii) The pairing-off frequency relation $\omega(k_3) = \omega(k_5)$, stemming from $k_5 = N-k_3$.\\
Because of these structural relations, one can visualise these quintets as in figure \ref{fig:quintets}, top left panel (applicable to the case $3\divides N$ and $6 \ndivides N$). Figure \ref{fig:quintets}, top centre panel, refers to the case $6 \divides N$, where the quintets ``double'', i.e. they come in pairs connected via $3$ common modes. The reason for this doubling can be seen by looking at equation (\ref{eq:5-wave-pair-off}): if $q$ is a solution then so is $q + N/2 \pmod N$.

\subsection{Case 0: $3 \ndivides N$.} In this case there are no $5$-wave resonances from our method.

\subsection{Case 1 (Simple Octahedra): $3 \divides N$ and $6 \ndivides N$.} 
\begin{figure}[h]
\begin{center}
\vspace{-0.1cm}
\includegraphics[width=0.15\textwidth]{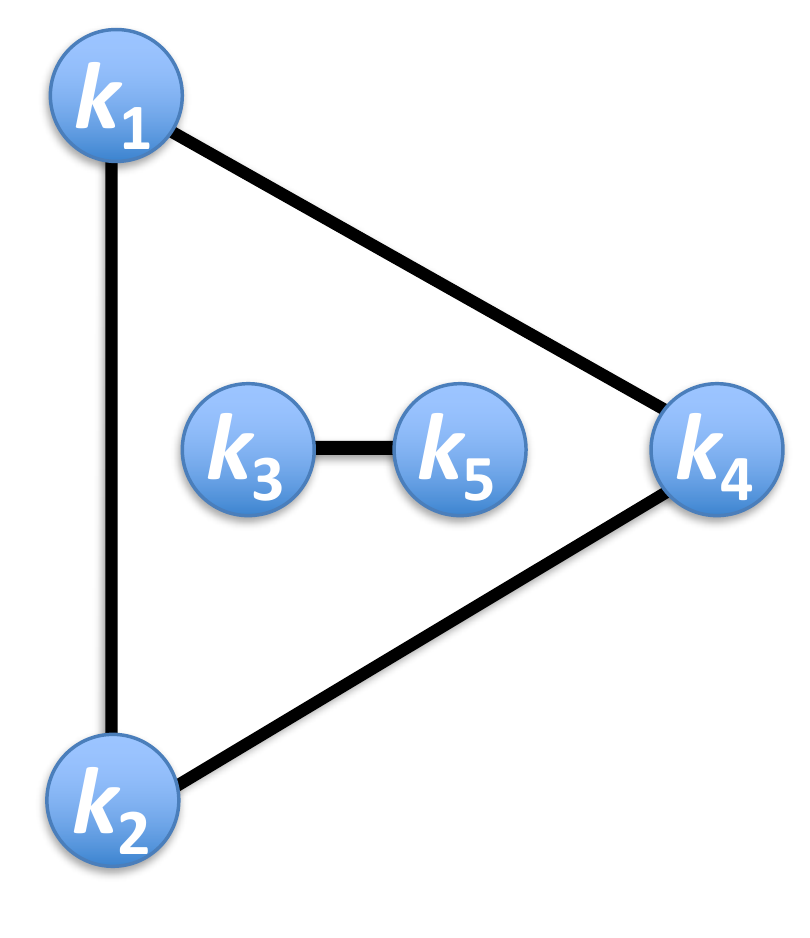}
\hfill
\includegraphics[width=0.15\textwidth]{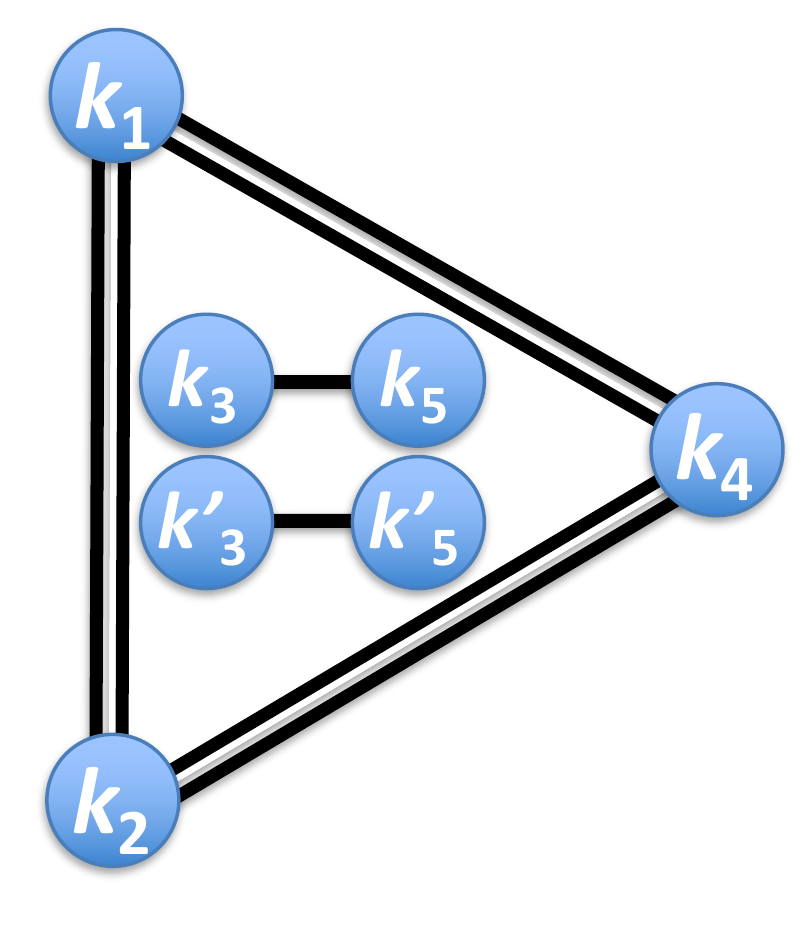}
\hfill
\includegraphics[width=0.17\textwidth]{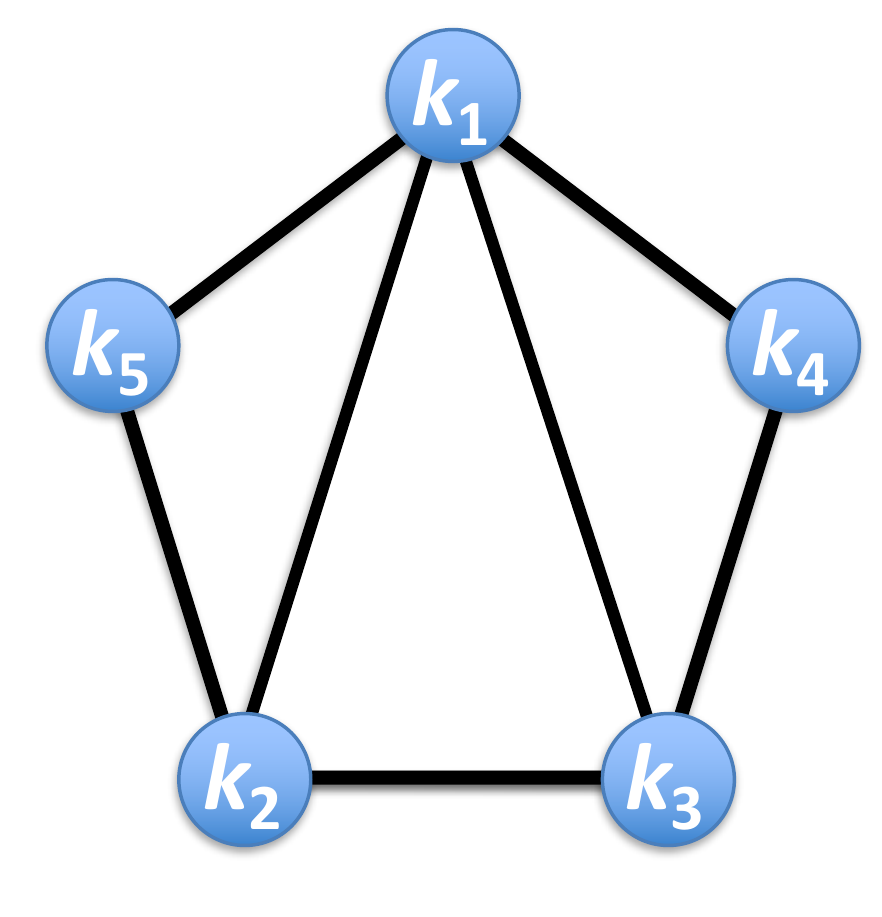}
\hfill
\hfill\\
\vspace{-0.8cm}
\hfill
\includegraphics[width=0.23\textwidth]{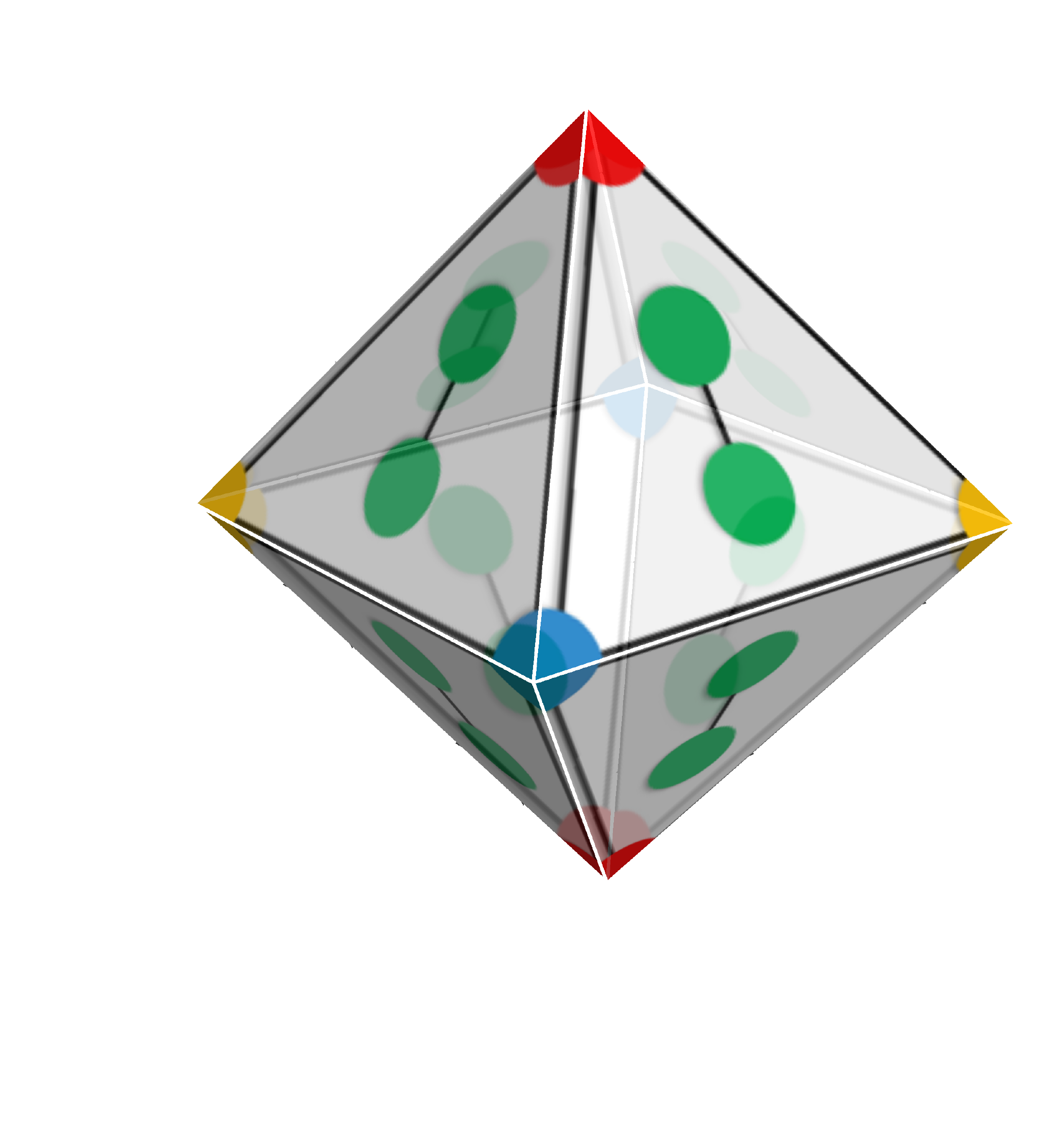}
\hfill
\includegraphics[width=0.23\textwidth]{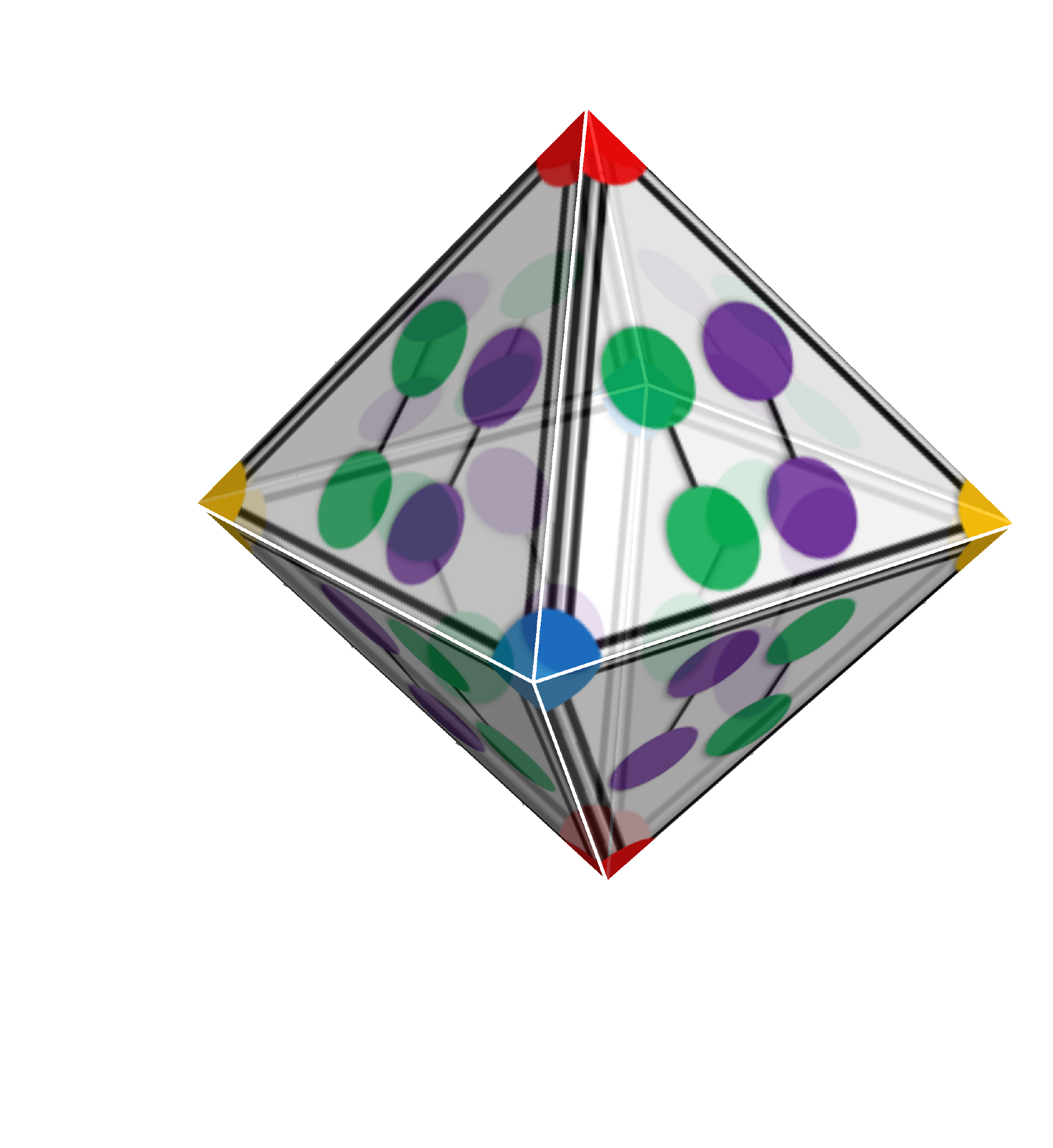}
\hfill
\vspace{-1.4cm}
\end{center}
\caption{\label{fig:quintets} Colour online. Schematic view of resonant quintets. Top left panel: A resonant quintet based on cyclotomic polynomials of length $3$. The plot encodes uniquely the relations $\omega(k_1)+\omega(k_2) = \omega(k_4)$ (triad), $k_5 = N - k_3$ (pair-off terms) and $k_1 + k_2 + k_3 = k_4 + k_5 \pmod N$ (quintet). Top centre panel:  When $6 \divides N$ the resonant quintets depicted on the left panel appear now in pairs, with $3$-common-mode connections. Notably, the inner pair-off terms satisfy the quartet resonance $k_3 + k'_5 = k'_3 + k_5 \pmod N$, which shows how $4$-wave resonances are embedded in the $5$-wave resonances. Top right panel: A resonant quintet based on cyclotomic polynomials of length $5$. The plot encodes the relations $k_1 + k_2 + k_3 = k_4 + k_5 \pmod N$ and $\omega(k_1) + \omega(k_2) + \omega(k_3) = \omega(k_4) + \omega(k_5) \pmod N$. Bottom left panel:  A typical FPUT resonant cluster (octahedron) made out of $8$ of the simple quintets that are based on cyclotomic polynomials of length $3$, depicted in the top left panel. Bottom right panel: In the case $6 \divides N$ the typical FPUT resonant clusters are octahedrons made out of $8$ of the ``double'' quintets depicted in the top centre panel.}
\end{figure}

We obtain that the resonant quintets are grouped in  $\lfloor\frac{N}{6}\rfloor$ octahedron clusters of $8$ quintets each, with the property that, within each cluster,  any quintet is connected to other 4 quintets via at least 2-common-mode connections and to other 3 quintets via at least 1-common-mode connections.  See figure \ref{fig:quintets} (top left panel and bottom left panel) for a depiction of a quintet and an $8$-quintet cluster (octahedron). The quintets within each octahedron cluster are given explicitly in terms of the even parameter $n$ that runs from $2$ to $\frac{N}{3}-1$:
\begin{eqnarray}
\nonumber
\left\{n,\frac{2N}{3}+n,\frac{N}{3}-\frac{n}{2}; \frac{N}{3}+n,\frac{2N}{3}+\frac{n}{2}\right\}\\
\nonumber
\left\{n,\frac{N}{3}-n,\frac{n}{2}; \frac{N}{3}+n,N-\frac{n}{2}\right\}\\
\nonumber
\left\{n,\frac{2N}{3}+n,N-\frac{3n}{2}; \frac{2N}{3}-n,\frac{3n}{2}\right\}\\
\nonumber
\left\{n,\frac{N}{3}-n,\frac{2N}{3}-\frac{n}{2}; \frac{2N}{3}-n,\frac{N}{3}+\frac{n}{2}\right\}\\
\nonumber
\left\{N-n,\frac{2N}{3}+n,\frac{N}{3}+\frac{n}{2}; \frac{N}{3}+n,\frac{2N}{3}-\frac{n}{2}\right\}\\
\nonumber
\left\{N-n,\frac{N}{3}-n,N-\frac{3n}{2}; \frac{N}{3}+n,\frac{3n}{2}\right\}\\
\nonumber
\left\{N-n,\frac{2N}{3}+n,N-\frac{n}{2}; \frac{2N}{3}-n,\frac{n}{2}\right\}\\
\nonumber
\left\{N-n,\frac{N}{3}-n,\frac{2N}{3}+\frac{n}{2}; \frac{2N}{3}-n,\frac{N}{3}-\frac{n}{2}\right\}\\
\label{eq:Case_1}
n=2,\ldots,\frac{N}{3}-1 \qquad (n \,\,\mathrm{even}).
\end{eqnarray}
Except for rare degenerate cases, the quintets within each octahedron cluster are connected via $2$ common modes: this is due to the connectivity of the triads $(k_1, k_2, k_4)$ that form the core of the quintets. Explicitly, the wavenumbers belonging to these core triads correspond to the vertices of the octahedron and appear in three ``conjugate'' pairs: $(n,N-n)$, $(n+N/3, 2N/3-n)$ and $(n+2N/3, N/3-n)$. A conjugate pair corresponds to opposite vertices (in the picture in figure \ref{fig:quintets}, bottom left panel, they have the same colour for identification purposes). The colouring of the pair-off modes depicted on the faces is for reference purposes only: while on any given face the wavenumbers are in pair-off form $(k_3, k_5=N-k_3)$, and opposite faces have the same wavenumbers, for non-opposite faces the wavenumbers are different. \\

\noindent \textbf{Non-interacting modes in Case 1 ($3 \divides N \land 6 \ndivides N$).} A direct inspection of equations (\ref{eq:Case_1}) shows that all wavenumbers in the range $1 \leq k \leq N-1$ appear, with the only exception of the two wavenumbers $N/3, 2N/3$, which do not appear in any octahedron cluster when $9 \ndivides N$. When $9 \divides N$, these wavenumbers appear in the cluster generated by $n=2N/9$ in equations (\ref{eq:Case_1}).
   
\subsection{Case 2 (Double Octahedra): $6 \divides N$ and $12 \ndivides N$.} In this case, due to the appearance of ``double quintets'' (expained at the beginning of Section \ref{sec:5-wave-resonances}), the clusters double in size (i.e. each cluster has now 16 quintets) and correspondingly the number of clusters is halved: there are $\frac{1}{2}\left(\frac{N}{6}-1\right)$ clusters now. See figure \ref{fig:quintets} (top centre panel and bottom right panel) for a depiction of a ``doubled'' quintet and a $16$-quintet cluster (``doubled'' octahedron).  The quintets within each cluster are given explicitly in terms of the even parameter $n$ that runs from $2$ to $\frac{N}{6}-1$:
\begin{eqnarray}
\nonumber
\mkern-72mu \left\{n,\frac{2N}{3}+n,\frac{N}{3}-\frac{n}{2}; \frac{N}{3}+n,\frac{2N}{3}+\frac{n}{2}\right\} &\quad& \left\{n,\frac{2N}{3}+n,\frac{5N}{6}-\frac{n}{2}; \frac{N}{3}+n,\frac{N}{6}+\frac{n}{2}\right\} \\
\nonumber
\mkern-72mu \left\{n,\frac{N}{3}-n,\frac{n}{2}; \frac{N}{3}+n,N-\frac{n}{2}\right\} &\quad& \left\{n,\frac{N}{3}-n,\frac{N}{2}+\frac{n}{2}; \frac{N}{3}+n,\frac{N}{2}-\frac{n}{2}\right\} \\
\nonumber
\mkern-72mu \left\{n,\frac{2N}{3}+n,N-\frac{3n}{2}; \frac{2N}{3}-n,\frac{3n}{2}\right\}&\quad& \left\{n,\frac{2N}{3}+n,\frac{N}{2}-\frac{3n}{2}; \frac{2N}{3}-n,\frac{N}{2}+\frac{3n}{2}\right\}\\
\nonumber
\mkern-72mu \left\{n,\frac{N}{3}-n,\frac{2N}{3}-\frac{n}{2}; \frac{2N}{3}-n,\frac{N}{3}+\frac{n}{2}\right\}&\quad& \left\{n,\frac{N}{3}-n,\frac{N}{6}-\frac{n}{2}; \frac{2N}{3}-n,\frac{5N}{6}+\frac{n}{2}\right\}\\
\nonumber
\mkern-72mu \left\{N-n,\frac{2N}{3}+n,\frac{N}{3}+\frac{n}{2}; \frac{N}{3}+n,\frac{2N}{3}-\frac{n}{2}\right\}&\quad& \left\{N-n,\frac{2N}{3}+n,\frac{5N}{6}+\frac{n}{2}; \frac{N}{3}+n,\frac{N}{6}-\frac{n}{2}\right\}\\
\nonumber
\mkern-72mu \left\{N-n,\frac{N}{3}-n,N-\frac{3n}{2}; \frac{N}{3}+n,\frac{3n}{2}\right\}&\quad& \left\{N-n,\frac{N}{3}-n,\frac{N}{2}-\frac{3n}{2}; \frac{N}{3}+n,\frac{N}{2}+\frac{3n}{2}\right\}\\
\nonumber
\mkern-72mu \left\{N-n,\frac{2N}{3}+n,N-\frac{n}{2}; \frac{2N}{3}-n,\frac{n}{2}\right\}&\quad& \left\{N-n,\frac{2N}{3}+n,\frac{N}{2}-\frac{n}{2}; \frac{2N}{3}-n,\frac{N}{2}+\frac{n}{2}\right\}\\
\nonumber
\mkern-72mu \left\{N-n,\frac{N}{3}-n,\frac{2N}{3}+\frac{n}{2}; \frac{2N}{3}-n,\frac{N}{3}-\frac{n}{2}\right\}&\quad& \left\{N-n,\frac{N}{3}-n,\frac{N}{6}+\frac{n}{2}; \frac{2N}{3}-n,\frac{5N}{6}-\frac{n}{2}\right\}\\
\label{eq:Case_2}
n=2,\ldots,\frac{N}{6}-1 \qquad (n \,\,\mathrm{even}).
\end{eqnarray}

\noindent \textbf{Non-interacting modes in Case 2 ($6 \divides N \land 12 \ndivides N$).} By inspection of equations (\ref{eq:Case_2}), all wavenumbers in the range $1 \leq k \leq N-1$ appear, with the following exceptions: \\
(i) The four wavenumbers $N/6, N/3, 2N/3, 5N/6$ do not appear in any double octahedron cluster when $9 \ndivides N$. When $9 \divides N$, these wavenumbers appear in the cluster generated by $n=N/9$ in equations (\ref{eq:Case_2}). \\
(ii) The wavenumber $N/2$ does not appear in any double octahedron cluster in equations (\ref{eq:Case_2})
.\\

\subsection{Case 3 (Squashed Octahedron): $12 \divides N$.} Here, like in Case 2, the clusters are formed out of $16$ quintets. In fact, exactly the same formulae as in equations (\ref{eq:Case_2}) hold, except that the even parameter $n$ runs from $2$ to $N/6$ now. However, the cluster corresponding to $n=N/6$, as can be easily checked by looking at the formulae, has many quintets repeated (in other words it is degenerate), so in the end that special cluster has $3$ different pairs of quintets only. These correspond to lines 3, 4, and 5 in equations (\ref{eq:Case_2}) and they are, explicitly:
\begin{eqnarray}
\nonumber
\mkern-72mu \left\{\frac{N}{6},\frac{5N}{6},\frac{3N}{4}; \frac{N}{2},\frac{N}{4}\right\} &\quad& \left\{\frac{N}{6},\frac{5N}{6},\frac{N}{4}; \frac{N}{2},\frac{3N}{4}\right\} \\
\nonumber
\mkern-72mu \left\{\frac{N}{6},\frac{N}{6},\frac{7N}{12}; \frac{N}{2},\frac{5N}{12}\right\} &\quad& \left\{\frac{N}{6},\frac{N}{6},\frac{N}{12}; \frac{N}{2},\frac{11N}{12}\right\} \\
\label{eq:Case_3}
\mkern-72mu \left\{\frac{5N}{6},\frac{5N}{6},\frac{5N}{12}; \frac{N}{2},\frac{7N}{12}\right\} &\quad& \left\{\frac{5N}{6},\frac{5N}{6},\frac{11N}{12}; \frac{N}{2},\frac{N}{12}\right\}\,. 
\end{eqnarray}

It is difficult to picture this degenerate cluster in the simple way we did for the octahedron clusters in figure \ref{fig:quintets}. While non-degenerate octahedron clusters are constructed out of $8$ triads, this degenerate cluster is based on \emph{three triads} only, which are generated by \emph{three wavenumbers} only: $N/6, 5N/6$ and $N/2$. Thus, the picture is like a ``squashed'' octahedron. \\

\noindent \textbf{Non-interacting modes in Case 3 ($12 \divides N$).}
By inspection of equations (\ref{eq:Case_2}) and (\ref{eq:Case_3}), all wavenumbers in the range $1 \leq k \leq N-1$ appear, with the exception of the two wavenumbers $N/3, 2N/3$, which do not appear in any double octahedron cluster when $9 \ndivides N$. When $9 \divides N$, these wavenumbers appear in the cluster generated by $n=N/9$ in equations (\ref{eq:Case_2}). \\

The four cases discussed above are exclusive and together they cover all possible quintets that are based on cyclotomic polynomials of length $3$. 

Finally, we provide one result stemming from the existence of cyclotomic polynomials of length $5$. Notice that this extra result can be used in combination with the above Cases 1, 2, and 3.

\subsection{Case 4 (Pentagons): $15 \divides N$.} Here there are just two extra quintets. These are:
\begin{eqnarray}
\label{eq:Case_4}
\mkern-72mu  \left\{\frac{N}{15},\frac{7N}{15},\frac{13N}{15}; \frac{2N}{3},\frac{11N}{15}\right\} &\quad& \left\{\frac{2N}{15},\frac{8N}{15},\frac{14N}{15}; \frac{4N}{15},\frac{N}{3}\right\}\,. 
\end{eqnarray}
By inspection these two quintets have no common modes (but they may have common modes with the octahedra found in Cases 1, 2, 3). To picture these quintets we refer to the pentagons in figure \ref{fig:quintets}, top right panel. \\

\noindent \textbf{Non-interacting modes in Case 4 ($15 \divides N$), in combination with the other cases.}
The wavenumbers $N/3, 2N/3$ appear once each in the pentagon clusters of equations (\ref{eq:Case_4}). As we recall from Cases 1, 2, and 3, these wavenumbers were exceptional. Thus, we summarise three combined cases:\\
(Case 4--1) When $15 \divides N \land 2 \ndivides N$, all wavenumbers in the range $1 \leq k \leq N-1$ appear in some cluster
.\\
(Case 4--2) When $30 \divides N \land 4 \ndivides N$, the wavenumber $N/2$ does not appear at all. All other wavenumbers in the range $1 \leq k \leq N-1$ appear in some cluster, except for the wavenumbers $N/6, 5N/6$, which appear if and only if $9 \divides N$. 
\\
(Case 4--3) When $60 \divides N$, all wavenumbers in the range $1 \leq k \leq N-1$ appear in some cluster. 
\\

We have checked by brute-force numerical search up to $N = 1155 \,(=3 \cdot 5 \cdot 7 \cdot 11)$ that the cases above (case 1 to case 4) comprise all resonant quintets.  Table \ref{tab:5-wave} summarises the results.
 \begin{table}[h]
\begin{center}
  \begin{tabular}{ | c | l | l | }
    \hline
  $\mathbf{N > 6}$   & $5 \ndivides N$ & $5 \divides N$ \\ \hline
    $3 \ndivides N$ & No Quintets. & No Quintets. \\ \hline
   & $\lfloor\frac{N}{6}\rfloor$ Clusters: $8$ Quintets Each; & $1$ Extra Cluster: $2$ Quintets;\\
    $3 \divides N  \land 6 \ndivides N$ & Formulae: Equations (\ref{eq:Case_1}).&  Formulae: Equations (\ref{eq:Case_1}) and (\ref{eq:Case_4}).\\
   &Total: $8 \lfloor\frac{N}{6}\rfloor$ Quintets. & Total: $8 \lfloor\frac{N}{6}\rfloor + 2$ Quintets.  \\ \hline
     & $\frac{1}{2}\left(\frac{N}{6} -1\right)$ Clusters: $16$ Quintets Each; & $1$ Extra Cluster: $2$ Quintets; \\   
   $6 \divides N \land 12 \ndivides N$ & Formulae: Equations (\ref{eq:Case_2}). & Formulae: Equations (\ref{eq:Case_2}) and (\ref{eq:Case_4}).\\
     &Total: $8(\frac{N}{6} -1)$ Quintets. & Total: $8 (\frac{N}{6} -1) + 2$ Quintets.  \\ \hline
    & $\frac{N}{12} -1$ Clusters: $16$ Quintets Each;  &  $1$ Extra Cluster: $2$ Quintets; \\ 
  $12 \divides N$   &  $1$ Cluster: $6$ Quintets; &Formulae: Equations (\ref{eq:Case_2}), (\ref{eq:Case_3}) \\ 
 & Formulae: Equations (\ref{eq:Case_2}) and (\ref{eq:Case_3}). & and (\ref{eq:Case_4}).  \\ 
    &  Total: $16 (\frac{N}{12} -1) + 6$ Quintets. & Total: $16 (\frac{N}{12} -1) + 6 + 2$ Quintets. \\ \hline
    \end{tabular}
    \caption{\label{tab:5-wave}
    Summary of cases of $5$-wave resonances for $N > 6$, regarding the counting of octahedron clusters and total number of quintets. }
\end{center}
\end{table}

\section{Super-clusters: Remarks about the connectivity of the clusters of $5$-wave resonances}
\label{sec:super-clusters}
The clusters found in the previous Section, depicted in figure \ref{fig:quintets} (bottom panels),  appear simple when considered separately. The basic octahedron cluster, made out of $8$ quintets (equation (\ref{eq:Case_1})), contains in the generic case a total of $14$ different modes, a count that is broken down as follows: the $6$ vertices contain $6$ different modes; the $8$ faces contain $2$ modes each but opposite faces have the same modes, so the faces contribute with $4\times 2 = 8$ modes, giving $6 + 8 = 14$ modes overall. The ``doubled'' octahedron, made out of $16$ quintets (equation (\ref{eq:Case_2})), contains in the generic case a total of $22$ different modes: $6$ modes from the vertices plus $4 \times 4=16$ modes from the faces, where now each face contributes with $4$ different modes but again opposite faces have the same modes. The ``squashed'' octahedron, made out of $6$ degenerate quintets (equation (\ref{eq:Case_3})), has a total of $9$ different modes: $3$ modes from the vertices plus $3 \times 2 = 6$ modes from its ``three faces''. Finally, the two simple ``pentagons'', equation (\ref{eq:Case_4}), depicted in figure \ref{fig:quintets}, top right panel, contain $5$ modes each.

Now, when we consider all the clusters together it is easy to see that they must be connected. Recall that most wavenumbers in the range $1 \leq k \leq N-1$ belong to at least one cluster, with a few exceptions depending on the divisibility of $N$. In the simplest case $3 \divides N \land 6 \ndivides N$ and a typical scenario of non-degenerate clusters, a direct counting of modes from the total number of clusters would give a total of $ 14 \times \left\lfloor\frac{N}{6}\right\rfloor$ modes, which is greater than $2N$, so many modes are repeated and thus some octahedron clusters are connected. This leads to the following

\begin{defi}[Super-cluster] 
\label{def:super-cluster}
Let $N$ be the number of particles of the FPUT system, with $N > 6$ and $3\divides N$. A super-cluster $\mathcal{S}_N$ is defined as an undirected graph whose vertices are either octahedron clusters, doubled octahedron clusters or pentagons, as appropriate, and where an edge connecting two vertices means that the clusters corresponding to the connected vertices have one or more common modes. We label the edge with a \emph{\bf cardinal} representing the number of common modes. Also, to identify the vertices properly, we label them with an \emph{\bf ordinal} representing their position in the succession appearing on the last lines of equations (\ref{eq:Case_1}), (\ref{eq:Case_2}), (\ref{eq:Case_3}) and (\ref{eq:Case_4}).
\end{defi}

\begin{figure}[h]
\begin{center}
\vspace{-0.1cm}
\includegraphics[width=0.45\textwidth]{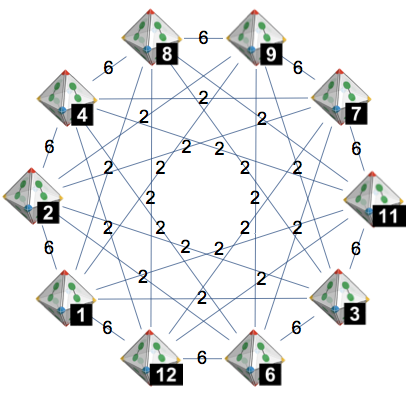}
\hfill
\includegraphics[width=0.54\textwidth]{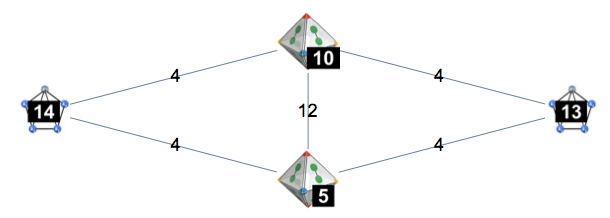}
\vspace{-1cm}
\end{center}
\caption{\label{fig:75} Colour online. Super-cluster $\mathcal{S}_{75}$ for $N = 3\cdot 5^2 = 75$. Using notation from Theorem \ref{thm:super-cluster}, all $10$ vertices in the component $\mathcal{S}_{75}^{(1)}$ (left) have 14 wavenumbers each. In the component $\mathcal{S}_{75}^{*}$ (right) the vertices numbered $5$ and $10$ have 12 wavenumbers each, which are strongly connected since their connecting edge has label $12$. The vertices numbered $13$ and $14$ have $5$ wavenumbers each.}
\end{figure}

\begin{exa}
Consider the case $N=3\cdot 5^2 =75$. This corresponds to the case $3 \divides N \land 6 \ndivides N$, along with $5 \divides N$, in table \ref{tab:5-wave}. The super-cluster $\mathcal{S}_{75}$ has a total of $14$ vertices, corresponding to the $12$ octagon clusters from equations (\ref{eq:Case_1}), where $n = 2$ corresponds to the $1$-st cluster (hence labeled with the ordinal $1$ in the figure), and $n=N/3 -1 = 24$ corresponds to the $12$-th cluster. The remaining $2$ clusters are the  ones appearing in equations (\ref{eq:Case_4}).  Figure \ref{fig:75} shows the super-cluster $\mathcal{S}_{75}$, where we can identify $2$ \emph{connected components} (termed ``component'' for simplicity). By definition, the wavenumbers in a given component interact via $5$-wave resonances exclusively with wavenumbers from that given component.  
\end{exa}

What is the reason for the decomposition of a super-cluster into disjoint connected components? For a given choice of $N$, the number of components of the super-cluster $\mathcal{S}_N$ can be predicted from simple arguments of divisibility. We first state a theorem that determines the common divisibility of wavenumbers inside any simple, double or squashed octahedron cluster. A second theorem states the special case of a pentagon cluster. Finally a third theorem states that a super-cluster is a disjoint union of sub-graphs such that within any sub-graph all wavenumbers are divisible by the same divisor of $N$. 

\begin{thm}[Octahedra divisibility]
\label{thm:divisibility}
Let the number of particles be $N = 2^{b_0} \cdot 3^{b_1} \cdot R$, where $b_0, b_1$ are integers with $b_0\geq 0$ and $b_1\geq1$, and $R$ is a positive integer with $2\ndivides R \land 3 \ndivides R$. Let $r$ be a positive integer such that $r \divides R$. 

Consider the multi-set $\mathcal{A}_n$ of allowed wavenumbers in a given octahedron (respectively double octahedron and ``squashed'' octahedron) cluster, defined as the multi-set of wavenumbers defined in equations (\ref{eq:Case_1}) if $b_0 = 0$ (respectively equations (\ref{eq:Case_2}) if $b_0 \geq 1$ and equations (\ref{eq:Case_3}) if $b_0 \geq 2$). Then 
$$r \divides k \quad \text{for \,\, some \,\,} k \in \mathcal{A}_n \iff r \divides k \quad \text{for \,\, all \,\,} k \in \mathcal{A}_n.$$
\end{thm}
\emph{Proof.} \\
(i) Case $b_0 = 0$. Regarding the simple octahedra, by looking at equations (\ref{eq:Case_1}) the allowed wavenumbers are in the multi-set $\mathcal{A}_n = \{n,N-n,n+N/3,2N/3-n,n+2N/3,N/3-n,N/3-n/2,2N/3+n/2,n/2,N-n/2,N-3n/2,3n/2,2N/3-n/2,N/3+n/2\}$ for some even integer $n$. So, by inspection, $r\divides n$ if and only if $r$ divides any other number in this multi-set, because $r \divides N, N/3$ and $2N/3$. Therefore $r \divides k$ for some $k \in \mathcal{A}_n$ if and only if $r \divides k$ for all $k \in \mathcal{A}_n$. \\
(ii) Case $b_0 > 0$. Regarding the double octahedra, including the squashed octahedron, by looking at equations (\ref{eq:Case_2}) the allowed wavenumbers are in the multi-set $\mathcal{A}_n = \{n,N-n,n+N/3,2N/3-n,n+2N/3,N/3-n,N/3-n/2,2N/3+n/2,n/2,N-n/2,N-3n/2,3n/2,2N/3-n/2,N/3+n/2,5N/6-n/2,N/6+N/2,N/2+n/2,N/2-n/2,N/2-3n/2,N/2+3n/2,N/6-n/2,5N/6+n/2\}$ for some even integer $n$. Again, by inspection, $r\divides n$ if and only if $r$ divides any other number in this multi-set, because $r \divides N, N/3, 2N/3, N/6, 5N/6$ and $N/2$. Therefore $r \divides k$ for some $k \in \mathcal{A}_n$ if and only if $r \divides k$ for all $k \in \mathcal{A}_n$.  
\qed

\begin{rem}
The restriction $2 \ndivides r$ is necessary because $2 \divides n$ even if $2 \ndivides \frac{n}{2}$, and the restriction $3 \ndivides r$ is necessary because $3 \divides \frac{3n}{2}$ even if $3 \ndivides n$.
\end{rem}

\begin{thm}[Pentagon divisibility]
\label{thm:pentagon}
Under the same hypotheses of 
 Theorem \ref{thm:divisibility}, if in addition $5 \divides R$ then 
one wavenumber from each pentagon quintet in Eqs.~(\ref{eq:Case_4}) $(2N/3$ and $N/3)$ is divisible by $R$. The remaining wavenumbers are divisible by $R/5$.  
\end{thm}
The proof is direct. \qed

\begin{rem}
Because the wavenumbers within a pentagon quintet have mixed divisibility properties, they will serve as connectors between clusters whose wavenumbers are divisible by $R$ and clusters whose wavenumbers are divisible by $R/5$ only.
\end{rem}

\begin{thm}[Super-cluster decomposition] 
\label{thm:super-cluster}
 Let the number of particles be $N = 2^{b_0} \cdot 3^{b_1} \cdot R$, where $b_0, b_1$ are integers with $b_0\geq 0$ and $b_1\geq1$, and $R$ is a positive integer such that $2\ndivides R,$ $3 \ndivides R$ and $R >2^{1-b_0} \cdot 3^{1-b_1}$ (i.e. $N>6$).

Consider the super-cluster $\mathcal{S}_N$ from Definition \ref{def:super-cluster}. Then:\\
(i) $\mathcal{S}_N$ is a disjoint union of graphs over the set of divisors of $R$:
\begin{equation}
\label{eq:S_N}
\mathcal{S}_N =\left(\bigoplus\limits_{\substack{d\,\divides R \\ d < R/5}} \mathcal{S}_N^{(d)}\right) \bigoplus  \mathcal{S}_N^{*}\,,
\end{equation}
where the graph $\mathcal{S}_N^{(d)}$ is defined as the disjoint union of all connected components $\,\mathcal{C}$ of $\mathcal{S}_N$ such that all wavenumbers $k$ in $\mathcal{C}$ satisfy $\gcd(k,R) = d$ (equation (\ref{eq:S_N^d})), and the graph $\mathcal{S}_N^{*}$ is defined as the connected component of $\mathcal{S}_N$ such that all wavenumbers $k$ in $\mathcal{S}_N^{*}$ satisfy $\gcd(k,R) = R$ or $\gcd(k,R) = R/5$ (if $5 \divides R$).\\
(ii) Provided $R\neq 1$ or $R\neq 5$, $\mathcal{S}_N^{(d)}$ exists and is non-empty for any divisor $d$ of $R$ such that $d < R/5$, and the set of wavenumbers appearing in $\mathcal{S}_N^{(d)}$ is given explicitly by
\begin{equation}
\label{eq:A_N^d}
\mathcal{A}_N^{(d)} =  \{d \cdot \ell:  \quad \ell \in \mathbb{N} \quad \land \quad 1\leq \ell < 2^{b_0}\cdot 3^{b_1} \cdot R/d \quad \land \quad \gcd(\ell,R/d)=1\}\,.
\end{equation}
(iii) The graph $\mathcal{S}_N^{*}$ is empty if and only if $b_0 \leq 1\,\, \land \,\, b_1 = 1\,\, \land \,\, R\neq 1 \,\, \land \,\, 5 \ndivides R$. When $\mathcal{S}_N^{*}$ is non-empty the set of wavenumbers appearing in $\mathcal{S}_N^{*}$ is given explicitly by
\begin{equation}
\label{eq:A_N^*}
\mathcal{A}_N^{*} = \{\ell \cdot R/5 \in \mathbb{N}:  \,\, \ell \in \mathbb{N} \,\, \land \,\, 1\leq \ell < 2^{b_0}\cdot 3^{b_1} \cdot 5\} \setminus \mathcal{F}_N\,,
\end{equation}
where the set of forbidden wavenumbers $\mathcal{F}_N$ is defined by 
$$\mathcal{F}_N = 
\begin{cases} 
\{N/6,N/3, N/2, 2N/3,5N/6\} & \text{if} \quad b_0=1 \land b_1 =1\,,\\
\{N/3, 2N/3\} & \text{if} \quad b_0\neq1 \land b_1 =1\,,\\
\{N/2\} & \text{if} \quad b_0=1 \land b_1 \geq 2\,,\\
\varnothing & \text{if} \quad  b_0 \neq 1 \land b_1 \geq 2\,.
\end{cases}
$$

Moreover, when $5 \divides R$ the two pentagon quintets are vertices of  $\mathcal{S}_N^{*}$, where they serve as the only connectors between vertices whose wavenumbers $k$ satisfy $\gcd(k,R) = R$ and vertices whose wavenumbers $k$ satisfy $\gcd(k,R) = R/5$. Finally, when $b_0 \geq 2$ the squashed octahedron cluster is a vertex of $\mathcal{S}_N^{*}$ and is in the subgraph of $\mathcal{S}_N^{*}$ made out of vertices whose wavenumbers $k$ satisfy $\gcd(k,R) = R$.
\end{thm}
\emph{Proof.}\\
(i) Consider any connected component $\mathcal{C}$ of $\mathcal{S}_N$, and any wavenumber $k$ in $\mathcal{C}$. Define \mbox{$d_{\mathcal{C}} = \gcd(k,R)$}. Assume $d_{\mathcal{C}}<R/5$, thus excluding precisely the cases $d_{\mathcal{C}}=R$ and $d_{\mathcal{C}}=R/5$, namely the only cases where the component $\mathcal{C}$ may contain the pentagon quintets as vertices (see Theorem \ref{thm:pentagon}). It is then easy to see that all wavenumbers $k'$ in $\mathcal{C}$ satisfy $d_{\mathcal{C}} = \gcd(k',R)$. This is because from Theorem \ref{thm:divisibility} all wavenumbers in the octahedron clusters (i.e., the vertices of $\mathcal{C}$) that contain $k$ share the common divisor $d_{\mathcal{C}}$, but this divisor is common to all vertices in $\mathcal{C}$ due to the fact that by definition there is a path that connects any two vertices in $\mathcal{C}$ via common-mode connections. So, in summary we have a way to classify a connected component $\mathcal{C}$ of $\mathcal{S}_N$ by the divisor
$$d_{\mathcal{C}} = \gcd(k,R) \qquad \text{for all} \,\, k \in \mathcal{C}.$$
To account for the possibility that several connected components of $\mathcal{S}_N$ have the same divisor $d_{\mathcal{C}}$, we define
\begin{equation}
\label{eq:S_N^d}
\mathcal{S}_N^{(d)} = \bigoplus\limits_{\substack{\mathcal{C} \subseteq \mathcal{S}_N\\ d_{\mathcal{C}} = d}} \mathcal{C}\,, \qquad \text{where} \quad d \divides R \quad \text{and} \quad d < R/5\,.
\end{equation}
Trivially, we have $\mathcal{S}_N^{(d)} \cap \mathcal{S}_N^{(d')} = \varnothing$ if $d \neq d'$, because the inequality $\gcd(k,R) \neq \gcd(k,R)$ cannot be satisfied for any wavenumber $k$. Thus we get preliminarily
$\mathcal{S}_N = \left(\bigoplus\limits_{\substack{d\,\divides R \\ d < R/5}} \mathcal{S}_N^{(d)}\right) \bigoplus  \mathcal{S}_N^{*}\,,$
where $\mathcal{S}_N^{*}$ is the remaining graph (to be determined in part (iii)) whose vertices contain wavenumbers that are divisible by $R/5$ (if $5\divides R$) and $R$. \\
(ii) When $d \divides R$ but $d < R/5$ (the condition $R\neq 1$ or $R\neq 5$ is necessary so that such a $d$ exists), the set $\mathcal{A}_N^{(d)}$ of wavenumbers in $\mathcal{S}_N^{(d)}$ is equal to the set of integers $k$ between $1$ and $N-1$ such that $\gcd(k,R)=d$. This is because, as demonstrated in Section \ref{sec:5-wave-resonances}, the set of all wavenumbers contained in the octahedron clusters include all integers between $1$ and $N-1$, except perhaps for the wavenumbers $k = N/6, N/3, N/2, 2N/3, 5N/6$ (when they exist, depending on the case), which satisfy $\gcd(k,R)=R$ so these exceptions do not apply when $d < R/5$. Now, explicitly:
$$\mathcal{A}_N^{(d)} 
= \{k \in \mathbb{N}: \,\, 1\leq k < N \,\, \land \,\, \gcd(k,R)=d\} 
= \left\{d \cdot \ell: \,\, \ell \in \mathbb{N} \,\, \land \,\, 1\leq \ell < \frac{N}{d} \,\, \land \,\, \gcd\left(\ell,\frac{R}{d}\right)=1\right\}\,.$$
(iii) By definition, the set $\mathcal{A}_N^{*}$ of wavenumbers in $\mathcal{S}_N^{*}$ is the union of the set of wavenumbers divisible by $R$ and the set of wavenumbers divisible by $R/5$, minus the set of non-interacting wavenumbers discussed in Section \ref{sec:5-wave-resonances}. Explicitly: 
$$\mathcal{A}_N^{*} = \left(\mathcal{A}_N^{(R/5)}  \,\cup\, \mathcal{A}_N^{(R)} \right)\,\setminus\, \mathcal{F}_N\,,$$
which simplifies to equation (\ref{eq:A_N^*}).\\
($\Longleftarrow$) 
If $b_0 \leq 1$ and $b_1 = 1$ then the set of wavenumbers divisible by $R$, $\mathcal{A}_N^{(R)} = \{k \in \mathbb{N}: \,\, 1\leq k < N \,\, \land \,\, \gcd(k,R)=R\}$, contains exactly the forbidden wavenumbers, i.e. the non-interacting wavenumbers discussed in Section \ref{sec:5-wave-resonances} under Cases 1 and 2. The condition $R\neq 1$ is kept for bookkeeping purposes only since it follows from the hypothesis $R>2^{1-b_0}\cdot 3^{1-b_1}$, which implies $R = 2^{1-b_0} > 1$. If $5 \ndivides R$ then the set $\mathcal{A}_N^{(R/5)}$ of wavenumbers divisible by $R/5$ is empty (in fact, it is not defined). Therefore 
$$\mathcal{S}_N^{*} = \varnothing \qquad \text{if} \quad b_0 \leq 1 \quad \text{and} \quad b_1 = 1 \quad \text{and} \quad R \neq 1 \quad \text{and} \quad 5 \ndivides N\,.$$
($\Longrightarrow$) For any other condition on $b_0$ and $b_1$ it is easy to see that the set of wavenumbers divisible by $R$ minus the set of non-interacting wavenumbers is non-empty by just calculating the set $\{k \in \mathbb{N}: \,\, 1\leq k < N \,\, \land \,\, \gcd(k,R)=R\} = \{R, 2R, \ldots, (2^{b_0}\cdot 3^{b_1} -1) R\}$ and subtracting from it the set of forbidden wavenumbers $\mathcal{F}_N \subseteq \{N/6, N/3, N/2, 2N/3, 5N/6\}$. When $b_0 \geq 2$, it follows $12 \divides N$ so the squashed octahedron, equation (\ref{eq:Case_3}), exists and all its wavenumbers are divisible by $R$, so the squashed octahedron is a vertex of $\mathcal{S}_N^{*}$. When $5 \divides N$ then $\mathcal{S}_N^{*}$ contains the two pentagon quintets so it is not empty. When $R = 1$ the hypothesis $N>6$ implies either $b_0 \geq 2 \,\,\land \,\, b_1=1$ or $b_0 \geq 0 \,\,\land \,\, b_1\geq 2$, which were covered at the beginning of this paragraph. Finally, when they exist, the pentagon clusters are the only vertices containing wavenumbers that have different common divisors with $R$: $R/5$ and $R$. Thus these vertices are the only possible connectors between subgraphs containing wavenumbers that are divisible by $R$ and subgraphs containing wavenumbers that are divisible by $R/5$ only. \qed

\begin{rem} 
\label{rem:counting}
The size of the sets of connected wavenumbers $\mathcal{A}_N^{(d)}$ and $\mathcal{A}_N^{*}$,  defined in points (ii) and (iii) above, can be calculated in terms of Euler's totient function $\phi(n)$, to be defined in Section \ref{sec:underpinnings}, equation (\ref{eq:totient}). The results are:
$$\#\mathcal{A}_N^{(d)} = 2^{b_0}\cdot 3^{b_1} \phi(R/d) \,, \qquad \#\mathcal{A}_N^{*} =- \#\mathcal{F}_N + \begin{cases} 
2^{b_0}\cdot 3^{b_1} \cdot 5 - 1 & \text{if} \quad 5 \divides R\,,\\
2^{b_0}\cdot 3^{b_1}-1 & \text{if} \quad  5 \ndivides R\,,
\end{cases} 
$$
where
$$\#\mathcal{F}_N = 
\begin{cases} 
5 & \text{if} \quad b_0=1 \land b_1 =1\,,\\
2 & \text{if} \quad b_0\neq1 \land b_1 =1\,,\\
1 & \text{if} \quad b_0=1 \land b_1 \geq 2\,,\\
0 & \text{if} \quad  b_0 \neq 1 \land b_1 \geq 2\,.
\end{cases}
$$
\end{rem}

\begin{exa}
Consider the case $N = 420 = 2^2 \cdot 3 \cdot 5 \cdot 7$, so $b_0 = 2, \,b_1 = 1$ and $R = 5 \cdot 7$. The set of divisors of $R$ that are less than $R/5$ is $\{1, 5\}$. From equation (\ref{eq:S_N}) the super-cluster has the following decomposition:
$$\mathcal{S}_N = \mathcal{S}_N^{(1)} \oplus \mathcal{S}_N^{(5)} \oplus \mathcal{S}_N^{*}\,,$$
where $\mathcal{S}_N^{*}$, containing by definition the wavenumbers that are divisible by $R (=35)$ or $R/5 (=7)$, is a disjoint union of two graphs (see figure \ref{fig:420}, top and bottom right graphs). Also in the figure, we see graph $\mathcal{S}_N^{(1)}$ (top left), made out of wavenumbers that are not divisible by $5$ nor by $7$, and graph $\mathcal{S}_N^{(5)}$ (bottom left), made out of wavenumbers that are divisible by $5$ but not by $7$. Vertices are mainly double octahedron clusters but we can observe the two pentagon quintets and the squashed octahedron cluster. As for the set of wavenumbers in each component of the supercluster, we have, using Remark \ref{rem:counting}: The set $\mathcal{A}_N^{(1)}$ of wavenumbers in $\mathcal{S}_N^{(1)}$, has a total of $\#\mathcal{A}_N^{(1)} = 2^{b_0}\cdot 3^{b_1} \phi(R) = 288$ wavenumbers. The set $\mathcal{A}_N^{(5)}$ of wavenumbers in $\mathcal{S}_N^{(5)}$, has a total of $\#\mathcal{A}_N^{(5)} = 2^{b_0}\cdot 3^{b_1} \phi(R/5) = 72$ wavenumbers. The set $\mathcal{A}_N^{*}$ of wavenumbers in $\mathcal{S}_N^{*}$, has a total of $\#\mathcal{A}_N^{*} = - 2 + 2^{b_0}\cdot 3^{b_1} \cdot 5 - 1= 57$ wavenumbers.  They total $417$ wavenumbers, namely two less than the universe of wavenumbers $(N-1)$ due to the fact that there are two forbidden wavenumbers: $\{N/3, 2N/3\}$.
\begin{figure}[h]
\begin{center}
\vspace{-0.1cm}
\includegraphics[width=0.80\textwidth]{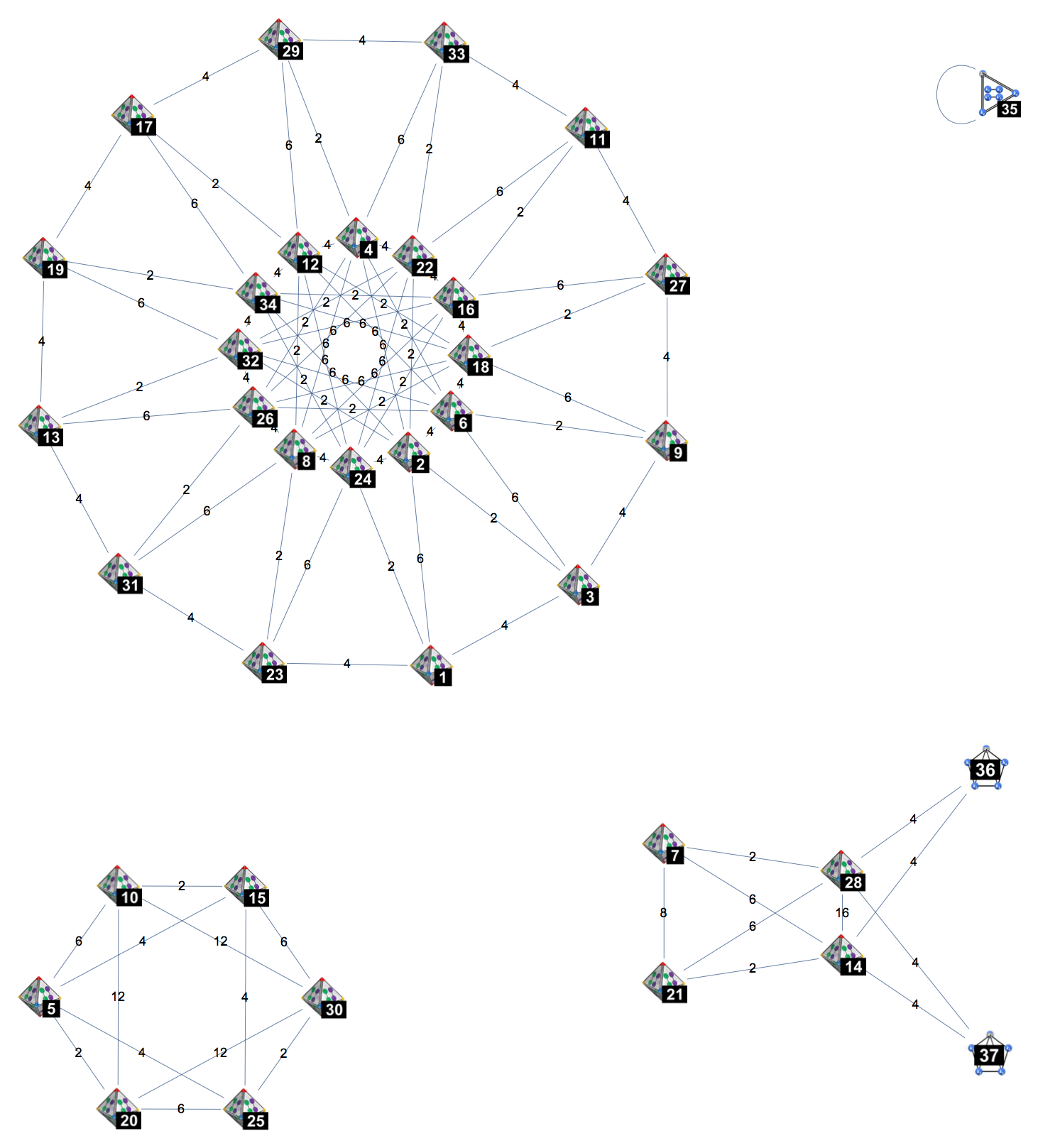}
\vspace{-0.9cm}
\end{center}
\caption{\label{fig:420} Colour online. Super-cluster $\mathcal{S}_{420}$ for $N = 2^2 \cdot 3 \cdot 5 \cdot 7 = 420$. All $24$ vertices in the component $\mathcal{S}_{420}^{(1)}$ (top left) and all $6$ vertices in the component $\mathcal{S}_{420}^{(5)}$ (bottom left) have 22 wavenumbers each. The component $\mathcal{S}_{420}^{*}$ is the disjoint union of the two subgraphs at the right of the figure. In this component, the vertices labeled $7$ and $21$ have 22 wavenumbers each, and the vertices labeled $28$ and $14$ have 20 wavenumbers each, which are strongly connected since their connecting edge has label $16$. They are connected to the pentagon quintets, (vertex labels $36$ and $37$) which have $5$ wavenumbers each. The isolated vertex with label $35$ is the squashed octahedron.}
\end{figure}

\end{exa}


\section{Route to thermalisation: combined analysis of $4$-, $5$- and $6$-wave resonances as a function of the number of particles $N$}
\label{sec:analysis}
The results of the previous Sections can be used to study how the lowest possible order of the resonances depends very sensitively on the number of interacting particles $N$. For example, as is well known, $4$-wave resonance solutions that are non-trivial (i.e. those that do not contribute to a Birkhoff normal form in the Hamiltonian, or in our notation those that are not composite resonances) exist for $N$ even only. These non-trivial $4$-wave resonances admit a so-called \emph{resonant Birkhoff normal form} \cite{henrici2008results}. Thus, if $N$ is odd, the lowest non-trivial order of a resonance is either $5$ or $6$. As we found in the last Sections, $5$-wave resonances exist if and only if $3 \divides N$ and $N>6$.  Thus, for $N$ odd, if $N\leq 6$ or $3 \ndivides N$ then the lowest non-trivial resonant order is $6$. \mbox{Table \ref{tab:lowest_order}} summarises the results as a function of $N$. In this table we always include the next order when $4$-wave resonances exist (i.e. for $N$ even) because, as is well known, if one truncates a system up to the $4$-wave resonances the result is an integrable system ({resonant Birkhoff normal form}) due to the fact that the resonant quartets, whose wavenumbers are defined in equation (\ref{eq:4-wave_sol_k}), are not connected: the wavenumbers in a given quartet do not belong to any other quartets. Thus, the next order of resonance must be included in order to obtain complex (and typically chaotic) dynamics characterised by energy mixing across all wavenumbers, which will lead to an understanding of the thermalisation process. We remark that our study produces irreducible $5$- and $6$-wave resonances, which by definition cannot be simplified in terms of $4$-wave resonant Birkhoff normal forms.

 \begin{table}[h]
\begin{center}
  \begin{tabular}{ | c | l | c | }
    \hline
  $N \leq 6$   & Lowest-order Resonances & Type of Resonance \\  \hline\hline
    $3$ & $6$-wave & Pairing-off \\ \hline
  $4$ & $4$-wave \& $6$-wave & Pairing-off \\ \hline
   $5$  &$6$-wave & Pairing-off \\ \hline
  $6$ & $4$-wave \& $6$-wave & Pairing-off \& Cyclotomic \\ \hline \hline\hline
  $N > 6$   & Lowest-order Resonances & Type of Resonance \\  \hline\hline
   $1 \pmod 6\quad (7, 13, 19, \ldots)$ &  $6$-wave & Pairing-off \\ \hline
   $2 \pmod 6\quad (8, 14, 20, \ldots)$ & $4$-wave  \& $6$-wave & Pairing-off  \\ \hline
   $\mathbf{3 \pmod 6\quad (9, 15, 21, \ldots)}$ &  $\mathbf{5}$\textbf{-wave} & \textbf{Cyclotomic}   \\ \hline
  $4 \pmod 6\quad (10, 16, 22, \ldots)$ & $4$-wave \& $6$-wave  & Pairing-off \\ \hline
   $5 \pmod 6\quad (11, 17, 23, \ldots)$ &  $6$-wave & Pairing-off \\ \hline
  $\mathbf{0 \pmod 6\quad (12, 18, 24, \ldots)}$ & $4$-wave \& $\mathbf{5}$\textbf{-wave} & Pairing-off \& \textbf{Cyclotomic} \\ 
  \hline 
    \end{tabular}
    \caption{\label{tab:lowest_order}
    Study of lowest order of FPUT irreducible resonances that are not of Birkhoff normal form (in other words, resonances that effectively exchange energy amongst modes), as a function of $N$. 
    The cases admitting $5$-wave resonances are highlighted in \textbf{boldface}. {$4$-wave resonances are always in so-called resonant Birkhoff normal form, which do not produce effective energy transfers throughout the whole spectrum of modes.} In contrast, $5$- and $6$-wave irreducible resonances cannot be simplified in terms of resonant Birkhoff normal forms, because they mix energies over a wide range of modes.
    }
\end{center}
\end{table}

\noindent \textbf{Super-clusters modify the thermalisation hypothesis.} Writing $N = 2^{b_0}\cdot 3^{b_1}\cdot R$ with $2\ndivides R$, $3\ndivides R$, $b_0 \geq 0$ and $b_1 \geq 1$, Theorem \ref{thm:super-cluster} establishes that the set of divisors $d$ of $R$ separates the universe of modes (with wavenumbers $1, \ldots, N-1$) into disjoint sets $\mathcal{S}_N^{(d)}$ (including $\mathcal{S}_N^{*}$) of interacting modes via $5$-wave resonances. Roughly speaking, wavenumbers in a given set must be divisible by $d$ but are not divisible by any $d' > d$. Physically, this result modifies the thermalisation hypothesis: modes whose wavenumbers belong to disjoint sets do not interact via $5$-wave resonances, but thermalisation is possible within each disjoint set $\mathcal{S}_N^{(d)}$.

What about $4$-wave resonances in combination with $5$-wave resonances? $4$-wave resonances that exchange energy across modes exists when $N$ is even only, or in our notation when $b_0 \geq 1$. Hence for $N$ odd the situation is as described in the previous paragraph. For $N$ even, the wavenumbers involved in a resonant quartet are shown in equation (\ref{eq:4-wave_sol_k}): $\{k_1, N/2-k_1, N-k_1, N/2+k_1\}$, where $k_1 =1, \ldots, \lfloor N/4\rfloor$. By inspection, given $k_1$ all four wavenumbers share the same divisors $d$ with $R$. Therefore energy-exchanging $4$-wave resonances occur within the same disjoint set $\mathcal{S}_N^{(d)}$ of the super-cluster. So, in terms of energy exchanges, for $N$ even the state space still separates as a direct sum of lower-dimensional sub-spaces, associated with each disjoint set $\mathcal{S}_N^{(d)}$. On each of these sub-spaces thermalisation may hold independently, but this is a matter to be studied further both numerically and analytically in a subsequent paper. 

It is worth mentioning that ``trivial'' or composite $4$-wave resonances (e.g. $k_1 + k_2 = k_1 + k_2$) exist for any $N$ and they do allow interactions across disjoint components of the super-clusters. However these interactions are subtle: they do not exchange energy but they mix up the Fourier phases across disjoint components. So, in terms of energy exchanges, the picture of independent ergodic components $\mathcal{S}_N^{(d)}$ still applies. In particular, energy must be conserved separately on each component. 

Therefore, in order to obtain global mixing we need to go to higher orders: pairing-off $6$-wave resonances will mix up the energies across disjoint components $\mathcal{S}_N^{(d)}$ of the super-cluster. To see this, consider the simple case $b_0 = 0, b_1 \geq 1$, so $N$ is odd and thus pairing-off sextuplet solutions are given in equation (\ref{eq:N_odd_6-wave_sol_k}): wavenumbers $k_1, k_2, k_3$ and their conjugate pairs $N-k_j$ interact if and only if $k_1 + k_2 + k_3 = N$. Therefore, while a wavenumber $k_j$ and its conjugate pair $N-k_j$ share the same greatest common divisor $d$ with $R$ (i.e. they belong to the same disjoint component $\mathcal{S}_N^{(d)}$ of the super-cluster), it is not necessary that the three wavenumbers $k_1, k_2, k_3$ share the same greatest common divisor with $R$. For example, we could take wavenumber $k_1$ from component $\mathcal{S}_N^{(d_1)}$ and wavenumber $k_2$ from component $\mathcal{S}_N^{(d_2)}$, where $d_2 \neq d_1$ so $\mathcal{S}_N^{(d_1)} \cap \mathcal{S}_N^{(d_2)}= \varnothing$. The third wavenumber, $k_3 = N - k_1-k_2$, will belong to some component $\mathcal{S}_N^{(d_3)}$, where  
$d_3$ is restricted by the condition $\gcd(d_1,d_2) \divides d_3$. If $\gcd(d_1,d_2)=1$ then this restriction is trivial. For example, consider $N = 3\cdot 5\cdot 7\cdot 13$ and take $k_1= 7 \in \mathcal{S}_N^{(7)}$,   $k_2= 13 \in \mathcal{S}_N^{(13)}$ and thus $k_3 = N - 7 - 13 = N - 20 \in \mathcal{S}_N^{(5)}$. 

In conclusion, to achieve full-scale mixing and thermalisation it is necessary to go to $6$-wave resonant interactions. If $3\divides N$ and $N>6$, then $5$-wave resonant interactions may provide partial mixing and thermalisation within separated regions of wavenumber space, the largest one being $\mathcal{S}_N^{(1)}$ (or $\mathcal{S}_N^{*}$ if $R=1$ or $R=5$), defined as the component of the super-cluster whose wavenumbers $k$ do not have any common factors with $N$, except for the factors $1, 2, 3$ (and $5$, if $R=5$) or products of powers thereof.\\

\noindent \textbf{The $5$-wave resonances  in the large-$N$ limit.} To see how results could depend sensitively on the number of particles, notice that all numerical studies we are aware of have been performed for the cases $N=2^b$, i.e. $N$ is a power of $2$. In such a case the only possible resonances are of pairing-off form, as demonstrated in Corollary \ref{cor:triv} and explicitly constructed in Section \ref{sec:pairing-off}. In such a case, $4$-wave and more importantly $6$-wave resonances will produce significant energy mixing across scales (and thus thermalisation) because the resonant sextuplets,  equations (\ref{eq:N_even_6-wave_sol_k_branch_1})--(\ref{eq:N_even_6-wave_sol_k_branch_3}), have $2$ free parameters so there are $\mathcal{O}(N^2)$ sextuplets, covering uniformly and surjectively the universe of available wavenumbers, of size $\mathcal{O}(N)$; therefore there will be a huge number of connections across resonant sextuplets, much like in the well-known case of triad interactions in the Euler fluid equations. The time scale of the thermalisation process is determined by the order of the resonance: $6$-wave resonant interactions in this case. 

However, changing the number of particles by just one unit can completely change the picture. Writing $N = 2^b-(-1)^b$ for integer $b\geq 4$, we can show $2 \ndivides N$, $3 \divides N$ and $N \geq 15$, so $5$-wave interactions would take place which could generate partial thermalisation within connected components of the super-cluster. These interactions would be a bit faster than $6$-wave interactions so they could dominate the dynamics for quite a while before $6$-wave interactions take over. It will be interesting to compare these near-power-of-two cases as the integer $b$ runs over a large interval: $b \in [4,200]$, corresponding to $N$ ranging from $15$ to about $1.6 \cdot 10^{60}$. Writing as usual $N=3^{b_1} \cdot R$ where $2\ndivides R$ and $3 \ndivides R$,  the quantities to study are:\\
(i) The total number of components in the super-cluster, $D(\mathcal{S}_N) \equiv \# \{d \in \mathbb{N}: \,\, d \divides R \,\, \land \,\, d < R/5\} $ ($+1$ if $\mathcal{S}_N^{*}$ is non-empty), which is roughly equal to the total number of divisors of $R$, denoted in the literature as $\sigma_0(R)$ or $d(R)$. In attention to the fact that $R/5$ and $R$ are gathered into one component $\mathcal{S}_N^{*}$, and that in some instances $\mathcal{S}_N^{*}$ is empty, the total number of components in the super-cluster may differ from $\sigma_0(R)$ by one unit in some instances.\\
(ii) The total amount of wavenumbers that belong to $\mathcal{S}_N^{(1)}$, the largest component of the super-cluster. This is equal to the size of the set $\mathcal{A}_N^{(1)}$ as defined in equation (\ref{eq:A_N^d}). In terms of $R$ this is equal to
$$ \#\mathcal{A}_N^{(1)} \,\,\,= \,\,\,\#\{\ell \in \mathbb{N} : \,\, 1\leq \ell < 3^{b_1} \cdot R \,\, \land \,\, \gcd(\ell,R)=1\} \,\,\,=\,\,\, 3^{b_1} \cdot \phi(R)\,,$$
where $\phi(n)$ is Euler's totient function, defined below in equation (\ref{eq:totient}).\\

\begin{figure}[h]
\begin{center}
\vspace{-0.1cm}
\includegraphics[width=0.45\textwidth]{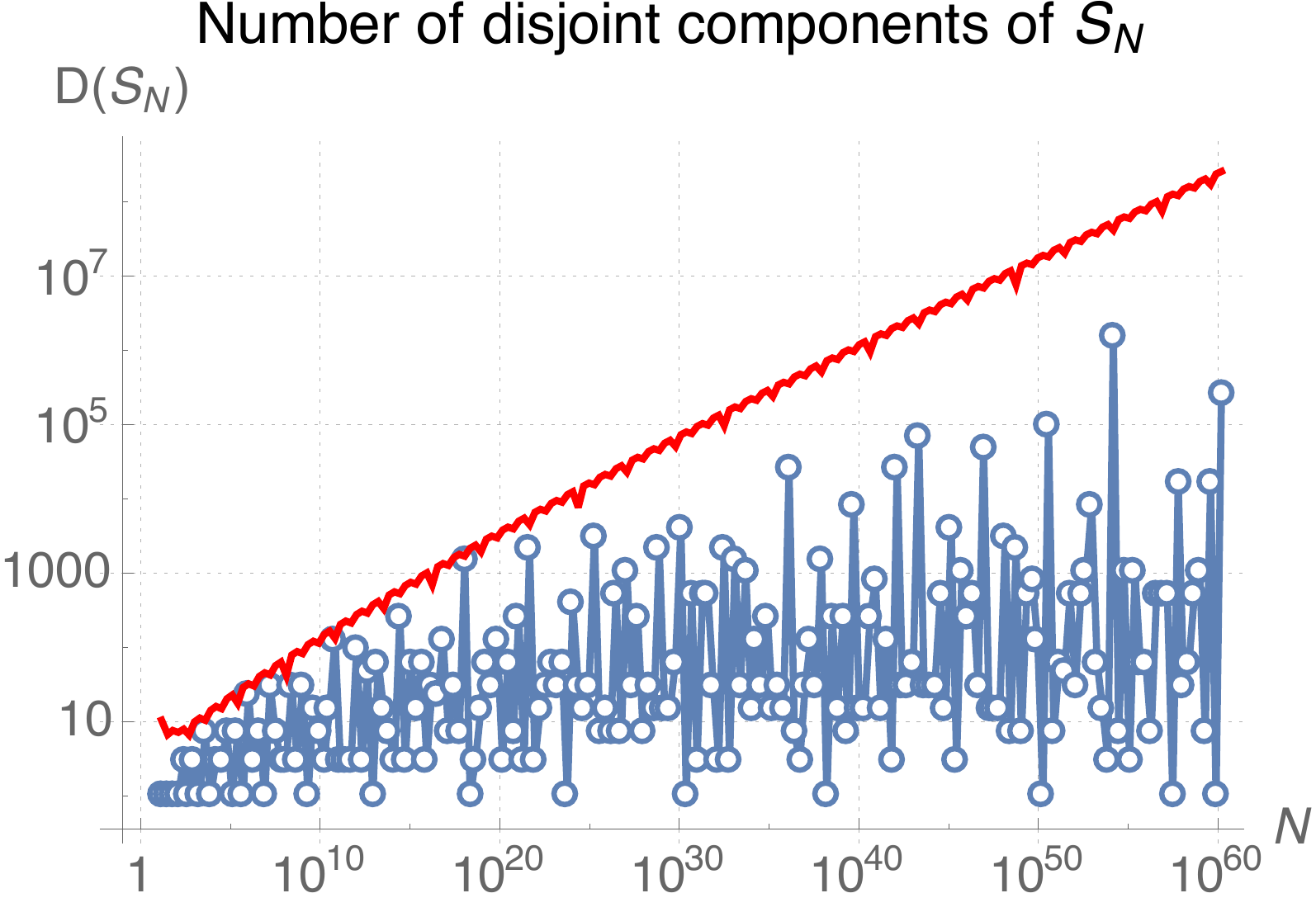}
\hfill
\includegraphics[width=0.45\textwidth]{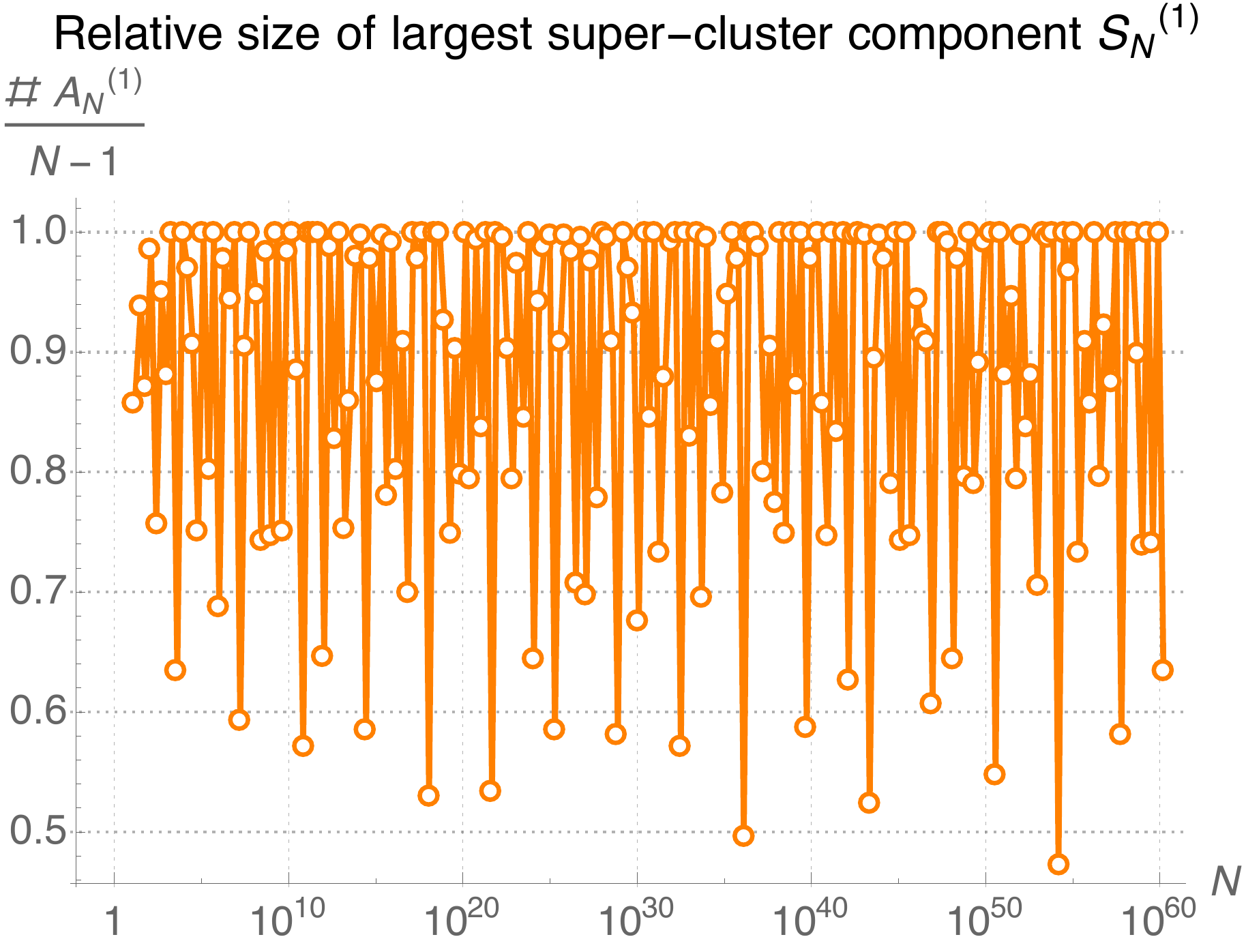}
\hfill
\includegraphics[width=0.45\textwidth]{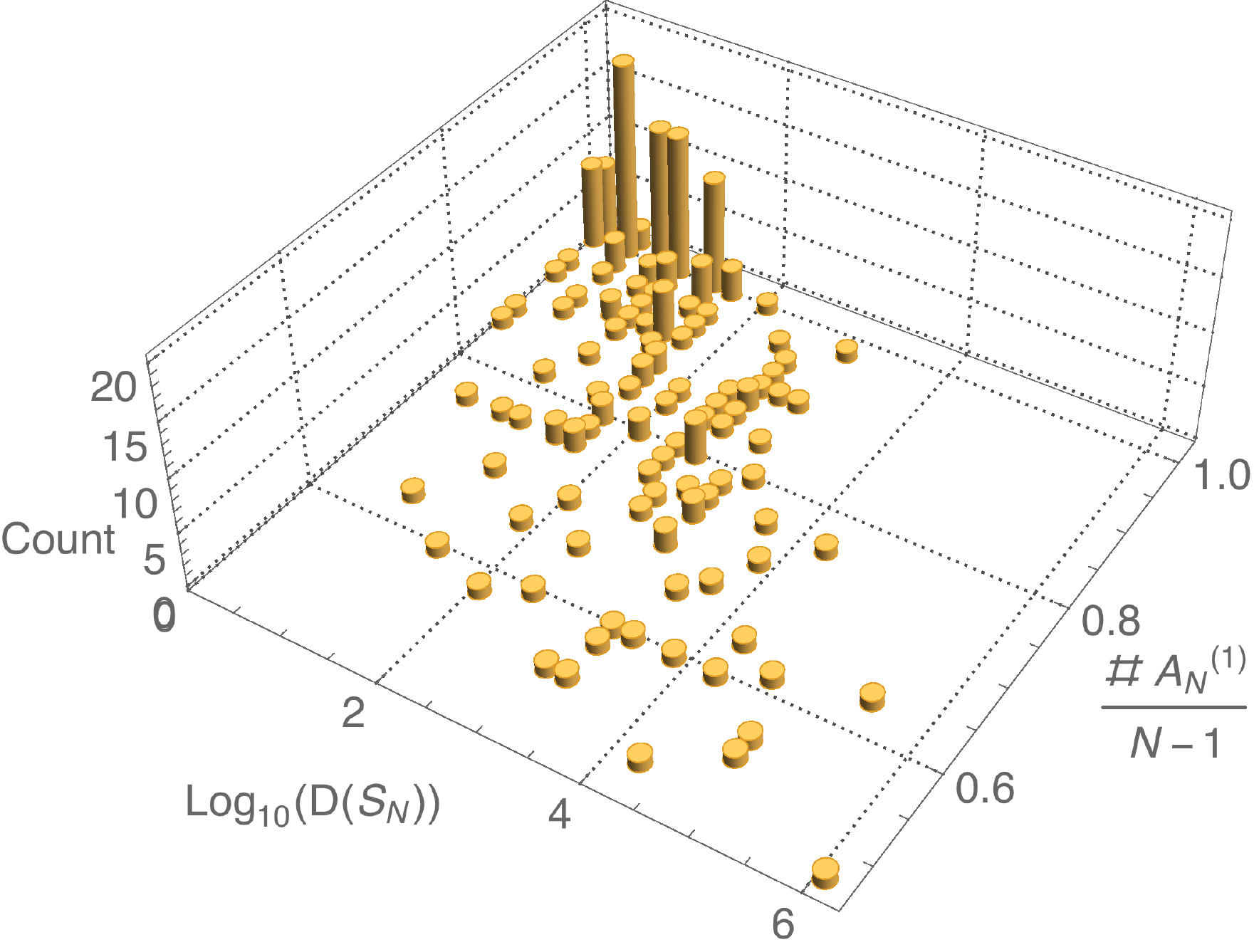}
\vspace{-0.9cm}
\end{center}
\caption{\label{fig:large-N} Colour online. Left panel: Log-log plot of $D(\mathcal{S}_N)$, the total number of disjoint components of the super-cluster $\mathcal{S}_N$, as a function of the number of particles $N$ of the form $N=2^b-(-1)^b$, for $b \in [4,200]$ (blue solid lines and open circles). The upper bound (red solid line) is Wigert's bound \cite{wigert1907ordre} which in our notation reads $D(\mathcal{S}_N) \approx \exp\left(\frac{\ln(2) \ln(R)}{\ln(\ln(R))}\right)$, where $N=3^{b_1} \cdot R$. Right panel: Log-linear plot of $\# \mathcal{A}_N^{(1)}/(N-1)$, namely the proportion of wavenumbers in the largest super-cluster component $\mathcal{S}_N^{(1)}$ (with respect to the total of available modes, $N-1$), as a function of the number of particles $N$ of the form $N=2^b-(-1)^b$, for $b \in [4,200]$.  Bottom panel: Combined 3D histogram of the two quantities $D(\mathcal{S}_N)$ and $\# \mathcal{A}_N^{(1)}/(N-1)$.}
\end{figure}

Figure \ref{fig:large-N} (left panel) plots in blue open circles the number of disjoint components of the super-cluster, $D(\mathcal{S}_N)$, for near-power-of-two values of $N$ ranging from $15$ to $1.6 \times 10^{60}$. The function oscillates wildly between a lower bound of $1$ component (corresponding to the case $N=3^{b_1}\cdot p$ with $p$ an odd prime) and an upper bound that grows as $N$ grows, reaching up to $\approx 1.57 \times 10^6$ at $N = 2^{180}-1\approx 1.5 \times 10^{54}$. In the same plot, an upper bound is shown (red solid line),  representing Wigert's  \cite{wigert1907ordre} maximal order of $\sigma_0(R)$, namely $\ln(2) \ln(R)/\ln(\ln(R))$, which satisfies
$$\limsup_{n\to\infty}\frac{\sigma_0(n)}{\ln(2) \ln(n)/\ln(\ln(n))} = 1.$$
 As evidenced in the figure, this bound on $\sigma_0(R)$ is nearly saturated only a couple of times at early values of $N$ ($6.7 \times 10^{10}$ and $1.15 \times 10^{18}$). 
 
Figure \ref{fig:large-N} (right panel) plots the relative size of the largest component of the super-cluster, for near-power-of-two values of $N$ ranging from $15$ to $1.6 \times 10^{60}$. First of all, for many values of $N$, and uniformly over the studied range, this ratio is close to $1$, meaning this single super-cluster component $\mathcal{S}_N^{(1)}$ dominates over all the other components. This is reinforced by figure \ref{fig:large-N} (bottom panel), which shows that cases with ratio $=1$ occur with high probability. Second, there is an evident spread of the ratio, from $1$ down to $0.473$ at $N = 2^{180}-1\approx 1.5 \times 10^{54}$. This point corresponds also to the largest instance of the number of components $D(\mathcal{S}_N)$, which is not a coincidence as figure \ref{fig:large-N} (bottom panel) demonstrates a significant correlation between low ratios and large number of components.

In conclusion, the large-$N$ limit of the FPUT system in the regimes dominated by $5$-wave resonances will have at most a probabilistic interpretation, in the sense that only averages over a given set of values of $N$ make sense. Any result will depend sensitively on the exact choice of the value of $N$. Even the way we take the limit will matter: the case $N=2^b, \,\,b\to \infty$ is dominated by $6$-wave resonances and the case $N=2^b \cdot 3, \,\, b\to \infty$ is dominated by $5$-wave resonances with a deceptively simple super-cluster given by just one component: $\mathcal{S}_N = \mathcal{S}_N^{*}$, because $R=1$. \\

{\noindent \textbf{The thermodynamic limit.} 
It is worth mentioning that in the thermodynamic limit, i.e. $N\rightarrow\infty$, $L\rightarrow\infty$, with $L$ the length of the chain, keeping constant the linear density of masses,  $4$-wave resonant interactions become the 
dominant process, see \cite{onorato2015route,pistone2018thermalization}. This is because the Fourier space  becomes dense and the re-scaled wavenumbers $k/N$ become real in the interval $(0, 1)$, allowing for a continuous resonant manifold.}

\section{Mathematical underpinnings of FPUT resonances: A basis for the real FPUT polynomials}
\label{sec:underpinnings}
In this Section we provide a rigorous mathematical construction of the general solution regarding $M$-wave resonances in the FPUT system. The main results are Theorems \ref{thm:realdim} and  \ref{thm:ker}, along with Corollaries \ref{cor:realdim} and \ref{thm:real}, 
 which establish a basis for the real FPUT polynomials defined in Section \ref{sec:RealPolynomials}. It is worth mentioning Corollary \ref{cor:triv}, which shows that if $N$ is a prime or a power of $2$, then the only possible resonances are of pairing-off form (discussed in Section \ref{sec:pairing-off}).

We begin with two definitions that will be essential in what follows. First, we recall Definition \ref{defi:Cyclotomic} of Cyclotomic polynomial $\Phi_{n}(x)$. Second, the following

\begin{defi}[Euler's totient function]
\label{defi:totient}
The degree of the cyclotomic polynomial $\Phi_{n}(x)$ is known as Euler's totient function $\phi(n),$ defined by
\begin{equation}
\label{eq:totient}
\phi(n) = n \prod_{p | n} \left(1-\frac{1}{p}\right)\,,
\end{equation}
where the product is over all primes $p$ that divide $n$. Notice that $\phi(n)\leq \frac{n}{2}$ when $n$ is even. If $n$ is a power of $2$, then $\phi(n) = \frac{n}{2}$. Also, $\phi(n)$ is even if $n \geq 3.$
\end{defi}

Although there is no general formula to obtain the cyclotomic polynomial $\Phi_{n}(x)$ for arbitrary $n \in \mathbb{N},$ it is possible to construct it algorithmically in a very fast way. For example, \emph{Mathematica} software as has a built-in function $\texttt{Cyclotomic[n,x]}$ which produces $\Phi_{n}(x)$.

For a given number of particles $N$, define as usual $\zeta = \exp(i \pi/N)$. Define also $\phi= \phi(2N)$, the Euler's totient function of $2N$. The basis for the real FPUT polynomials $\rho(x)$ has dimension $N-\phi/2-1$ and consists of two parts: 
\begin{enumerate}
\item The basis $\mathcal{K}_N$ of the kernel of the map $\rho(x) \to \rho(\zeta)$, with dimension $N-\phi$.
\item Any complementary basis $\mathcal{P}_N$, made out of pairing-off binomials, with dimension $\phi/2-1$. For example we can choose
$$\mathcal{P}_N = \{x-x^{N-1},\ldots, x^{\phi/2-1}-x^{N-\phi/2+1}\}\,.$$
\end{enumerate}

\subsection{Integer basis for real polynomials}
As above, $\zeta\in \mathbb{C}$ denotes a primitive $2N$-th root of unity.

We wish to construct or enumerate all real polynomials; i.e. all integer polynomials 
\begin{equation}\nonumber
\rho(x)=\sum_{k=1}^{N-1}\rho_kx^k
\end{equation}
which evaluate to real numbers when $\zeta$ is substituted for $x$.

Let $P$ denote the additive group
 of all polynomials in $x$ of degree at most $N-1$ with integer coefficients. By definition,
every element of $P$ has a unique expression of the form $\sum_{k=0}^{N-1}\rho_kx^k$ with 
$\rho_k\in \mathbb{Z}$; we say that $\{ 1,x,\ldots, x^{N-1}\}$ is a $\mathbb{Z}$-basis of $P$. 
(Since it has a $\mathbb{Z}$-basis with $N$ elements, $P$ is said 
to be a (free abelian) group of \emph{rank} $N$.)

Let $C$ denote the 
additive subgroup of $\mathbb{C}$ consisting of all polynomials with integer coefficients
 of any degree, in $\zeta$.

The following (well-known) theorem 
establishes the dimension and appropriate bases of the group $C$ and its subgroup of real elements.

\begin{thm}\label{thm:cyc}

(i) $\zeta^r\in C$ for all $r\in \mathbb{Z}$.  Furthermore, for any $r\in \mathbb{Z}$ the set
\begin{equation}\nonumber
\{ \zeta^r,\zeta^{r+1},\ldots,\zeta^{r+\phi-1}\}
\end{equation}
 is a $\mathbb{Z}$-basis of $C$.\\
(ii) Let $C^+:=C\cap\mathbb{R}$. Then $C^+$ is a subgroup of $C$ of rank $\phi/2$. The set
\begin{equation}\nonumber
\{ 1,\zeta+\zeta^{-1},\ldots, \zeta^{\phi/2-1}+\zeta^{1-\phi/2}\}
\end{equation}
is a $\mathbb{Z}$-basis of $C^+$.
\end{thm}
\emph{Proof}\\
(i) First note that since $\zeta^{2N}=1$, we have $\zeta^{r}=\zeta^{r+2tN}$ for any $t$ and thus 
$\zeta^r\in C$ even for negative $r$. 

It follows that multiplication by $\zeta^r$ is an isomorphism from $C$ to itself (whose inverse is
multiplication by $\zeta^{-r}$). So it is enough to show that $1,\zeta,\ldots, \zeta^{\phi-1}$ is a 
$\mathbb{Z}$-basis of $C$. 

Now $\Phi_{2N}(x)\in P$ is an integer polynomial of degree $\phi$ with leading term $x^\phi$. Since 
$\Phi_{2N}(\zeta)=0$, we obtain a relation $\zeta^{\phi}= \sum_{i=0}^{\phi-1}a_i\zeta^i$ so that 
$\zeta^\phi$ lies in the integer span of $\{ \zeta^0,\zeta^{1},\ldots,\zeta^{\phi-1}\}$. By induction, 
suppose that $\zeta^t$, $t\geq \phi$, lies in this span. Then $\zeta^{t+1}=\zeta\cdot \zeta^t$ lies 
in the span of $\{\zeta^{1},\ldots,\zeta^{\phi}\}$ and hence also in the span of 
$\{ \zeta^0,\zeta^{1},\ldots,\zeta^{\phi-1}\}$.

Finally, suppose that $\sum_{i=0}^{\phi-1}a_i\zeta^i=0$. Then $a_i=0$ for all $i$, since $\zeta$ is not a root 
of any nonzero rational polynomial of degree less than $\phi$. \\
(ii) Since $\zeta^{1-\phi/2},\zeta^{2-\phi/2}\ldots, 1,\zeta,\ldots, \zeta^{\phi/2-1},\zeta^{\phi/2}$ is 
a $\mathbb{Z}$-basis of $C$ by part 1, it follows that the real numbers 
\begin{equation}\nonumber
1,\zeta+\zeta^{-1},\ldots, \zeta^{\phi/2-1}+\zeta^{1-\phi/2}
\end{equation} 
are linearly independent over $\mathbb{Z}$. So the span of these elements is contained in $C^+$ 
and is linearly 
independent. We need to show that this span is all of $C^+$:

Suppose that 
\begin{equation}\nonumber
\sigma=a_{1-\phi/2}\zeta^{1-\phi/2}+\cdots + a_0+a_1\zeta+\cdots + a_{\phi/2-1}\zeta^{\phi/2-1}+a_{\phi/2}\zeta^{\phi/2}\in 
C
\end{equation}
is real; i.e. suppose that $\sigma\in C^+$. Then, since $\bar{\zeta^i}=\zeta^{-i}$, we have  
\begin{equation}\nonumber
\sigma=\bar{\sigma}= a_{\phi/2}\zeta^{-\phi/2}+a_{\phi/2-1}\zeta^{1-\phi/2}+\cdots a_{1-\phi/2}\zeta^{\phi/2-1}
\end{equation}
and hence, comparing terms, we get  
\begin{equation}\nonumber
a_i=a_{-i}\mbox{ for } 0\leq i\leq \phi/2, \mbox{ and } a_{\phi/2}=0
\end{equation}
by the linear independence of part 1. 

It follows that $\sigma=a_0+\sum_{i=1}^{\phi/2-1}a_i(\zeta^i+\zeta^{-i})$ lies in the span of 
\begin{equation}\nonumber
\{ 1,\zeta+\zeta^{-1},\ldots, \zeta^{\phi/2-1}+\zeta^{1-\phi/2}\}
\end{equation}
 as required. \qed\\

From the fact that $\zeta^N = -1$ we can rewrite this last $\mathbb{Z}$-basis of $C^+$ in terms of 
positive powers of $\zeta:$
\begin{equation}
\label{eq:B+_basis}
B_+ = \{ 1,\zeta-\zeta^{N-1},\ldots, \zeta^{\phi/2-1}-\zeta^{N-\phi/2+1}\}\,.
\end{equation}
Theorem \ref{thm:cyc} implies that the individual powers of $\zeta$ appearing in the above basis, 
along with the element $\zeta^{N-\phi/2},$ form a $\phi$-dimensional $\mathbb{Z}$-basis of $C:$
\begin{equation}\nonumber
B = \{ 1,\zeta,\ldots, \zeta^{\phi/2-1}; \zeta^{N-\phi/2},\zeta^{N-\phi/2+1},\zeta^{N-1}\}\,.
\end{equation}

From the fact that the elements of $B$ are linearly independent it follows that a polynomial supported by 
the basis $B$ is real if and only if it is supported by the basis $B_+.$ In particular, 
the coefficient of $\zeta^{N-\phi/2}$ of a real polynomial supported by the basis $B$ is identically zero:

\begin{cor} The expression
\begin{equation}\nonumber
a_0+ \sum_{k=1}^{\phi/2-1}a_k\zeta^k + \sum_{k=N-\phi/2}^{N-1}a_k\zeta^k\in C
\end{equation}
is real if and only if $a_k=a_{N-k}$ for $k=1,\ldots, \phi/2-1$ and $a_{N-\phi/2}=0$: i.e.
 if and only if the terms pair off (see definition of pairing off in Section \ref{sec:pairing-off}).
\end{cor}

\subsection{Finding real polynomials}
\subsubsection{Trivially real polynomials and pairing-off resonances}
\newcommand{\kernel}{\mathcal{K}}
\newcommand{\real}{\mathcal{R}}
Let $\real$ denote the additive subgroup of $P$ consisting of real polynomials with integer 
coefficients; 
i.e. the group of polynomials 
$\rho(x)$ such that $\rho(\zeta)\in \mathbb{R}$.

Our interest 
is in calculating the real polynomials which do not arise from pairing 
off terms. The key to doing this is to understand the kernel of the evaluation map 
$$P\to C,\  \alpha(x)\mapsto \alpha(\zeta).$$
We denote this kernel by $\kernel$.

This evaluation map  is a homomorphism of groups (even of rings).
By Theorem \ref{thm:cyc} it is a surjective map. It follows that its
 kernel, $\kernel$,  is a free abelian group of rank $N-\phi$. 

We thus have:

\begin{thm}\label{thm:realdim} 
Let $B$ be any $\mathbb{Z}$-basis of $\kernel$. Then 
\begin{equation}\nonumber
B\cup\{ 1,x-x^{N-1},\ldots, x^{\phi/2-1}-x^{N-\phi/2+1}\}
\end{equation}
is a $\mathbb{Z}$-basis of the group of real polynomials $\real$. 

In particular, the group of real polynomials has rank $N-\phi/2$.
\end{thm}
\emph{Proof.} Suppose that $\alpha(x)\in \real$. Then $\alpha(\zeta)\in \mathbb{R}$. By Theorem \ref{thm:cyc} (ii), there exist unique $\lambda_k\in\mathbb{Z}$ such that 
\begin{equation}\nonumber
\alpha(\zeta)= \lambda_0+ \sum_{k=1}^{\phi/2-1}\lambda_k(\zeta^k-\zeta^{N-k}).
\end{equation}  
Let $\beta(x):=\lambda_0 +\sum_{k=1}^{\phi/2-1}\lambda_k(x^k-x^{N-k})$. Then $\alpha(x)-\beta(x)\in\kernel$.
Hence there exist $\mu_p\in \mathbb{Z}$ such that 
\begin{equation}\nonumber
\alpha(x)-\beta(x)= \sum_{p(x)\in B}\mu_pp(x).
\end{equation}
Thus 
$\alpha(x)= \sum_{p(x)\in B}\mu_pp(x)+ \lambda_0 +\sum_{k=1}^{\phi/2-1}\lambda_k(x^k-x^{N-k})
$. \qed\\

On the other hand, let us say that $\alpha(x)\in \real$ is \emph{trivially real} if it is of the form 
\begin{equation}\nonumber
\alpha(x)=\lambda_0+ \sum_{k=1}^{\lfloor \frac{N-1}{2}\rfloor}\lambda_k(x^k-x^{N-k}).
\end{equation}
Thus a trivially real polynomial is precisely one which gives rise to a \emph{pairing-off resonance} (see Section \ref{sec:pairing-off}).
We observe that the group of trivially real polynomials has basis 
\begin{equation}\nonumber
\{ 1,x-x^{N-1},\ldots, x^{\lfloor \frac{N-1}{2}\rfloor}-x^{N-\lfloor \frac{N-1}{2}\rfloor}\}.
\end{equation}
In particular this is a free abelian group of rank $\lfloor \frac{N-1}{2}\rfloor+1=\lceil N/2\rceil$.

\begin{cor}\label{cor:triv} 
Every real polynomial is trivially real if and only if 
\begin{equation}\nonumber
\phi/2=N-\lceil N/2\rceil.
\end{equation}
This happens if and only if either $N$ is a power of $2$ or an odd prime number $p$. 
\end{cor}
\emph{Proof.} We begin by noting that $\phi/2\leq N-\lceil N/2\rceil$ with equality 
if and only if either $N$ is a power of $2$ or an odd prime number $p$.

For if $N=2^ap_1^{b_1}\cdots p_t^{b_t}$ then $\phi=\phi(2N)=N\cdot (1-1/p_1)\cdots(1-1/p_t)$. Thus 
$\phi\leq N$ with equality only if $t=0$; ie. $N$ is a power of $2$.  

When $N$ is even, we have $N-\lceil N/2\rceil=N/2$ and $\phi/2\leq N/2$ with equality if and only if 
$N$ is a power of $2$.

When $N$ is odd, we have $N-\lceil N/2\rceil=(N-1)/2$ and $\phi/2\leq (N-1)/2$  with equality if and 
only if $\phi=N-1$; i.e. if and only if $N=p_1$ is prime. 

Let $\mathcal{T}\subset \real$ be the group of trivially real polynomials. We have 
rank$(\mathcal{T})= \lceil N/2\rceil$ and rank$(\real)=N-\phi/2$, so that 
$\mathcal{T}\subset \real$ with equality only if  $N=2^b$ or $N=p$. 

Observe that if $N=2^b$, then $\phi=N$ so that $\kernel=\{ 0\}$ and $\mathcal{T}=\real$ by 
the proof of Theorem \ref{thm:realdim}, while if $N=p$ then $N-\phi=1$ and $\{ \Phi_{2p}(x)\}$ is 
a basis of $\kernel$.
However 
\begin{equation}\nonumber
\Phi_{2p}(x)=1-x+x^2-\cdots -x^{p-1}=1-(x-x^{p-1})+\cdots \pm(x^{(p-1)/2}-x^{(p+1)/2})\in \mathcal{T}.
\end{equation} \qed

\begin{rem} In general we have 
\begin{equation}\nonumber
\mathrm{rank}(\real)-\mathrm{rank}(\mathcal{T})=N-\phi/2-\lceil N/2\rceil 
=\left\{
\begin{array}{ll}
\frac{N-\phi}{2},& N \mbox{ even }\\
\frac{N-\phi-1}{2},& N \mbox{ odd }\\
\end{array}
\right.
\end{equation}
This is thus the number of independent real polynomials we expect to find which do not correspond to 
pairing-off resonances.
\end{rem}

In fact, for our purposes in this paper, we are interested in the additive group $\real_0$ of 
real polynomials with constant term zero. Theorem \ref{thm:realdim} can easily be 
amended to apply to this group:

\begin{cor}\label{cor:realdim} For any polynomial $\rho=\rho(x)$ we let 
$\tilde{\rho}:= \rho(x)-\rho(0)$.  Let $B$ be any $\mathbb{Z}$-basis of $\kernel$. Let 
$\tilde{B}:= \{ \tilde{\rho}\ |\ \rho\in B\}$. Then 
\begin{equation}\nonumber
\tilde{B}\cup\{ x-x^{N-1},\ldots, x^{\phi/2-1}-x^{N-\phi/2+1}\}
\end{equation}
is a $\mathbb{Z}$-basis of the group $\real_0$ of real polynomials with zero constant term.  

\end{cor}

\subsubsection{A basis for the kernel of the evaluation map}
\label{sec:basis_of_kernel}
The results of this Subsection have motivated the construction of $5$-wave resonances via the cyclotomic method (Section  \ref{sec:cyclotomic_method}). 
By Theorem \ref{thm:realdim}, in order to find all real polynomials we must construct a  basis 
of the kernel $\kernel$.  Here is one such basis: 
\begin{prop}\label{prop:ker}
Suppose that $N$ is divisible by an odd prime. Then 
$\phi<N$. Let $F_i(x):= x^i\Phi_{2N}(x)$ for $i=0,\ldots, N-\phi-1$. 
The set 
$\{ F_i(x)\}_{i=0}^{N-\phi-1}$ is a basis for $\mathcal{K}$. 
\end{prop}
\emph{Proof.} For each $i$, $F_i(\zeta)=\zeta^i\Phi_{2N}(\zeta)=0$. Furthermore the degree of $F_i(x)$ is 
$\phi+i$. Thus the $F_i(x)$ belong to $\kernel$ and they are linearly independent since their 
degrees all differ. Since there are $N-\phi$ of them and since they are monic polynomials, 
it follows that they form a basis of the kernel space. \qed
\begin{rem} 
If, on the other hand,  $N=2^r$, then $N=\phi$ and $\kernel=0$. 
\end{rem}
\vskip 10pt

However, the polynomials $F_i(x)$ are generally complicated and long, since the general cyclotomic 
polynomial can be complicated and unpredictable.
For this reason, we now construct a basis such that the  polynomials belonging to it are 
particularly 
short and easy to write down explicitly. 
Each of them has length $p_i$ for some prime $p_i$ dividing $N$.

Before describing this basis of $\kernel$, we make some 
observations and introduce some notation:
 
Suppose that the prime $p$ divides $2N$. Then $\zeta_p:=\zeta^{2N/p}$ is a primitive $p$th root of unity 
and hence 
$0=\Phi_p(\zeta_p)=\Phi_p(\zeta^{2N/p})=1+\zeta^{2N/p}+\zeta^{2\cdot (2N/p)}+\cdots + \zeta^{(p-1)\cdot (2N/p)}
$. It follows that $\Phi_p(x^{2N/p})\in \kernel$. 

More generally, any polynomial in $P$ of the form $x^s\Phi_p(x^{2N/p})$ will lie in $\kernel$. We will 
describe a basis for $\kernel$ in terms of these polynomials.  Note that if  $\alpha = x^s\Phi_p(x^{2N/p})$ then $\mathrm{deg}(\alpha) =s+ \frac{2N(p-1)}{p}\geq N$ so 
that $\alpha(x)\not\in P$ since its 
degree is greater than $N-1$. However, for convenience, 
we will interpret such a polynomial as an element of 
$P$ in the following way: if the coefficient of $x^k$ in $\alpha$ is nonzero,
 and if $k=tN+r$, with $r\leq N-1$, then 
replace $x^k$ with $(-1)^tx^r$.  In fact, strictly speaking, 
we are working in the ring $\mathbb{Z}[x]/\langle x^N+1\rangle$ and choosing a lowest degree representative:
see \ref{app:ker}. 

\begin{thm}\label{thm:ker}
Suppose that $2N=2^{b_0}p_1^{b_1}\cdots p_t^{b_t}$ where the $p_i$ are distinct odd primes. 

Let 
$x_i=x^{2N/p_i^{b_i}}$ for $i=0,\ldots, t$ (where $p_0=2$). Let $\Phi_i:= \Phi_{p_i}(x^{2N/p_i})$. 

Then $\kernel=\{ 0\}$ if $N$ is a power of $2$.

Otherwise,
a $\mathbb{Z}$-basis of $\kernel$ is $E=E_1\cup\cdots\cup E_t$ where 
\begin{equation}\nonumber
E_i:=\{ x_0^{m_0}x_1^{m_1}\cdots x_t^{m_t}\Phi_i\ |\ 0\leq m_j\leq N_{j,i}\,\,, \quad j=0,\ldots, t\}
\end{equation} 
where 
\begin{equation}\nonumber
N_{j,i}:=\left\{
\begin{array}{ll}
 p_j^{b_{j}-1} - 1,& j=0,i\\
p_j^{b_j}-1,& 1\leq j<i\\
p_j^{b_j}-p_j^{b_j-1}-1,&i < j \leq t\,.\\
\end{array}
\right.
\end{equation}
\end{thm}

\emph{Proof}\
See \ref{app:ker} below.

As an immediate 
corollary to Theorems \ref{thm:realdim} and  \ref{thm:ker} and Corollary \ref{cor:realdim} 
we obtain:
\begin{cor}\label{thm:real}
\item
Any real polynomial with zero constant term 
$\alpha(x)\in \real_0$ has a unique representation of the form 
\begin{equation}\nonumber
\alpha(x)=\sum_{p(x)\in E}\mu_p(p(x)-p(0))+\sum_{k=1}^{\phi/2-1}\lambda_k(x^k-x^{N-k})
\end{equation}
with $\mu_p,\lambda_k\in \mathbb{Z}$. 
\end{cor}

\begin{rem} It is not hard to see that the polynomials $\Phi_i$ belong to the group $\mathcal{T}$ of 
trivially real polynomials; i.e. their terms pair off. 

However this is usually not the case for the polynomials $x^m\Phi_i$ for $m\geq 1$. 
\end{rem}

\begin{exa}  
We illustrate the theorem in the case $N=15=3\cdot 5=p_1\cdot p_2$.

Then $\Phi_1=\Phi_3(x^{10})=1+x^{10}+x^{20}=1-x^5+x^{10}$. So 
\begin{equation}\nonumber
E_1=\{ \Phi_1,x_2\Phi_1,x_2^2\Phi_1, x_2^3\Phi_1\}.
\end{equation}
Here $x_2=x^6$ and hence 
\begin{eqnarray*}
x_2\Phi_1=x^6\Phi_1&=& x^6-x^{11}+x^{16}= x^6-x^{11}-x\\
x_2^2\Phi_1=x^{12}\Phi_1&=& x^{12}-x^{17}+x^{22}= x^{12}+x^{2}-x^7\\
x_2^3\Phi_1=x^{18}\Phi_1&=& x^{18}-x^{23}+x^{28}=-x^3+x^8-x^{13}\\
\end{eqnarray*}

Thus
\begin{equation}\nonumber
E_1=\{1-x^5+x^{10}, x^6-x^{11}-x, x^{12}+x^{2}-x^7,-x^3+x^8-x^{13}\}.
\end{equation}

Similarly, $\Phi_2=\Phi_5(x^6)=1+x^6+x^{12}+x^{18}+x^{24}=1+x^6+x^{12}-x^3-x^9$. We have 
\begin{equation}\nonumber
E_2=\{\Phi_2,x_1\Phi_2,x_1^2\Phi_2\}=\{1+x^6+x^{12}-x^3-x^9, x^{10}-x-x^7-x^{13}+x^4, 
-x^5-x^{11}+x^2+x^8+x^{14}\}. 
\end{equation}

Of course, our basis $E=E_1\cup E_2$ consists of $N-\phi=15-8=7$ polynomials.  
\end{exa}

\begin{exa} \label{exa:simplest}
We will construct the simplest example of an irreducible resonance which 
is not a pairing-off resonance. 

We take $N=2\cdot 3=6$ (so that $\mathrm{rank}(\real)-\mathrm{rank}(\mathcal{T})=1$). This 
is of course the smallest value of $N$ which is neither prime nor a power of $2$. 
Then $\phi=\phi(12)=4$ and so $\kernel$ has rank $2$. 

A basis for $\kernel$ is $\{ \Phi_1, x^3\Phi_1\}$ where $\Phi_1=1+x^4+x^8=1+x^4-x^2$. 

The first of these is trivially real. 
The second of these gives the real (indeed, vanishing) -- but not trivially real -- polynomial 
\begin{equation}\nonumber
\sum_{k=1}^5\rho_k\zeta^k:= \zeta-\zeta^3+\zeta^5
\end{equation}
which gives the resonance 
\begin{equation}\nonumber
(\zeta-\zeta^{-1})+(\zeta^5-\zeta^{-5})=(\zeta^3-\zeta^{-3})
\end{equation}
with $M=3$. However this does not satisfy the momentum conservation condition 
since $\sum_kk\rho_k=3\not\equiv 0\pmod{6}$. 

We multiply by $2$ to obtain an irreducible resonance which is not a pairing-off resonance:
\begin{equation}\nonumber
2(\zeta-\zeta^{-1})+2(\zeta^5-\zeta^{-5})=2(\zeta^3-\zeta^{-3})
\end{equation}
\end{exa}
\vskip 10pt

\subsection{The case $N=2p$}\label{sec:2p}
By Corollary \ref{cor:triv} above if $N$ is either a power of $2$ or a prime, all wave resonances 
are of pairing-off type and hence the total number of interacting waves $M$ will always be even.

When $N=2p$ where $p$ is an odd prime, Corollary \ref{cor:triv} tells us that there 
are wave resonance solutions which are not of pairing-off type. 
Nevertheless, we can show that in this case also the total number of interacting waves $M$ must be even.    

\begin{prop}\label{prop:2p} Let $N=2p$ where $p$ is an odd prime number. Let $\rho(x)\in\real_0$ 
be a real polynomial with constant term $0$. Suppose that $\rho'(1)\equiv 0\pmod{N}$. 
Then the total number of interacting waves $M(\rho)$ is even. 

In other words, when $N=2p$ then any wave resonance solution must have even total number of interacting waves $M$. 
\end{prop}
\emph{Proof.} Let $\rho=\sum_{k=0}^{d}\rho_k\in \mathbb{Z}[x]$.
For any integer $m$ we have $m\equiv |m|\pmod{2}$. It follows that 
\begin{equation}\nonumber
M(\rho)=\sum_{k=1}^{d}|\rho_k|\equiv \sum_{k=1}^{d}\rho_k=\rho(1)\pmod{2}.
\end{equation}
 In particular, 
the map $\rho\mapsto M(\rho)\pmod{2}$ is linear. 

Furthermore, note that if $k=N+s$ for some $s\geq 0$, replacing $x^k$ by $-x^s$ does not 
affect $M(\rho)\equiv\rho(1)\pmod{2}$. Also, since $N$ is even, $\rho'(1)\pmod{2}$ is unchanged. 

Now suppose given a real polynomial $\rho=\sum_{k=1}^{N-1}\rho_k\in \real_0$. We \emph{Claim} that 
\begin{equation}\nonumber
\rho(1)\equiv\rho'(1)\pmod{2}.
\end{equation} 
 
By linearity, it is enough to prove this claim on a basis of $\real_0$. By Corollary 
\ref{cor:realdim}
and Theorem \ref{thm:ker} a basis is 
\begin{equation}\nonumber
\{ x^i-x^{N-i}\ | \ i=1,\ldots, p-2\} \cup \{ \Phi_1(x)-1, x^p\Phi_1(x)\}
\end{equation} 
where $\Phi_1(x)=\Phi_p(x^4)=1+x^4+\cdots + x^{4(p-1)}$. 

When $\rho=x^i-x^{N-1}$ then 
$\rho(1)=0$ while $\rho'(1)=2i-N\equiv 0\pmod{2}$. When $\rho= \Phi_1(x)-1$, 
$\rho(1)=p-1\equiv 0\pmod{2}$ and $\rho'(1)=4(\sum_{k=1}^{p-1}k)\equiv 0\pmod{2}$. 
When $\rho= x^p\Phi_1(x)$, $\rho(1)=\Phi_1(1)=p\equiv1\pmod{2}$ and 
$\rho'(1)\equiv \Phi_1'(1)+\textcolor{blue}{p}\,\Phi_1(1)\equiv 0+1\equiv 1\pmod{2}$. This proves the 
\emph{Claim}.

Thus if $\rho'(1)\equiv 0\pmod{N}$ then $\rho'(1)\equiv 0\pmod{2}$, since $N$ is even. Thus 
$M(\rho)\equiv \rho(1)\equiv 0\pmod{2}$ by the \emph{Claim}. \qed
\begin{rem}
If $N$ is odd but not prime, there are always wave resonance solutions with odd wave number: 
There will be a basis vanishing polynomial of the form $\rho=x^i\Phi_p(x^k)$ with $i\geq 1$ and 
$p$ a prime divisor of $N$. Then $M(\rho)\equiv \rho(1)\equiv p\equiv 1\pmod{2}$ so that 
$M(\rho)$ is odd.  Then $N\rho$ is a solution of the momentum conservation condition 
whose wave number $N\cdot M(\rho)$ is again odd.
\end{rem} 

\begin{rem} The argument in the proof of  Proposition \ref{prop:2p} 
does not extend to the case that 
$N=2K$ where $K$ is odd but not a prime. In this case $K=p_1^{b_1}\cdots p_t^{b_t}$ where either 
$b_1\geq 2$ or $t\geq 2$. Either way, by Theorem \ref{thm:ker} there is a basis element 
of $\kernel$ of the form $\alpha(x)=x^i\Phi_p(x^k)$ where $p$ is a prime dividing $K$ and $i\geq 2$ 
is even. Then $\alpha'(1)$ is even but $\alpha(1)=p$ is odd. It follows that 
$\rho=K\alpha$ satisfies the momentum conservation condition $\rho'(1)\equiv 0\pmod{N}$ and has 
odd wave number.
\end{rem}
\begin{rem}
Likewise, the argument of Proposition \ref{prop:2p} does not extend to the case where 
$N=4p$ with $p$ an odd prime. In this case, by Theorem \ref{thm:ker} there is a basis polynomial 
of $\real_0$ of the form $\alpha(x)=x^{2p}\Phi_p(x^8)$. So $\alpha(1)=p$ but $\alpha'(1)$ is of the 
form $2p^2+8s\equiv 2\pmod{4}$. If we let $\beta(x)=\alpha(x)+(x-x^{N-1})$ then $\beta(1)$ is odd, 
but $\beta'(1)=\alpha'(1)+(2-N)$ is divisible by $4$. Hence $\rho(x)=p\beta(x)$ satisfies the 
momentum conservation condition and has odd wave  number $M(\rho)$.
 \end{rem}

We now find a simplified basis of $\real_0$ in the case $N=2p$.

\begin{lem} \label{lem:x2m}
Let $m\geq 1$ be any odd number. Then \\
(i) $x^4+x^8+\cdots +x^{4(m-1)}\equiv \sum_{i=1}^{\frac{m-1}{2}}(-1)^i(x^{2i}-x^{2(m-i)})\pmod{
x^{2m}+1}$.\\
(ii) $x^{m+4}+x^{m+8}+\cdots + x^{m+4(m-1)}\equiv (-1)^{\frac{m+1}{2}}\sum_{i=1}^{\frac{m-1}{2}}
(-1)^i(x^{2i-1}+x^{2m-(2i-1)})\pmod{x^{2m}+1}$.
\end{lem}
\emph{Proof.} Recall that for any $s\geq 1$, $x^s= (-1)^{\lfloor s/2m\rfloor}x^r\pmod{x^{2m}+1}$ where 
$s =r\pmod{2m}$ and $0\leq r <2m$. \\
(i) If $t\leq \frac{m-1}{2}$ then $4t<2m$ and $\lfloor 4t/2m\rfloor=0$. In this case 
$x^{4t}=x^{2i}$ where $i=2t$ is even.

If $\frac{m+1}{2}\leq t\leq m-1$ then $2m < 4t<4m$ and $\lfloor 4t/2m\rfloor=1$. In this case 
$x^{4t}=-x^{4t-2m}=-x^{2j}\pmod{x^{2m}+1}$ where $j=2t-m$ is odd. Thus $j= m-i$ where $i=2(m-t)$ is 
even.\\
(ii) By (i), 
\begin{equation}
x^{m+4}+x^{m+8}+\cdots + x^{m+4(m-1)}=
\sum_{i=1}^{\frac{m-1}{2}}(-1)^ix^m(x^{2i}-x^{2(m-i)})=
\sum_{i=1}^{\frac{m-1}{2}}(-1)^i(x^{m+2i}-x^{3m-2i})\pmod{
x^{2m}+1}.
\end{equation}
For $i\leq \frac{m-1}{2}$, $m+2i<2m$ and $x^{m+2i}=x^{2j-1}$ where $j= i+\frac{m+1}{2}$. 
It follows that $(-1)^i=(-1)^{\frac{m+1}{2}}(-1)^j$. 
Likewise, for  $i\leq \frac{m-1}{2}$ we have $2m<3m-2i<3m$ so that 
$x^{3m-2i}=-x^{m-2i}=-x^{2m-(2j-1)}$ where $j= i+\frac{m+1}{2}$. \qed

\begin{cor}\label{cor:2pbasis}
 Let $N=2p$ where $p$ is an odd prime number. Then a basis for 
$\real_0$ is $\{ x^i-x^{N-i}\}_{i=1}^{p-1}\cup \{ \Psi(x)\}$ where 
\begin{equation}
\Psi(x)=x^p+(-1)^{\frac{p+1}{2}}\sum_{i=1}^{\frac{p-1}{2}}(-1)^i2x^{2i-1}.
\end{equation}
\end{cor}
\emph{Proof.} By Corollary \ref{thm:real} a basis for $\real_0$ is 
$\{ x^i-x^{2p-i}\}_{i=1}^{p-2}\cup \{ \Phi-1, x^p\Phi\}$ where $\Phi=\Phi_p(x^4)= 
1+x^4+\cdots +x^{4(p-1)}$. By Lemma \ref{lem:x2m} (i),
\begin{equation}
\Phi-1= \sum_{i=1}^{\frac{p-1}{2}}(-1)^i(x^{2i}-x^{2p-2i}).
\end{equation} 
We can replace $\Phi-1$ with 
$(\Phi-1)-\sum_{i=1}^{\frac{p-3}{2}}(-1)^i(x^{2i}-x^{2p-2i})= (-1)^{\frac{p-1}{2}}(x^{p-1}-x^{p+1}).
$

By Lemma \ref{lem:x2m} (ii), we have 
\begin{equation}
x^p\Phi=x^p+(-1)^{\frac{p+1}{2}}\sum_{i=1}^{\frac{p-1}{2}}(-1)^i(x^{2i-1}+x^{2p-(2i-1)}).
\end{equation}
We can thus replace $x^p\Phi$ with 
$x^p\Phi+(-1)^{\frac{p+1}{2}}\sum_{i=1}^{\frac{p-1}{2}}(-1)^i(x^{2i-1}-x^{2p-(2i-1)})=\Psi. 
$ \qed

\begin{prop} Let $N=2p$ where $p$ is an odd prime number. Let $\rho(x)\in\real_0$ satisfy 
$\rho'(1)\equiv 0\pmod{N}$. If $\rho$ is not trivially real then $M(\rho)\geq N$. 

In other words, when $N=2p$ any wave resonance solution which is not of pairing-off type 
has a total of at least $N$ interacting waves. 
\end{prop}
\emph{Proof.} If $\rho\in \real_0$ then Corollary \ref{cor:2pbasis} tells us that 
$\rho$ is uniquely expressible in the form 
$\rho= \sum_{i=1}^{p-1}\lambda_i(x^i-x^{N-i})+\mu\Psi(x)$. Thus 
\begin{eqnarray*}
M(\rho)&=& |\mu|+\sum_{i=1}^{\frac{p-1}{2}}(|\lambda_{2i-1}\pm 2\mu|+|\lambda_{2i-1}|)+ 
\sum_{i=1}^{\frac{p-1}{2}}2|\lambda_{2i}|\\
&\geq& |\mu|+\sum_{i=1}^{\frac{p-1}{2}}(|\lambda_{2i-1}\pm 2\mu|+|\lambda_{2i-1}|)\\
&\geq& |\mu|+\frac{p-1}{2}|2\mu|= p|\mu|.\\
\end{eqnarray*}
If $\rho'(1)\equiv 0\pmod{N}$ then Proposition \ref{prop:2p} tells us that $M(\rho)$ is 
even. But $M(\rho)\equiv \rho(1)\equiv \mu\pmod{2}$. Thus $\mu$ is even. 

Now $\rho$ is trivially real if and only if $\mu=0$. Therefore, if $\rho$ is not trivially real 
we have $|\mu|\geq 2$ and therefore $M(\rho)\geq p|\mu|\geq 2p=N$.  \qed

\begin{rem} On the other hand, whenever $N=2K$ there are many pairing-off $4$-wave resonances, such as 
$\rho= (x-x^{N-1})+(x^{K-1}-x^{K+1})$ (see Section \ref{sec:pairing-off}).
\end{rem}

\section{Conclusions}
\label{sec:conclusions}
{The numerical results observed by Fermi, Pasta, Ulam and Tsingou have triggered the discovery of new scientific paths, such as soliton dynamics and integrability in partial differential equations. However, the explanation of the numerical observations has remained elusive, and still is in some aspects,  after over 50 years. 
In this paper we have concentrated on the role played by exact resonant interactions on the dynamics that leads to thermalisation; at the moment such a role is only a conjecture, but it finds a solid foundation in the limit of weak nonlinearity via the Wave Turbulence Theory which is formally developed in the thermodynamic limit. In discrete low-dimensional chains the role played by exact discrete resonances seems 
to be very important (see \cite{onorato2015route,lvov2018double,pistone2018thermalization}); the results contained in these papers have stimulated us to investigate the possible discrete resonances allowed by the linear dispersion relation 
of the $\alpha$- and $\beta$-FPUT systems. In this regard we show that $N$, the number of particles, plays a crucial role via the set of its divisors. Up to date this result has remained unnoticed, possibly because, for computational reasons, most of the simulations in the literature have been performed with $N$ equal to a power of $2$. For extensive numerical/analytical studies over many values of $N$ (all of them powers of two) see, amongst others, \cite{giorgilli2005local, paleari2005equipartition, benettin2013fermi} and references therein.

The results and the methodology used in the present paper are very articulated:
Table \ref{tab:lowest_order} shows an overview of the most important results obtained. The 6-wave resonant interactions play an important role for many values of $N$. However, it turns out that, for some specific values of $N$ (namely when $N>6$ and $N$ is divisible by $3$), 5-wave resonances do exist and they may in principle lead to fast equipartition due to common-mode connections amongst $5$-wave resonances, forming a web of interconnected modes, which we term super-cluster (Definition \ref{def:super-cluster}). A quantitative assessment of this proposed equipartition using numerical simulations of the FPUT system is still unexplored.}
It would be interesting to observe the fast $5$-wave dominated dynamics as opposed to the slow $6$-wave dynamics, in order to gain insights on the route to thermalisation. Also, it would be interesting to study whether the multi-component nature of the super-cluster is evident in numerical simulations. Theoretically, one important study concerns the calculation of the maximal number of conservation laws (apart from the energy) for a given component of the super-cluster. Such extra conservation laws are known to arise, at any order, in a variety of resonant systems such as the Charney-Hasegawa-Mima equation, see \cite{harper2013quadratic}.

Finally, from a number-theoretical point of view there is still an open question: Does the cyclotomic method described in Section \ref{sec:cyclotomic_method}, to construct $5$-wave resonances out of real FPUT polynomials of length $3$, fulfil all possible $5$-wave resonances? We have checked by brute-force search that the answer is positive for $N \leq 1155 \,( = 3\cdot5\cdot7\cdot11)$, but for larger $N$ the wild nature of cyclotomic polynomials as the number of divisors of $N$ grows does not rule out the possibility that real FPUT polynomials of various lengths may combine to produce new resonant FPUT polynomials of length $5$.\\

{\bf Acknowledgments} 
M. O. has been funded by Progetto di Ricerca d'Ateneo CSTO160004. M.O. was supported by the ``Departments of Excellence 2018-2022'' Grant awarded by the Italian Ministry of Education, University and Research (MIUR) (L.232/2016).

\appendix

\section{Interaction matrix} \label{app:int}

After removing non-resonant three wave interactions, the matrix in the Zakharov equation is given by
 (see \cite{falkovich1992kolmogorov}):
 \begin{equation}
\begin{split}
 \bar T_{1,2,3,4}= &T_{1,2,3,4} -V_{1,3,1-3}V_{4,2,4-2}
\left[ \frac{1}{\omega_3+\omega_{1-3}-\omega_1}+
\frac{1}{\omega_2+\omega_{4-2}-\omega_4}\right] \\
& -V_{2,3,2-3}V_{4,1,4-1}
\left[ \frac{1}{\omega_3+\omega_{2-3}-\omega_2}+
\frac{1}{\omega_1+\omega_{4-1}-\omega_4}\right] \\  
& -V_{1,4,1-4}V_{3,2,3-2}
\left[ \frac{1}{\omega_4+\omega_{1-4}-\omega_1}+
\frac{1}{\omega_2+\omega_{3-2}-\omega_3}\right] \\    
& -V_{2,4,2-4}V_{3,1,3-1}
\left[ \frac{1}{\omega_4+\omega_{2-4}-\omega_2}+
\frac{1}{\omega_1+\omega_{3-1}-\omega_3}\right]  \\    
& -V_{1+2,1,2}V_{3+4,3,4}
\left[ \frac{1}{\omega_{1+2}-\omega_{1}-\omega_2}+
\frac{1}{\omega_{3+4}-\omega_{3}-\omega_4}\right] \\       
&  -V_{-1-2,1,2}V_{-3-4,3,4}
\left[ \frac{1}{\omega_{1+2}+\omega_{1}+\omega_2}+
\frac{1}{\omega_{3+4}+\omega_{3}+\omega_4}\right].
\label{coupl_coeff_zakh}     
\end{split}
\end{equation}

\section{Constructing $6$-wave resonances for arbitrary $N (\geq 6)$, which are not of pairing-off form}
\label{app:6-wave-cyclotomic}
We provide results of combining sums or subtractions of two real FPUT polynomials of length $3$, to obtain FPUT resonant polynomials of length six, i.e. $6$-wave resonances or, in short, sextuplets. These are not of pairing-off form so they correspond to truly different resonances with respect to the ones discussed in Section \ref{sec:6_wave_pairing_off}. As in the case of $5$-wave resonances, there will be a number of relevant cases.

\subsection{Case 0: $3 \ndivides N$.} In this case there are no $6$-wave resonances (which are not of pairing-off form) from our method.

\subsection{Case 1: $3 \divides N$ and $N \geq 6$.} 
In this case there are 5 one-parameter families of sextuplets. These are defined in table \ref{tab:6-waves} below.

\begin{table}[h]
\begin{center}
  \begin{tabular}{ | c | c | l | }
    \hline
     & Form of Sextuplets: \,\, $\{k_1,\ldots,k_S; k_{S+1},\ldots,k_{S+T}\}$ & Comments \\ \hline
&&$S=4, T=2$ \\
    Family & $\left\{n, {\frac{N}{3}-n}, {n+\frac{2 N}{3}}, {N-n}; {n+\frac{N}{3}}, {\frac{2 N}{3}-n}\right\}\,, $ & $3 \divides N \land N\geq 6$\\
   1  & $n=1, \ldots, \lfloor N/6 \rfloor$ & Total Sextuplets: $\lfloor N/6 \rfloor$.\\ \hline
&&$S=4, T=2$ \\
    & $\left\{n, {\frac{N}{3}-3n}, 3n, {n+\frac{2 N}{3}}; {3n+\frac{N}{3}}, {\frac{2 N}{3}-n}\right\}\,, $ & $3 \divides N \land N\geq 12$\\
 Family &&Complex-conjugate pairs: \\
  2   & $\left\{{N-n}, {3n +\frac{2 N}{3}}, {N-3n}, {\frac{N}{3}-n};  {\frac{2 N}{3}-3n}, {n+\frac{N}{3}}\right\}\,,
$& $k_j \to N-k_j, \,\,j=1, \ldots, 6$.\\
     & $n = 1, \ldots, \lceil N/9 \rceil - 1$& Total Sextuplets: $2\lceil N/9 \rceil - 2$.\\ \hline
&&$S=3, T=3$ \\
    & $\left\{n, {\frac{2N}{3}-3n}, {n+\frac{2 N}{3}}; {\frac{2 N}{3}-n}, N-3n, {3n+\frac{2N}{3}}\right\}\,, $ & $3 \divides N \land N\geq 15$\\
 Family &&Complex-conjugate pairs: \\
  3   & $\left\{{N-n}, {3n + \frac{N}{3}}, {\frac{N}{3}-n};  {n+\frac{N}{3}}, 3n, {\frac{N}{3}-3n}\right\}\,,
$& $k_j \to N-k_j, \,\,j=1, \ldots, 6$.\\
     & & Total Sextuplets: $2\lceil N/9 \rceil - 2$.\\
     &$n = 1, \ldots, \lceil N/9 \rceil - 1, \quad n\neq N/12$&(2 Sextuplets less if $12 \divides N$) \\\hline
&&$S=4, T=2$ \\
    & $\left\{{3n - \frac{N}{3}}, {\frac{N}{3}-n}, {3n + \frac{N}{3}}, N - n; 3n, {n+\frac{N}{3}}\right\}\,, $ & $3 \divides N \land N\geq 15 \land N \neq 18$\\
 Family &&Complex-conjugate pairs: \\
  4   & $\left\{{\frac{4N}{3}-3n}, {\frac{2N}{3}+n}, {\frac{2N}{3}-3n}, n; N-3n, {\frac{2N}{3}-n}\right\}\,,
$& $k_j \to N-k_j, \,\,j=1, \ldots, 6$.\\
     & & Total Sextuplets: \\
     &$n = \lfloor N/9 \rfloor + 1, \ldots, \lceil N/6 \rceil - 1$&$2\lceil N/6 \rceil - 2\lfloor N/9 \rfloor - 2$.\\\hline
&&$S=3, T=3$ \\
    & $\left\{{3n - \frac{N}{3}}, {\frac{2N}{3}-n}, {3n + \frac{N}{3}}; n, 3n, {n+\frac{2N}{3}}\right\}\,, $ & $3 \divides N \land N\geq 15 \land N \neq 18$\\
 Family &&Complex-conjugate pairs: \\
   5  & $\left\{{\frac{4N}{3}-3n}, {\frac{N}{3}+n}, {\frac{2N}{3}-3n}; N-n, N-3n, {\frac{N}{3}-n}\right\}\,,
$& $k_j \to N-k_j, \,\,j=1, \ldots, 6$.\\
     & & Total Sextuplets: \\
     &$n = \lfloor N/9 \rfloor + 1, \ldots, \lceil N/6 \rceil - 1$&$2\lceil N/6 \rceil - 2\lfloor N/9 \rfloor - 2$.\\\hline  \end{tabular}
\end{center}
\caption{\label{tab:6-waves} Summary of $6$-wave resonances obtained via linear combinations of two real FPUT polynomials of length $3$.}
\end{table}

Regarding `complex conjugation' operation, i.e. $k_j \to N - k_j, \,\, j=1, \ldots, 6$ (related to pairing-off terms), notice that the sextuplets in Family 1 are invariant under this operation. The other families' sextuplets come in conjugate pairs.

Regarding the clustering of the sextuplets in these families,  we do not provide a thorough discussion but just mention the following: (i) By simple wavenumber counting, within a given family we expect very few connections in families 1, 4 and 5. (ii) By looking at their explicit parameterisations, it is evident that families 1, 2 and 3 can be grouped via 3-common-mode connections, and  families 4 and 5 can be grouped via 3-common-mode connections as well. Connections between these groups of families 1-2-3 and families 4-5 are not evident from the parameterisation but they occur too, and they range from 2- to 4-common-mode connections.

We consider some examples:

\begin{exa}
$N=6$ or $N=9$. In this case we have only one sextuplet, coming from Family 1. Explicitly, when $N=6$ we get the sextuplet $\{1,1,5,5;3,3\}$, already discussed in an earlier Section. The new case is $N=9$, which gives the sextuplet $\{1,2,7,8;4,5\}$, corresponding to the identity
$\sin(\pi/9) + \sin(2\pi/9) = \sin(4\pi/9)$. It is remarkable that an `incomplete resonant triad' (i.e., a triad where frequency resonance is satisfied but momentum condition is not) is always involved in these resonances.
\end{exa}

\begin{exa} $N = 12$. In this case Family 1 provides the sextuplets
$$n=1: \quad \left\{1, {3}, {9}, {11}; {5}, {7}\right\}\,,$$
$$n=2: \quad \left\{2, {2}, {10}, {10}; {6}, {6}\right\}\,.$$

Family 2 provides sextuplets in conjugate pairs:
$$\{1,1,3,9;7,7\} \qquad \mathrm{and} \qquad \{11,11,9,3;5,5\}\,.$$
\end{exa}

We now present and discuss the construction of exceptional sextuplets, made by combining cyclotomic polynomials of length 3 and length 5, in such a way that one monomial from each polynomial cancel out, resulting in a FPUT resonant polynomial of length 6.

Unlike the sextuplets obtained above, these exceptional sextuplets do not come in one-parameter families; rather, there is a fixed number of them depending on the divisibility of the number of particles $N$.

\subsection{Case 2: $15 \divides N$ and $2 \ndivides N$.} In this case, in addition to the five families of sextuplets summarised in table \ref{tab:6-waves}, there are four conjugate pairs of resonant sextuplets. The first two pairs convert $4$ to $2$ waves and the last two pairs convert $3$ to $3$ waves:
\begin{eqnarray}
\nonumber
\mkern-72mu \left\{{\frac{N}{15}}, {\frac{N}{15}}, {\frac{2 N}{15}}, {\frac{7 N}{15}}; {\frac{N}{3}}, {\frac{2 N}{5}}\right\} &\quad& \left\{{\frac{14 N}{15}}, {\frac{14 N}{15}}, {\frac{13 N}{15}}, {\frac{8 N}{15}}; {\frac{2 N}{3}}, {\frac{3 N}{5}}\right\}\\
\nonumber
\mkern-72mu \left\{{\frac{N}{15}}, {\frac{4 N}{5}}, {\frac{13 N}{15}}, {\frac{13 N}{15}}; {\frac{N}{3}}, {\frac{4 N}{15}}\right\} &\quad& \left\{{\frac{14 N}{15}}, {\frac{N}{5}}, {\frac{2 N}{15}}, {\frac{2 N}{15}}; {\frac{2 N}{3}}, {\frac{11 N}{15}}\right\}\\
\nonumber
\mkern-72mu  \left\{{\frac{N}{15}}, {\frac{8 N}{15}}, {\frac{8 N}{15}}; {\frac{N}{5}}, {\frac{2 N}{3}}, {\frac{4 N}{15}}\right\} &\quad& \left\{{\frac{14 N}{15}}, {\frac{7 N}{15}}, {\frac{7 N}{15}}; {\frac{N}{3}}, {\frac{4 N}{5}}, {\frac{11 N}{15}}\right\} \\
\label{eq:6_wave_Case_2}
\mkern-72mu \left\{{\frac{3 N}{5}}, {\frac{2 N}{15}}, {\frac{7 N}{15}}; {\frac{2 N}{3}}, {\frac{4 N}{15}}, {\frac{4 N}{15}}\right\} &\quad& \left\{{\frac{2 N}{5}}, {\frac{13 N}{15}}, {\frac{8 N}{15}}; {\frac{N}{3}}, {\frac{11 N}{15}}, {\frac{11 N}{15}}\right\}\,.
\end{eqnarray}

\subsection{Case 3: $30 \divides N$ and $4 \ndivides N$.} In this case the four conjugate pairs of sextuplets from equations (\ref{eq:6_wave_Case_2}) still apply, but there are additional sextuplets: one sextuplet that is invariant under complex conjugation and converts $4$  to $2$ waves: 
\begin{equation}
\label{eq:6_wave_Case_3_1}
\left\{{\frac{N}{10}}, {\frac{N}{6}}, {\frac{5 N}{6}}, {\frac{9 N}{10}}; {\frac{3 N}{10}}, {\frac{7 N}{10}}\right\},
\end{equation}
and seven conjugate pairs of sextuplets, of which three convert $4$ to $2$ waves and four convert $3$ to $3$ waves:
\begin{eqnarray}
\nonumber
\mkern-72mu \left\{{\frac{N}{30}}, {\frac{13 N}{30}}, {\frac{5 N}{6}}, {\frac{9 N}{10}}; {\frac{17 N}{30}}, {\frac{19 N}{30}}\right\} &\quad& \left\{{\frac{29 N}{30}}, {\frac{17 N}{30}}, {\frac{N}{6}}, {\frac{N}{10}}; {\frac{13 N}{30}}, {\frac{11 N}{30}}\right\}\\
\nonumber
\mkern-72mu \left\{{\frac{N}{30}}, {\frac{N}{6}}, {\frac{23 N}{30}}, {\frac{9 N}{10}}; {\frac{7 N}{30}}, {\frac{19 N}{30}}\right\} &\quad&\left\{{\frac{29 N}{30}}, {\frac{5 N}{6}}, {\frac{7 N}{30}}, {\frac{N}{10}}; {\frac{23 N}{30}}, {\frac{11 N}{30}}\right\}\\
\nonumber
\mkern-72mu \left\{{\frac{N}{30}}, {\frac{N}{2}}, {\frac{9 N}{10}}, {\frac{9 N}{10}}; {\frac{19 N}{30}}, {\frac{7 N}{10}}\right\} &\quad& \left\{{\frac{29 N}{30}}, {\frac{N}{2}}, {\frac{N}{10}}, {\frac{N}{10}}; {\frac{11 N}{30}}, {\frac{3 N}{10}}\right\}\\
\nonumber
\mkern-72mu \left\{{\frac{N}{30}}, {\frac{13 N}{30}},{\frac{N}{2}}; {\frac{11 N}{30}}, {\frac{23 N}{30}}, {\frac{5 N}{6}}\right\} &\quad&\left\{{\frac{29 N}{30}}, {\frac{17 N}{30}}, {\frac{N}{2}}; {\frac{19 N}{30}}, {\frac{7 N}{30}}, {\frac{N}{6}}\right\}\\
\nonumber
\mkern-72mu \left\{{\frac{N}{30}}, {\frac{7 N}{30}}, {\frac{3 N}{10}}; {\frac{N}{6}}, {\frac{13 N}{30}}, {\frac{29 N}{30}}\right\} &\quad& \left\{{\frac{29 N}{30}}, {\frac{23 N}{30}}, {\frac{7 N}{10}}; {\frac{5 N}{6}}, {\frac{17 N}{30}}, {\frac{N}{30}}\right\}\\
\nonumber
\mkern-72mu \left\{{\frac{N}{10}}, {\frac{N}{2}}, {\frac{17 N}{30}}; {\frac{7 N}{10}}, {\frac{7 N}{10}}, {\frac{23 N}{30}}\right\} &\quad& \left\{{\frac{9 N}{10}}, {\frac{N}{2}}, {\frac{13 N}{30}}; {\frac{3 N}{10}}, {\frac{3 N}{10}}, {\frac{7 N}{30}}\right\}\\
\label{eq:6_wave_Case_3_2}
\mkern-72mu \left\{{\frac{N}{6}}, {\frac{11 N}{30}}, {\frac{17 N}{30}}; {\frac{19 N}{30}}, {\frac{7 N}{10}}, {\frac{23 N}{30}}\right\} &\quad& \left\{{\frac{5 N}{6}}, {\frac{19 N}{30}}, {\frac{13 N}{30}}; {\frac{11 N}{30}}, {\frac{3 N}{10}}, {\frac{7 N}{30}}\right\}\,.
\end{eqnarray}

\subsection{Case 4: $60 \divides N$.} In this case all sextuplets from equations (\ref{eq:6_wave_Case_2}), (\ref{eq:6_wave_Case_3_1}) and (\ref{eq:6_wave_Case_3_2}) still apply, but there are four additional conjugate pairs of sextuplets, all converting $3$ to $3$ waves: 
\begin{eqnarray}
\nonumber
\mkern-72mu \left\{{\frac{N}{60}}, {\frac{41 N}{60}}, {\frac{3 N}{4}}; {\frac{N}{20}}, {\frac{11 N}{20}}, {\frac{17 N}{20}}\right\} &\quad& \left\{{\frac{59 N}{60}}, {\frac{19 N}{60}}, {\frac{N}{4}}; {\frac{19 N}{20}}, {\frac{9 N}{20}}, {\frac{3 N}{20}}\right\}\\
\nonumber
\mkern-72mu \left\{{\frac{N}{20}}, {\frac{29 N}{60}}, {\frac{11 N}{20}}; {\frac{11 N}{60}}, {\frac{N}{4}}, {\frac{13 N}{20}}\right\} &\quad& \left\{{\frac{19 N}{20}}, {\frac{31 N}{60}}, {\frac{9 N}{20}}; {\frac{49 N}{60}}, {\frac{3 N}{4}}, {\frac{7 N}{20}}\right\}\\
\nonumber
\mkern-72mu \left\{{\frac{7 N}{60}}, {\frac{31 N}{60}}, {\frac{7 N}{12}}; {\frac{N}{4}}, {\frac{17 N}{60}}, {\frac{41 N}{60}}\right\} &\quad& \left\{{\frac{53 N}{60}}, {\frac{29 N}{60}}, {\frac{5 N}{12}}; {\frac{3 N}{4}}, {\frac{43 N}{60}}, {\frac{19 N}{60}}\right\}\\
\label{eq:6_wave_Case_4}
\mkern-72mu \left\{{\frac{3 N}{20}}, {\frac{11 N}{20}}, {\frac{37 N}{60}}; {\frac{N}{4}}, {\frac{7 N}{20}}, {\frac{43 N}{60}}\right\} &\quad& \left\{{\frac{17 N}{20}}, {\frac{9 N}{20}}, {\frac{23 N}{60}}; {\frac{3 N}{4}}, {\frac{13 N}{20}}, {\frac{17 N}{60}}\right\}\,.
\end{eqnarray}

\section{Basis for vanishing sums: 
Proof of Theorem \ref{thm:ker}}
\label{app:ker}

The approach here is motivated by the arguments of Lam and Leung  \cite{lam2000vanishing}.  
An additive group $A$ is \emph{free} with basis $E=\{ e_1,\ldots, e_r\}\subset A$ if every element 
in $A$ is uniquely expressible as a sum $\sum_{i=1}^rm_ie_i$ with $m_i\in \mathbb{Z}$. Here $r$ is the 
\emph{rank} of $A$. If $E$ decomposes as a disjoint union $E_1\cup E_2$ then $A=A_1\oplus A_2$ where 
each $A_i$ is free with basis $E_i$.

If $f(x)\in \mathbb{Z}[x]$ is a polynomial with integer coefficients then we can form the 
ring $\mathbb{Z}[x]/\langle f(x) \rangle$ whose elements are polynomials with integer coefficients 
and we identify the polynomials $h(x)$ and $g(x)$ whenever $h(x)=g(x) + f(x)m(x)$ for some 
$m(x)\in \mathbb{Z}[x]$. 

If $f(x)$ is \emph{monic} of degree $d$, then $\mathbb{Z}[x]/\langle f(x) \rangle $ is easily seen to 
be a free additive group with basis $\{ 1,x,\ldots, x^{d-1}\}$.

Now for any $n\geq 1$ we let $C_n$ denote the subring of $\mathbb{C}$ consisting of all 
polynomials in $\zeta_n:=e^{2\pi i/n}$. Then the natural map 
$\mathbb{Z}[x]/\langle \Phi_n(x)\rangle \to C_n$ sending $p(x)$ to $p(\zeta_n)$ is an isomorphism 
of rings: it sends the basis $\{ 1,x, \ldots, x^{\phi(n)-1}\}$ to the basis 
$\{ 1,\zeta_n,\ldots, \zeta_n^{\phi(n)-1}\}$. (See Theorem \ref{thm:cyc} (i).)

For any $n\geq 1$ we let $P_n$ denote the ring $\mathbb{Z}[x]/\langle x^n-1 \rangle$ and we let 
$D_n= \mathbb{Z}[x]/\langle x^n+1 \rangle$. Note that $\{ 1,x,\ldots, x^{n-1}\}$ is a basis in each 
case and that $D_n \cong P_n$ (via $x\mapsto -x$) when $n$ is odd. Furthermore, 
$D_{2^n}=C_{2^{n+1}}$ since $\Phi_{2^{n+1}}(x)=x^{2^n}+1$.

A subgroup $B$ of a free group $A$ is always free, and $\mathrm{rank}(B)\leq \mathrm{rank}(A)$,
 but -- unlike the situation for vector spaces -- 
$B$ is not usually a direct summand of $A$. For example, if $A$ is free with basis 
$\{ e_1,\ldots, e_r\}$ then $B:=2A=\{ 2a\ | a\in A\}$ is a free subgroup with basis 
$\{ 2e_1,\ldots, 2e_r\}$, but is not a direct summand. 

However, we do have the following:

\begin{lem} \label{lem:free}
Let $\phi:A\to B$ be a surjective homomorphism of free additive groups. 
Then there exists a submodule $Q$ such that $A=\ker(\phi)\oplus Q$.  
\end{lem}
\emph{Proof.} Let $\{ e_1,\ldots, e_s\}$  be a basis of $\ker(\phi)$. Let $\{ f_1,\ldots, f_t\}$ be a basis 
of $B$. Let $e_{s+1},\ldots, e_{s+t}$ be any elements of $A$ satisfying $\phi(e_{s+j})=f_j$ for 
$1\leq j\leq t$.  

Then $E=\{ e_1,\ldots, e_{s+t}\}$ is a basis of $A$: Let $a\in A$. Then 
$\phi(a)=\sum_{j=1}^tm_jf_j$ for some (unique) $m_j\in \mathbb{Z}$. So 
$\phi(a)= \phi(\sum_{j=1}^tm_je_{s+j})$ and thus $a-\sum_{j=1}^tm_je_{s+j}\in \ker(\phi)$. Hence 
$a-\sum_{j=1}^tm_je_{s+j}=\sum_{i=1}^sn_ie_i$ for some (unique) $n_i\in \mathbb{Z}$ and so 
$a= \sum_{i=1}^sn_ie_i+ \sum_{j=1}^tm_je_{s+j}$. 

It follows that $A=\ker(\phi)\oplus Q$ where $Q$ is free with basis $\{ e_{s+1},\ldots, e_{s+t}\}$. \qed

\begin{lem}\label{lem:pin}
Let $\pi_n$ denote the surjective homomorphism $P_n\to C_n$ sending $p(x)$ to 
$p(\zeta_n)$. Then \\
(i) $\ker(\pi_n)$ has basis $\{ x^i\Phi_n(x)\}_{i=0}^{n-\phi(n)-1}$. \\
(ii) $P_n=\ker(\pi_n)\oplus Q_n$ where $Q_n$ has basis $\{ 1,x,\ldots, x^{\phi(n)-1}\}$. 
\end{lem}
\emph{Proof}\\
(i)  Let $p(x)\in \ker(\pi_n)$. We can assume that $p(x)$ has degree at most $n-1$. 
Then $p(\zeta_n)=0$ and hence $p(x)=\Phi_n(x)r(x)$ for some $r(x)\in \mathbb{Z}[x]$. Clearly 
$r(x)$ has degree at most $n-\phi(n)-1$. So $r(x)=\sum_{i=1}^{n-\phi(n)-1}m_ix^i$ and hence 
$p(x)=\sum_{i=1}^{n-\phi(n)-1}m_ix^i\Phi_n(x)$. This expression is unique since the degrees of the 
$x^i\Phi_n(x)\in P_n$ are all distinct. \\
(ii) Since $\pi_n(x^i)=\zeta_n^i$ for $i\leq \phi(n)-1$, the proof of Lemma \ref{lem:free} 
implies the result. \qed

For any additive groups $A$ and $B$, the tensor product $A\otimes B$ is defined. If $A$ and 
$B$ are rings, so is $A\otimes B$. 

If $A$ and $B$ are free additive groups with bases $\{ e_1,\ldots, e_r\}$ and $\{ f_1,\dots f_t\}$ 
respectively, then   $A\otimes B$ is the free additive group with basis $\{ e_i\otimes f_j\}$. 
If $a=\sum_im_ie_i\in A$ and $b=\sum_jn_jf_j\in B$ then $a\otimes b$ denotes the element 
$\sum_{i,j}m_in_j(e_i\otimes f_j)\in A\otimes B$.  In particular, $A\otimes B$ is free 
of rank $\mathrm{rank}(A)\cdot\mathrm{rank}(B)$. 

If $\phi:A\to A'$ and $\psi:B\to B'$ are homomorphisms (of groups or rings), there is a well-defined 
homomorphism $\phi\otimes \psi :A\otimes B \to A'\otimes B'$ sending $a\otimes b$ to 
$\phi(a)\otimes \psi(b)$. 

\begin{prop}\label{prop:tensor} Let $n,m\geq 1$ be  \emph{relatively prime} integers. 
Then the following are isomorphisms of rings:\\
(i) $P_n\otimes P_m\to P_{nm}$, $p(x)\otimes q(x)\to p(x^m)q(x^n)$\\
(ii)  $D_n\otimes D_m\to D_{nm}$, $p(x)\otimes q(x)\to p(x^m)q(x^n)$\\
(iii) $C_n\otimes C_m \to C_{nm}$, $p(\zeta_n)\otimes q(\zeta_m)\to p(\zeta_{nm}^m)q(\zeta_{nm}^n)$. 
\end{prop}
\emph{Proof.} One easily verifies that the maps described (when extended linearly to any element 
of the domain) are ring homomorphisms.\\
(i) This map sends the basis elements $x^i\otimes x^j$ of $P_n\otimes P_m$ to the elements 
$x^{mi+nj}$ of $P_{nm}$. In $P_{nm}$, $x^{mi+nj}=x^r$ where $r< nm-1$ and $r$ 
is the remainder of $mi+nj$ on division by $mn$. 

Suppose that $mi+nj=mi'+nj'$ for some 
$0\leq i,i'\leq n-1$, $0\leq j,j'\leq m-1$. Then $m(i-i')=n(j'-j)$. Thus $n$ divides 
$m(i-i')$ and since $\mathrm{gcd}(n,m)=1$ it follows that $n$ divides $i-i'$. Since 
$0\leq i,i'\leq n-1$ this forces $i=i'$. Similarly $j=j'$. 

So the induced map from the basis $\{ x^i\otimes x^j\}$ of $P_n\otimes P_m$ to the basis 
$\{ x^r\ |\ 0\leq r\leq mn-1\}$ is injective and hence bijective. It follows that our 
ring homomorphism is an isomorphism.\\
(ii) An almost identical argument applies in this case: the only difference is that 
$x^{mi+nj}=\pm x^r$ in $D_{nm}$ (the sign depending on the parity of $\lfloor \frac{mi+nj}{mn}
\rfloor$). \\
(iii) This is a less obvious fact, but is well known from the theory of cyclotomic integers in number 
theory. For a proof, see for example [CITE Marcus, Number Fields, p36]. \qed

Combining this result with an induction argument we deduce:
\begin{cor}\label{cor:tensor} Let $n=p_1^{a_1}\cdots p_t^{a_t}$ where $p_1<\cdots <p_t$ are primes. 
Let $n_i=n/p_i^{a_i}$ for $1\leq i\leq t$.  Then there are ring isomorphisms
\begin{enumerate}
\item $P_{p_1^{a_i}}\otimes \cdots \otimes P_{p_t^{a_t}}\cong P_n$, 
$p_1(x)\otimes\cdots \otimes p_t(x)\mapsto p_1(x^{n_1})\cdots p_t(x^{n_t})$. 
 \item $D_{p_1^{a_i}}\otimes \cdots \otimes D_{p_t^{a_t}}\cong D_n$, 
$p_1(x)\otimes\cdots \otimes p_t(x)\mapsto p_1(x^{n_1})\cdots p_t(x^{n_t})$.
\item $C_{p_1^{a_i}}\otimes \cdots \otimes C_{p_t^{a_t}}\cong C_n$, 
$p_1(\zeta_{p_1^{a_1}})\otimes\cdots \otimes p_t(\zeta_{p_t^{a_t}})\mapsto 
p_1(\zeta_n^{n_1})\cdots p_t(\zeta_n^{n_t})$. 
\end{enumerate}
\end{cor}

Now let $n=2N=2^{b_0}p_1^{b_1}\cdots p_t^{b_t}$ where $2<p_1<\cdots <p_t$ are primes (and 
$b_0\geq1$, of course). Thus $D_N=\mathbb{Z}[x]/\langle x^N+1\rangle$ is the ring of 
polynomials modulo the relation $x^k=(-1)^tx^r$ where $k=tN+r$. 

Our goal is to calculate a basis for the kernel of the homomorphism 
$\rho_n:D_N\to C_n$ sending $p(x)$ to $p(\zeta)$ where $\zeta=\zeta_n=e^{\pi i/N}$.

By Corollary \ref{cor:tensor} we have tensor decompositions 
\begin{equation}\nonumber
D_N\cong D_{2^{b_0-1}}\otimes P_{p_1^{b_1}}\otimes \cdots \otimes P_{p_t^{b_t}}
\end{equation} 
and
\begin{equation}\nonumber
C_n\cong C_{2^{b_0}}\otimes C_{p_1^{b_1}}\otimes \cdots \otimes C_{p_t^{b_t}}.
\end{equation}

With these identifications the map $\rho_n:D_N\to C_n$ becomes the map 
\begin{equation}\nonumber
D_{2^{b_0-1}}\otimes P_{p_1^{b_1}}\otimes \cdots \otimes P_{p_t^{b_t}}\to 
C_{2^{b_0}}\otimes C_{p_1^{b_1}}\otimes \cdots \otimes C_{p_t^{b_t}}
\end{equation}
given by the formula
\begin{equation}\nonumber
p_0(x)\otimes p_1(x)\otimes \cdots \otimes p_t(x)\mapsto 
p_0(\zeta_{2^{b_0}})\otimes p_1(-\zeta_{p_1^{b_1}})\otimes \cdots \otimes p_t(-\zeta_{p_t^{b_t}}).
\end{equation}
Up to a sign, which does not affect the kernel, this is just the map 
$\rho_{2^{b_0}}\otimes \pi_{p_1^{b_1}}\otimes \cdots\otimes \pi_{p_t^{b_t}}$. 

\begin{lem}\label{lem:tensorkernel}
Let $\phi_1:A_1\to B_1$ and $\phi_2:A_2\to B_2$ be surjective homomorphisms of free additive 
groups. Suppose that $A_i=\ker(\phi_i)\oplus Q_i$ for $i=1,2$. Then 
\begin{equation}\nonumber
\ker(\phi_1\otimes \phi_2)= \left(\ker(\phi_1)\otimes Q_2\right) \oplus 
\left(A_1\otimes \ker(\phi_2)\right)
\end{equation}
and 
\begin{equation}\nonumber
A_1\otimes A_2= \ker(\phi_1\otimes \phi_2)\oplus (Q_1\otimes Q_2).
\end{equation}
\end{lem}
\emph{Proof.}\ Let $r_i=\mathrm{rank}(A_i)$, $s_i=\mathrm{rank}(B_i)$ and 
$t_i=r_i-s_i=\mathrm{rank}\ker(\phi_i)$ for $i=1,2$. Let $\{ e_1,\ldots, e_{t_1}\}$ and
$\{ f_1,\ldots, f_{t_2}\}$ be bases of $\ker(\phi_1)$ and $\ker(\phi_2)$ respectively. 
Let $\{ e_{t_1+1},\ldots,e_{r_1}\}$ and $\{ f_{t_2+1},\ldots, f_{r_2}\}$ be bases of $Q_1$ and $Q_2$ 
respectively. 

Let 
\begin{equation}\nonumber
x= \sum_{i=1}^{r_1}\sum_{j=1}^{r_2}m_{ij}(e_i\otimes f_j)\in \ker(\phi_1\otimes \phi_2).
\end{equation}   

Then 
\begin{equation}\nonumber
y:= \sum_{i=1}^{r_1}\sum_{j=1}^{t_2}m_{ij}(e_i\otimes f_j)\in \ker(\phi_1\otimes \phi_2)
\end{equation}
since $f_j\in \ker(\phi_2)$ for $j\leq t_2$. Note that $y\in A_1\otimes \ker(\phi_2)$.

Thus 
\begin{equation}\nonumber
z=x-y = \sum_{i=1}^{r_1}\sum_{j=t_2+1}^{r_2}m_{ij}(e_i\otimes f_j)
\end{equation}
lies in $\ker(\phi_1\otimes \phi_2)$. We rewrite this as 
\begin{equation}\nonumber
z=\sum_{j=t_2+1}^{r_2}w_{j}\otimes f_j
\end{equation}
where $w_{j}=\sum_{i=1}^{r_1}m_{ij}e_i\in A_1$ for $j\geq t_2+1$. Thus 
\begin{equation}\nonumber
0=(\phi_1\otimes \phi_2)(z)= \sum_{j=t_2+1}^{r_2}\phi_1(w_j)\otimes \phi_2(f_j).
\end{equation}
Since $\{ \phi_2(f_j)\ |\ j=t_2+1,\ldots, r_2\}$ is a basis of $B_2$, it follows that 
$\phi(w_j)=0$ for all $j\geq t_2+1$. Thus $w_j\in \ker(\phi_1)$ for $j\geq t_2+1$. 
It follows that $w_j=\sum_{i=1}^{t_1}m_{ij}e_i\otimes f_j$ and that 
$z\in \ker(\phi_1)\otimes Q_2$. This proves the first statement.

The second statement follows immediately from the proof of Lemma \ref{lem:free} since 
$\{ (\phi_1\otimes \phi_2)(e_i\otimes f_j)\}_{i>t_1,j>t_2}
=\{ \phi_1(e_i)\otimes \phi_2(f_j)\}_{i>t_1,j>t_2}$ 
is a basis of $B_1\otimes B_2$. \qed\\

Combining this lemma with a straightforward induction argument we deduce:

\begin{cor}\label{cor:tensorkernel}
Let $\phi_i:A_i\to B_i$ be surjective homomorphisms of free additive groups, for $i=1,\ldots, t$. 
Suppose that $A_i=\ker(\phi_1)\oplus Q_i$ for each $i$. 

Then $\ker(\phi_1\otimes \cdots \otimes \phi_t)= \oplus_{i=1}^t K_i$ where 
\begin{equation}\nonumber
K_i= A_1\otimes \cdots \otimes A_{i-1}\otimes \ker(\phi_i)\otimes Q_{i+1}\otimes \cdots Q_t.
\end{equation} 

Thus let $F_i$ be a basis of $A_i$ for each $i$ and suppose 
that $F_i= \mathcal{K}_i\cup G_i$ where 
$\mathcal{K}_i$ is a basis of $\ker(\phi_i)$ and $G_i$ is a basis of $Q_i$. Then 
$E=E_1\cup \cdots \cup E_t$ is a basis of $\ker(\phi_1\otimes \cdots \otimes \phi_t)$ where 
\begin{equation}\nonumber
E_i=\left\{ x_1\otimes \cdots \otimes x_t \ |\   
x_j\in \left[ 
\begin{array}{ll}
E_j,& j < i\\
\mathcal{K}_i,&j=i \\
 G_j,& i < j\leq t \\
\end{array}
\right.\right\}\,.
\end{equation}
\end{cor}

Combining all of this we get:\\[10pt]

\noindent \emph{Proof of Theorem \ref{thm:ker}.}  Under the isomorphism 
\begin{equation}\nonumber
D_{2^{b_0-1}}\otimes P_{p_1^{b_1}}\otimes \cdots \otimes P_{p_t^{b_t}}\cong D_N
\end{equation}
the kernel of $\rho_{2N}:D_N\to C_{2N}$  corresponds to 
$\ker(\rho_{2^{b_0}}\otimes \pi_{p_1^{b_1}}\otimes \cdots \otimes \pi_{p_t^{b_t}})$.

Since $\rho_{2^{b_0}}:D_{2^{b_0-1}}\to C_{2^{b_0}}$ is an isomorphism, by 
Corollary \ref{cor:tensorkernel} this kernel is $\oplus_{i=1}^t K_i$ where 
\begin{equation}\nonumber
K_i= D_{2^{b_0-1}}\otimes P_{p_1^{b_1}}\otimes \cdots \otimes P_{p_{i-1}^{b_{i-1}}} \otimes 
\ker(\pi_{p_i^{b_i}})\otimes Q_{p_{i+1}^{b_{i+1}}}\otimes \cdots \otimes Q_{p_t^{b_t}}
\end{equation}
where $Q_{p^a}$ is as described in Lemma \ref{lem:pin}. 

A basis of $D_{2^{b-1}}$ is $\{ x^i\ |\ 0\leq i\leq 2^{b-1}-1\}$. A basis of 
$P_{p^b}$ is $\{ x^i\ |\ 0\leq i\leq p^b-1\}$. 

By Lemma \ref{lem:pin}, a basis for $\ker(\pi_{p^b})$ is 
$\{ x^i\Phi_{p^b}(x)\ |\ i=0, \ldots, p^b-\phi(p^b)-1= p^{b-1}-1\}$ and a basis for 
$Q_{p^b}$ is $\{ x^i\ |\ i=0,\ldots, \phi(p^b)-1=p^b-p^{b-1}-1\}$. 

Theorem \ref{thm:ker} now follows from Corollary \ref{cor:tensorkernel}.  \qed


\bibliographystyle{unsrt}

\bibliography{references}{}

\end{document}